\documentclass[useAMS,usenatbib]{mn2e}

\usepackage{graphicx,bm,amsmath,multirow,subfigure,longtable,float}

\usepackage[T1]{fontenc}
\usepackage{lmodern}
\newcommand{\Msol}{\mbox{$M_{\sun}$}}
\newcommand{\Msun}{\mbox{$M_{\sun}$}}

\newcommand{\tr}{\mbox{$\mathrm{tr}$}}

\title[Weak lensing analysis of SZ-selected clusters]{Weak lensing analysis of SZ-selected clusters of galaxies from the SPT and \emph{Planck} surveys}
\author[D. Gruen et al.]
{\parbox{\textwidth}{D. Gruen$^{1,2}$\thanks{E-mail:
dgruen@usm.uni-muenchen.de (DG)},
S. Seitz$^{1,2}$,
F. Brimioulle$^{1,2,3}$,
R. Kosyra$^{1,2}$,
J. Koppenhoefer$^{2,1}$,
C.-H. Lee$^{2,1}$,
R. Bender$^{1,2}$,
A. Riffeser$^{1,2}$,
T. Eichner$^{1,2}$,
T. Weidinger$^{1,2}$ and
M. Bierschenk$^{1,2}$}\vspace{0.4cm}\\
\parbox{\textwidth}{$^{1}$University Observatory Munich, Scheinerstrasse 1, 81679 Munich, Germany\\
$^{2}$Max Planck Institute for Extraterrestrial Physics, Giessenbachstrasse, 85748 Garching, Germany\\
$^{3}$Observat\'{o}rio Nacional, Rua Gal. Jos\'{e} Cristino, 20921-400, Rio de Janeiro, Brazil}}
\begin{document}

\date{}

\pagerange{\pageref{firstpage}--\pageref{lastpage}} \pubyear{2013}

\maketitle

\label{firstpage}

\begin{abstract}
We present the weak lensing analysis of the \textbf{W}ide-Field \textbf{I}mager \textbf{S}Z \textbf{C}luster of galax\textbf{y} (WISCy) sample, a set of 12 clusters of galaxies selected for their Sunyaev-Zel'dovich (SZ) effect. After developing new and improved methods for background selection and determination of geometric lensing scaling factors from absolute multi-band photometry in cluster fields, we compare the weak lensing mass estimate with public X-ray and SZ data. We find consistency with hydrostatic X-ray masses with no significant bias, no mass dependent bias and less than 20\% intrinsic scatter and constrain $f_{\mathrm{gas},500c}=0.128^{+0.029}_{-0.023}$. We independently calibrate the South Pole Telescope significance-mass relation and find consistency with previous results. The comparison of weak lensing mass and Planck Compton parameters, whether extracted self-consistently with a mass-observable relation (MOR) or using X-ray prior information on cluster size, shows significant discrepancies. The deviations from the MOR strongly correlate with cluster mass and redshift. This could be explained either by a significantly shallower than expected slope of Compton decrement versus mass and a corresponding problem in the previous X-ray based mass calibration, or a size or redshift dependent bias in SZ signal extraction.
\end{abstract}

\begin{keywords}
gravitational lensing: weak -- galaxies: clusters: general -- cosmology: observations
\end{keywords}

\section{Introduction}

As the end product of hierarchical structure formation, clusters of galaxies are particularly sensitive to the cosmological interplay of dark matter and dark energy. Studies of individual clusters and, even more so, large surveys have for this reason been considered a valuable cosmological probe for several decades (see \citealt{2011ARA&A..49..409A} for a recent review).

The framework for cosmological interpretation of cluster surveys consists, on the theoretical side, of a halo mass function that predicts the dependence of the number density of clusters as a function of mass and redshift on cosmology \citep[e.g.][]{1974ApJ...187..425P,1999MNRAS.308..119S,2008ApJ...688..709T}. The observational task consists in providing an ensemble of clusters detected with a well-determined selection function and measurements of an observable that can be related to their mass. 

Most observations that allow a sufficiently high signal-to-noise ratio (S/N) detection of sufficiently many clusters to date relate to the minority of cluster matter that is of baryonic origin (but see \citealt{2007A&A...462..459G,2007ApJ...669..714M,2007A&A...462..875S} for lensing-detected surveys). In particular, use has been made of the density of red galaxies (e.g. \citealt{2005ApJS..157....1G}, \citealt{2007ApJ...660..239K} or \citealt{2013arXiv1303.3562R} for observations and \citealt{2010ApJ...708..645R} or \citealt{2013MNRAS.434..684M} for cosmological interpretation) or the hot gas in the intra-cluster medium (ICM) that can be detected by its X-ray emission (e.g. \citealt{2010arXiv1007.1916P} for a meta-catalogue and \citealt{2009ApJ...692.1060V} or \citealt{2010MNRAS.406.1759M} for cosmological interpretation).

Another observable effect is due to the inverse Compton scattering of cosmic microwave background (CMB) photons by the ICM, the  \citet[][hereafter SZ]{1972CoASP...4..173S} effect. The scattering distorts the CMB spectrum such that below (above) a global null-point frequency of approximately 220~GHz, a decrease (an increase) in microwave flux density is observed in galaxy clusters. The effect at any point scales with the integrated electron pressure $P$ along the line of sight, which defines the dimensionless Compton parameter $y$,
\begin{equation}
 y=\frac{\sigma_{\rm T}}{m_{\rm e} c^2}\int P\; dl \; ,
\end{equation}
where $\sigma_T$ is the Thomson cross-section and $m_{\rm e} c^2$ the rest energy of electrons.

Integration of $y$ over the angular extent of the cluster, $Y=\int y \; d\Omega$, yields a volume integral of electron pressure,
\begin{equation}
 D_{\rm A}^2 Y=\frac{\sigma_{\rm T}}{m_{\rm e} c^2}\int P\; dV \; \,
\end{equation}
with the angular diameter distance $D_{\rm A}$ used to convert apparent angles to physical scales. Note that the volume integral of pressure equals the thermal energy and is therefore expected to be closely related to cluster mass.

Large surveys of the SZ sky have been and are currently being performed by the \emph{Planck} Satellite (e.g. \citealt{2013arXiv1303.5089P}), the South Pole Telescope (SPT; \citealt{2011PASP..123..568C}) and the Atacama Cosmology Telescope (ACT; e.g. \citealt{2011ApJ...737...61M}).

\subsection{Calibration of the SZ mass-observable relation}

As a connection between cosmological models and SZ surveys, it is necessary to establish a mass-observable relation (MOR) between SZ observable and cluster mass. As important as the mean relation is the intrinsic scatter of the MOR \citep{2005PhRvD..72d3006L}, since the steepness of the halo mass function causes preferential up-scatter of the (more numerous) less massive haloes. There are a number of ways of achieving this calibration.

External mass calibration is not strictly required for the cosmological interpretation of an SZ survey, since large surveys can determine both cosmological parameters, a parametrized MOR and the intrinsic scatter simultaneously \citep{2003PhRvD..67h1304H,2004ApJ...613...41M,2005PhRvD..72d3006L} by requiring that cluster counts as a function of observable be consistent with the halo mass function of the respective cosmology. The fewer assumptions about the form and evolution of the MOR are made, however, the less well-constrained cosmological parameters become in such a scheme.

Previous studies have used astrophysical modelling (e.g. \citealt{2011ApJ...728L..35M}) or X-ray mass estimates for SZ-selected systems \citep{2011AuA...536A..11P,p2013cosmology} to constrain the MOR. These approaches require assumptions about the astrophysical state of clusters, e.g. virialization and hydrostatic equilibrium (HSE), and are complicated by the variety of evolutionary states clusters are in fact found to be in. One important example of this is the question of hydrostatic mass bias, i.e. a mean underestimation of true mass by X-ray analyses based on HSE \citep{2007ApJ...655...98N,2008A&A...491...71P}, which has been investigated with controversial results (e.g. \citealt{2008MNRAS.384.1567M,2010ApJ...711.1033Z,2013ApJ...767..116M,2013AuA...550A.129P}).

Weak lensing (WL) constraints on the MOR are a complement to these approaches. A moderate number of accurately measured masses greatly reduces the uncertainty of self-calibration schemes \citep[e.g.][]{2004ApJ...613...41M}. WL can measure masses and mass-observable scatters for samples of clusters selected according to the respective survey with the important advantage that it is sensitive to all matter regardless of its astrophysical state. 

In practice, however, WL also faces observational challenges. Biases in WL measurements of the mass due to, for instance, shape measurement bias (e.g. Young et al., in preparation), cluster orientation \citep{2007MNRAS.380..149C} or uncertain determination of source redshifts \citep{2012arXiv1208.0605A} have been explored. Increased uncertainty of observed mass due to unrelated projected structures \citep{2001A&A...370..743H,2003MNRAS.339.1155H,2012MNRAS.420.1384S} or deviations of individual systems from the common assumption of spherical, isolated \citet*[][hereafter NFW]{1997ApJ...490..493N} haloes \citep{2010arXiv1011.1681B,2011MNRAS.416.1392G} is an issue of similar importance. Despite the need of reducing and quantifying these effects, gravitational lensing remains the best candidate for an unbiased mass measurement of galaxy clusters to date. For this reason, the WL analysis of SZ selected samples of clusters has been the focus of a number of recent studies [cf. \citealt{2009MNRAS.399L..84M} (3 SPT systems), \citealt{2009ApJ...701L.114M,2012ApJ...754..119M} (a total of 29 systems with pointed SZ observations at $z=0.15\ldots0.3$, 25 of which are also detected by Planck), \citealt{2012ApJ...758...68H} (5 SPT systems), \citealt{2012MNRAS.419.2921A} (6 systems with pointed SZ observations) and \citealt{2012MNRAS.427.1298H} (a total of 30 systems with SZ observations, mostly at $z=0.15,\ldots,0.3$, 18 of which are also detected by \citealt{2011AuA...536A...8P} in the early data release)].

This work aims to be complementary to the aforementioned studies. Our sample of 12 clusters with an overlap of only one is a significant addition in terms of statistics to the present list of SZ clusters with WL measurements. We probe a wide range of $0.10<z<0.69$, extending to higher redshift than typical previous studies. Seven systems from our sample can be compared to the 2013 \emph{Planck} release, and this is the first study to compare the 2013 \emph{Planck} catalogue and MOR to independent measurements of cluster mass with lensing. Finally, we pay particular attention to the aspects of shape measurement calibration, background selection and modelling of neighbouring structures, improving upon methods commonly used to date and reducing potential biases resulting from the incomplete treatment of these effects.

This paper is structured as follows. Section~\ref{sec:sample} gives an overview of our sample and the data reduction procedure up to photometric catalogues. Section~\ref{sec:background} describes our background galaxy selection including an improved method based on multi-band photometry without photometric redshifts. Our methodology for the measurement and interpretation of the WL signal is laid out in Section~\ref{sec:shapemeas}. Our use of SZ data, including the calculation of self-consistent Planck SZ masses, is detailed in Section~\ref{sec:psz}. Section~\ref{sec:individual} contains individual analyses for each cluster of the Wide-Field Imager SZ Cluster of Galaxy (WISCy) sample. The main result of this work is presented in Section~\ref{sec:combined}, where we compare our WL measurements to the SZ observables and X-ray mass estimates. We conclude in Section~\ref{sec:conclusions}.

In this paper we adopt a $\Lambda$ cold dark matter ($\Lambda$CDM) cosmology with $H_0=70$~km s$^{-1}$ Mpc$^{-1}$ and $\Omega_{\rm m}=1-\Omega_{\Lambda}=0.3$. To accommodate different conventions used in the literature, we consistently mark published masses with $h=H_0/(100$~km s$^{-1}$ Mpc$^{-1})$ or $h_{X}=H_0/(X$~km s$^{-1}$ Mpc$^{-1})$. We denote the radii of spheres around the cluster centre with fixed overdensity as $r_{\Delta \rm m}$ and $r_{\Delta \rm c}$, where $\Delta$ is the overdensity factor of the sphere with respect to the mean matter density $\rho_{\rm m}$ or critical density $\rho_{\rm c}$ at the cluster redshift. The mass inside these spheres is labelled and defined correspondingly as $M_{\Delta \rm m}=\Delta\times\frac{4\pi}{3}r_{\Delta \rm m}^3\rho_{\rm m}$ and $M_{\Delta \rm c}=\Delta\times\frac{4\pi}{3}r_{\Delta \rm c}^3\rho_{\rm c}$. When comparing masses, we convert according to differences in $h$ but ignore small differences in $\Omega_{\rm m}$ (which would require to assume a density profile to be corrected). All magnitudes quoted in this work are given in the AB system. All maps and images assume a tangential coordinate system where north is up and east is left.

\section{Sample and Data}
\label{sec:sample}

Our sample contains 12 clusters of galaxies. Of these, five and seven are detected by SPT and \textit{Planck}, respectively, of which four and two are in fact discovered by their SZ signal in these surveys.

The set of clusters was selected from four parent samples. We selected four objects (SPT-CL J0551--5709, SPT-CL J0509--5342, SPT-CL J2332--5358, SPT-CL J2355--5056) from the detection-limited sample of 2008 SPT observations \citep{2010ApJ...722.1180V} based on their visibility in the observing time allocated to us at the 2.2m MPG/ESO telescope. By a similar selection, we added two systems (PLCKESZ G287.0+32.9, and PLCKESZ G292.5+22.0) from the Planck early SZ catalog \citep{2011AuA...536A...8P} and two known strong lensing systems (MACS J0416.1--2403 and RXC J2248.7--4431) from the Cluster Lensing And Supernova survey with Hubble (CLASH; \citealt{2012ApJS..199...25P}), for which we expected later SZ detection by Planck at the time. Note that both of these were recently selected as \emph{Hubble Space Telescope} (HST) Frontier Fields.\footnote{Observations for MACS J0416.1--2403 have already been performed while for RXC J2248.7--4431 they are scheduled for year 3 of the survey and contingent on results from the previously observed fields.} Finally, we added all systems detected by Planck in the 2013 catalogue \citep{2013arXiv1303.5089P} that were covered serendipitously by the Canada-France-Hawaii Telescope Legacy Survey (CFHTLS) in its final public data release (PSZ1 G168.02--59.95, PSZ1 G230.73+27.70, PSZ1 G099.84+58.45, PSZ1 G099.48+55.62).

The sample spans a wide dynamic range. In terms of mass it reaches from  $~1\times10^{14}\Msol$ to several $10^{15}\Msol$, in terms of redshift from the almost local Universe at $z\approx0.1$ close to the limit of feasible ground-based WL at $z\approx0.7$. Table~\ref{tbl:list} gives an overview of the sample.

\begin{table*}
\begin{center}
\begin{tabular}{|l|l|c|r|r|r|c|c|}
\hline
\# & SZ name & Assoc. names & $z$ & RA & Dec & WFI & CFHTLS \\ \hline
1 & SPT-CL J0509--5342 & ACT-CL J0509--5341 & 0.4626$^a$ & 05:09:21 & -53:42:18 & \textit{B\textbf{VR}I} & - \\ \hline
2 & SPT-CL J0551--5709 & - & 0.4230$^a$ & 05:51:36 & -57:09:22 & \textit{B\textbf{R}I} & - \\ \hline
3 & SPT-CL J2332--5358 & SCSO J233227--535827 & 0.4020$^b$ & 23:32:27 & -53:58:20 & \textit{B\textbf{R}I} & - \\ \hline
4 & SPT-CL J2355--5056 & - & 0.3196$^c$ & 23:55:49 & -50:56:13 & \textit{B\textbf{R}I} & - \\ \hline
5 & PLCKESZ G287.0+32.9 & - & 0.3900$^d$ & 11:50:51 & -28:04:09 & \textit{V\textbf{RI}} & - \\ \hline
6 & PLCKESZ G292.5+22.0 & - & 0.3000$^d$ & 12:01:00 & -39:51:35 & \textit{\textbf{RI}} & - \\ \hline
7 &- & MACS J0416.1--2403 & 0.3970$^e$ & 04:16:09 & -24:04:04 & \textit{\textbf{BVRI}} & - \\ \hline
\multirow{3}{*}{8} & \multirow{3}{*}{SPT-CL J2248--4431} & ACO S 1063 &  \multirow{3}{*}{0.3475$^f$} & \multirow{3}{*}{22:48:44} & \multirow{3}{*}{-44:31:48} & \multirow{3}{*}{\textit{UB\textbf{VRI}Z}} & \multirow{3}{*}{-} \\ &  & PLCKESZ G349.46-59.94 \\ & & RXC J2248.7--4431 \\ \hline
\multirow{3}{*}{9} & \multirow{3}{*}{PSZ1 G168.02--59.95} & ACO 329 & \multirow{3}{*}{0.1456$^g$} & \multirow{3}{*}{02:14:41} & \multirow{3}{*}{-04:33:22} & \multirow{3}{*}{-} & \multirow{3}{*}{\textit{ugr\textbf{i}z}} \\ & & RXC J0214.6--0433 \\ & & RCC J0214.6--0433 \\ \hline
\multirow{2}{*}{10} & \multirow{2}{*}{PSZ1 G230.73+27.70} & MaxBCG J135.43706-01.63946 & \multirow{2}{*}{0.2944$^h$} & \multirow{2}{*}{09:01:30} & \multirow{2}{*}{-01:39:18} & \multirow{2}{*}{-} & \multirow{2}{*}{\textit{ugr\textbf{i}z}} \\ & & XCC J0901.7-0138 \\ \hline
11 & PSZ1 G099.84+58.45 & SL2S J141447+544703 & 0.6900$^i$ & 14:14:47 & +54:47:04 & - & \textit{ugr\textbf{i}z}  \\ \hline
\multirow{2}{*}{12} & \multirow{2}{*}{PSZ1 G099.48+55.62} & ACO 1925 & \multirow{2}{*}{0.1051$^j$} & \multirow{2}{*}{14:28:26} & \multirow{2}{*}{+56:51:36} & \multirow{2}{*}{-} & \multirow{2}{*}{\textit{ugr\textbf{i}z}}  \\ & &  RXC J1428.4+5652 \\ \hline
\end{tabular}
\end{center}
\caption{Overview of WISCy sample. The first and second columns give an ID and the name given by the respective SZ survey, both to be used in the rest of this work. Additional names are shown in the third column. Available photometric bands are shown in the last two columns, marking bands used for shape measurement in bold print. Redshift references:  (a) \citealt{2010ApJ...723.1736H}, (b) \citealt{2012ApJ...761...22S}, (c) \citealt{2013ApJ...763..127R}, (d) \citealt{2011AuA...536A...9P}, (e) \citealt{2014ApJS..211...21E}, (f) \citealt{2004A&A...425..367B}, (g) Mirkazemi et al. (in preparation), (h) photometric redshift of \citet{2007ApJ...660..239K}, (i) median photometric redshift of 32 visually selected cluster member galaxies (this work), (j) \citealt{1999ApJS..125...35S}, citing \citealt{1991AISAO..31...91L}.}
\label{tbl:list}
\end{table*}

\subsection{Data reduction and photometry}
\label{sec:dr}
\label{sec:photometry}
For eight of the clusters, observations were made with the Wide-Field Imager (WFI; \citealt{1999Msngr..95...15B}) on the 2.2 m MPG/ESO telescope at La Silla. Seeing and photometric depth is typically best in \emph{R} band, although in some cases additional bands can be used for shape measurement of galaxies (see Table~\ref{tbl:list}). The raw images are de-biased, flat-fielded and bad pixels are masked in all bands and fringe patterns are corrected in the \emph{I} and \emph{Z} band using the \textsc{astro-wise}\footnote{\texttt{http://www.astro-wise.org/}} framework \citep{2007ASPC..376..491V}. Background subtraction, final astrometry and co-addition of suitable frames is done with custom scripts using \textsc{scamp}\footnote{\texttt{http://www.astromatic.net/software/scamp}} \citep{2006ASPC..351..112B} and \textsc{swarp}\footnote{\texttt{http://www.astromatic.net/software/swarp}} \citep{2002ASPC..281..228B}. For fields highly contaminated with bright star ghost images, we use the outlier masking method of \citet[][see their Fig.~8 for an example of an \emph{R} band stack of SPT-CL J0551--5709]{stacking} to remove artefacts from the stack.

Observations of the fields in photometric nights together with fields of standard stars are used to fix the photometric zero-points in \emph{R} band of all WFI clusters except PLCKESZ G287.0+32.9 and PLCKESZ G292.5+22.0. Due to the unavailability of standard star observations during the nights in which the latter two were observed, we use 2MASS \citep{2003tmc..book.....C} \emph{JHK} infrared magnitudes of stars in their field of view and the stellar colours of the \citet{1998PASP..110..863P} library to fix their \emph{R} and \emph{I} zeropoints. Comparison with \emph{V} band magnitudes of a single exposure of PLCKESZ G287.0+32 shows that the R-I colour is recovered correctly by this procedure. The same stellar locus method is also used to fix the zeropoints of the remaining bands of the WFI fields. This is verified against standard star observations in additional filters in the field of SPT-CL J2248--4431 \citep[][their Section~2]{2013MNRAS.tmp.1221G}. Extinction corrections of \citet{1998ApJ...500..525S} are applied consistently. We note that for the relatively low galactic latitude fields of PLCKESZ G287.0+32.9 and PLCKESZ G292.5+22.0, significant uncertainty (comparing the Galactic extinction models of \citealt{1998ApJ...500..525S} and \citealt{2011ApJ...737..103S}) and spatial variation of extinction of $\approx$~0.15~mag likely cause systematic offsets of our magnitudes in those fields.

For the four clusters in our sample covered by the CFHTLS, we use photometric redshifts from the \citet{fabrice} pipeline, re-run on the latest publicly available stacks.\footnote{cf. \citet{2012arXiv1210.8156E}, \texttt{http://www.cfhtlens.org/astronomers/data-store}} Since no public shape catalogue is complete in a region of sufficient diameter around the two clusters, we process these co-added images with our own shape pipeline (see Section~\ref{sec:shapemeas}).

The multi-band aperture photometry of the WFI and Canada-France-Hawaii Telescope (CFHT) fields is extracted using the procedure described in \citet{fabrice} and \citet{2013MNRAS.tmp.1221G}.

\section{Background selection}
\label{sec:background}
The overall lensing signal at a given lens redshift $z_d$ scales with a factor
\begin{equation}
\frac{D_d(z_d)D_{ds}(z_s,z_d)}{D_s(z_s)}=D_d(z_d)\times\beta(z_s,z_d) \, ,
\label{eqn:beta}
\end{equation}
where $D$ are angular diameter distances. The subscripts $d$ and $s$ denote deflector and source position and $ds$ the distance between the two.
Since lens redshifts are known accurately in the case of cluster lensing, we have factored out the dependence on source redshift in $\beta(z_s,z_d)$, which is zero for $z_s<z_d$, then rises steeply before it approaches the asymptotic value at $z_s\rightarrow\infty$. This fact requires an accurate estimation of $\beta$ for galaxies in the shape catalogue in order to select a suitable background sample and correctly scale the signal in the lensing analysis.

Precise photometric source redshifts can, when available, be simply inserted in the above equation to get the scaling of the lensing effect. For the CFHTLS fields and SPT-CL J2248--4431 the photometric information in 5 and 6 bands, respectively, allows fitting the redshifted galaxy spectral energy distribution (SED) of individual sources over a wide wavelength range. In these cases, we therefore use photometric redshifts as provided by \citet{2013MNRAS.tmp.1221G} and a re-run of the \citet{fabrice} pipeline on the latest CFHTLenS data reduction \citep{2012arXiv1210.8156E}.

The fewer bands exist, however, the less well determined any single object's redshift becomes. Limiting, for the purpose of testing this effect, the number of bands used for the template fitting in the case of SPT-CL J2248--4431, the field with the otherwise best wavelength coverage, we verify that estimates of $\beta$ are in fact biased when using simply the best fitted but noisy photometric $z_s$, and in particular when selecting by them. We note that this is a natural consequence of the non-linear propagation of errors from redshifts to geometric scaling factors in equation (\ref{eqn:beta}) (cf. \citealt{2012arXiv1208.0605A} and Section~\ref{sec:zcomp}).

\subsection{$\beta$ from Limited Photometric Information}
\label{sec:beta}
Even when photometric redshifts are not feasible, use of the full photometric information allows for an optimal background selection. While in some studies magnitude cuts in a single band have been used (cf., e.g., \citealt{2000A&A...355...23E,2010AuA...514A..88R,2012A&A...546A..79I}), the inclusion of additional bands can greatly improve the separability of foreground and background objects (see, for instance, \citealt{2012ApJ...758...68H} for a comparison of two-band versus three-band information). We will show that using not only the colour but also including the apparent magnitude of objects can be beneficial (see also Fig.~\ref{fig:bribeta} and discussion below).

We therefore develop a probabilistic method of calculating the appropriate $\beta$ factor to use for a galaxy with limited photometric information in this section. The basis of our method is the position of the galaxy in magnitude space, which we compare to a deeper reference catalogue with accurate redshift information. In this way we use not just the available magnitudes of galaxies in our cluster field, but also the empirical distribution of unavailable magnitudes for galaxies similar to them, to estimate the geometrical scaling of the WL signal. We note that this method shares some characteristics with the one described by \citet{2008MNRAS.390..118L} and \citet{2009MNRAS.396.2379C}.

The reference catalogue and basics of the method are explained in Sections~\ref{sec:refcat} and \ref{sec:betamagbasic}. In addition, we correct for contamination with cluster member galaxies (described in Section~\ref{sec:fcl}) and calculate the optimal minimal $\beta$ (Section~\ref{sec:minbeta}) above which galaxies should be used as sources in a WL analysis.

\subsubsection{Reference catalogue}

\label{sec:refcat}
\begin{figure}
\centering
\includegraphics[width=0.48\textwidth]{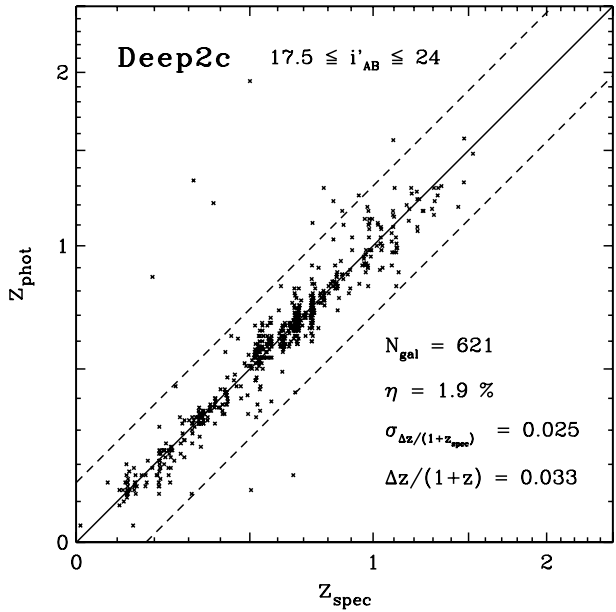}
\caption{Photometric and spectroscopic redshifts of objects in the ESO-DPS Deep2c field. For the matched 621 objects, the outlier rate $\eta$, scatter $\sigma_{\Delta z/(1+z)}$ and bias $\Delta z/(1+z)$ (as defined in \citealt{fabrice}, equations 38-40) are excellent down to the depth of the spectroscopic survey at $m_R\approx24$.}
\label{fig:photspec}
\end{figure}

The method used in this work requires a catalogue with magnitudes and accurate photometric redshifts to which sources in our cluster fields can be compared. In our case, this catalogue is extracted from stacks of the ESO Deep Public Survey \citep[ESO-DPS;][]{2005AN....326..432E,2006AuA...452.1121H}. The optical data are taken with the WFI camera on the MPG/ESO 2.2 m telescope, the primary instrument also used in the WISCy sample. We use six pointings (named Deep1a, 1b, 2b, 2c, 3a and 3b in \citealt{2006AuA...452.1121H}) for which photometric information is complete in the WFI filters \emph{UBVRI} and data is available in \emph{JK} from an additional infrared survey with NTT/SOFI \citep{2006AuA...452..119O,2006AuA...456..881O}. These yield approximately 200 000 galaxies at similar or significantly better depth than our cluster fields. Photometric redshifts are calculated with the template-fitting algorithm of \citet{2001defi.conf...96B} as described in \citet{fabrice}. Fig.~\ref{fig:photspec} shows a comparison of photometric redshifts of objects in Deep2c to spectroscopic measurements from the overlapping VIMOS VLT Deep Survey \citep{2005AuA...439..845L}. The quality of photometric redshifts, as a result of wavelength coverage, depth and calibration, is excellent, a prerequisite for using the catalogue as a reference. 

Any selection effects from photometric objects in our cluster fields into the catalogue of successful shape measurements that are potentially redshift dependent must be done similarly when selecting objects from the reference field into a reference catalogue for the redshifts of lensing sources. While magnitude dependent selection is taken into account automatically by the scheme described below, we therefore apply a size-dependent cut as in the cluster fields (cf. Section~\ref{sec:shapes}) to our reference catalogues. This has a small but significant effect, particularly for faint galaxies (cf. Fig.~\ref{fig:bribeta}).

\subsubsection{$\beta$ in Magnitude Space}

\label{sec:betamagbasic}
\begin{figure}
\centering
\includegraphics[width=0.48\textwidth]{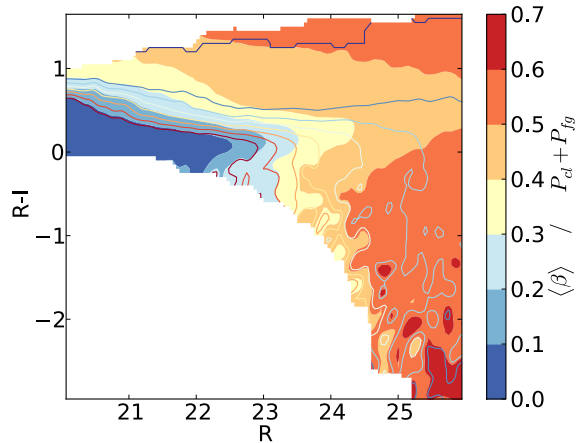}
\caption{Mean value of $\beta=D_{ds}/D_s$ as a function of \emph{R} and \emph{I} magnitude for a hypothetical lens redshift $z_{\rm cl}=0.4$ (coloured area). Note that while regions of high and low mean $\beta$ are easy to separate, contamination with objects in the foreground or close to the cluster redshift is problematic (see the contour lines giving the combined probabilities for both cases in the field) and will have to be corrected in the central part of the cluster where such galaxies are overabundant relative to the field.}
\label{fig:ribeta}
\end{figure}

Given a set of magnitudes $\bm{m}=\lbrace m_i\rbrace$, in our case $i=R,I(,B,V)$, consider the spherical volume in $2-4$ dimensional magnitude space centred on $\bm{m}$ with a radius of $|\Delta\bm{m}|=0.1$ and select a reference sample from that volume. For any fixed cluster redshift, the mean $\beta(\bm{m})$ of the reference sample and, in addition, the fraction of objects from the reference sample which are in the foreground ($P_{\rm fg}(\bm{m})$ for $z<z_{\rm cl}-0.06\times(1+z_{\rm cl})$) or near the cluster redshift ($P_{\rm cl}(\bm{m})$ for $|z-z_{\rm cl}|\leq0.06\times(1+z_{\rm cl})$) can be calculated and assigned to the objects of interest. For the case of two available bands, this is visualized in Fig.~\ref{fig:ribeta}.

\begin{figure}
\centering
\includegraphics[width=0.48\textwidth]{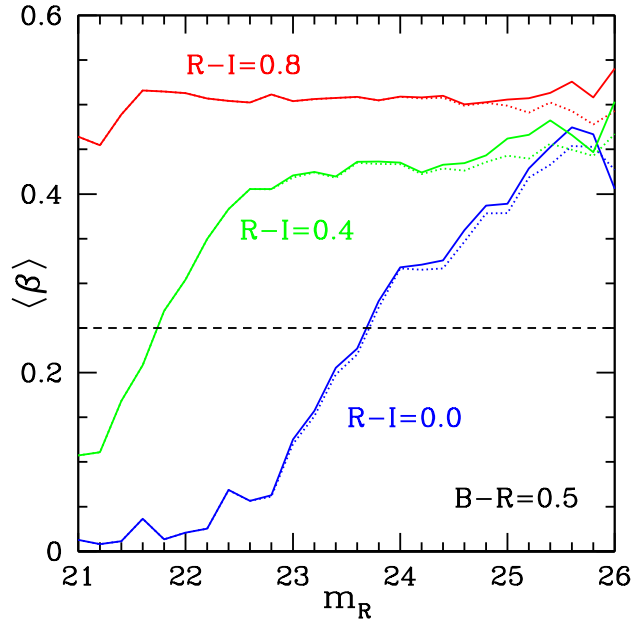}
\caption{Mean value of $\beta$ as a function of \emph{R} magnitude in bins of \emph{R}-\emph{I} colour for fixed \emph{B}-\emph{R}~=~0.5 at a hypothetical $z_{\rm cl}=0.4$. The horizontal dashed line indicates a typical level of $\beta$ below which objects add more noise than signal (see equation \ref{eqn:betamin}). While for the reddest bin $\beta$ is determined from the B-R and R-I colours and virtually independent of $m_R$, full magnitude information improves the background selection in the other colour bins (R-I~=~0.0, 0.4). Here, only by including $m_R$ we can discriminate objects with low $\beta$ that add excess noise from objects with high $\beta$ that yield an improved signal, which are both contained in the same colour bin. The solid and dotted lines show results for a reference sample selected for our lensing catalogue size cut at 0.5~arcsec \texttt{FLUX\_RADIUS} and for the complete reference sample, respectively, indicating that at magnitudes fainter than $m_R\approx24$ this is a relevant effect.}
\label{fig:bribeta}
\end{figure}

We note that it is common in WL analyses to use only colour information, i.e. the difference of the magnitudes of a galaxy in different bands (cf. e.g., \citealt{2010MNRAS.405..257M}, \citealt{2012ApJ...758...68H} or \citealt{2013ApJ...769L..35O} for recent examples). Magnitude cuts are used, too, yet typically in cases where only one or two bands are available (cf., e.g., \citealt{2009ApJ...697.1793N}, \citealt{2010AuA...514A..88R} or \citealt{2011ApJ...741..116O}). When only colours are used, one discards the magnitude offset that corresponds to a scaling of apparent flux for background selection. In contrast, our method includes the complete information. Fig.~\ref{fig:bribeta} shows that, depending on the position in colour space, the apparent magnitude does indeed help to discriminate low and high redshift objects. While for sufficiently many bands the colour information might constrain the source redshift well enough, for few bands and certain regions in colour colour space, therefore, complete magnitude information should be used in order to exploit the full power of photometric information for background estimation.

\subsubsection{Correction for Cluster Members}
\label{sec:fcl}
The treatment presented above would be sufficient if the distribution of galaxies in our fields was similar to the reference field. However, our fields contain rich clusters of galaxies, and the excess of cluster members, which is also a strong function of the separation from the cluster centre, has not been considered so far. As has typically also been done in previous studies, positional information must be used to correct the background sample for a dilution with cluster members.

Where additional galaxies at the cluster redshift exist, $\beta$ is overestimated by assuming the distribution of galaxies in magnitude-redshift space to be equal to an average field, when in fact a larger than usual fraction of galaxies is situated at the cluster redshift with $\beta=0$.\footnote{This misestimation of course can only happen in regions of magnitude space that are populated with galaxies at the cluster redshift. Due to the presence of non-red cluster members particularly at larger cluster redshifts and in the deeper data used in WL studies, however, these regions are larger than the commonly excised red sequence.} Estimates of cluster member density based on counts alone inevitably come with high uncertainty and systematic problems due to blending, masked areas and the intrinsic clustering of galaxies, at least in the case of single clusters. 

Our method of determining the cluster member contamination is based on decomposing the distribution of $\beta$ that we measure in an annulus around the cluster centre into the known distributions of $\beta$ for galaxies at the cluster redshift and in fields without excess galaxies at the cluster redshift. The coefficients of both components then give the proportion of excess galaxies at the cluster redshift in that region.

\begin{figure}
\centering
\includegraphics[width=0.48\textwidth]{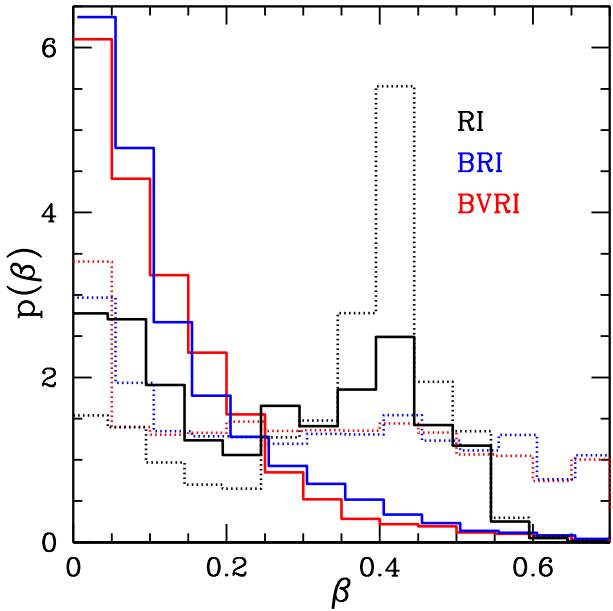}
\caption{Distributions of $\beta$ determined by magnitude space position for galaxies at a hypothetical lens redshift $z_{\rm cl}=0.4$ (solid lines) and field galaxies (dotted lines), in both cases assuming a lens redshift $z_l=0.4$, illustrated using subsets of the \emph{BVRI} photometry in the field of SPT-CL J0509--5342. When photometry in \emph{B}, \emph{R} and \emph{I} is available, the distributions are well discriminable. Adding \emph{V} band does not improve the separation much. Limiting the information to \emph{R} and \emph{I} only makes the situation significantly worse, although the two components are still distinguishable.}
\label{fig:pbeta}
\end{figure}

In this we make only the following weak assumption, namely that the measured distribution of $\beta(\bm{m})$ of galaxies at the cluster redshift and in the field is constant over the image. We can calculate $p_c(\beta)$, the distribution of $\beta$ in a redshift slice around the cluster, and $p_f(\beta)$, the distribution of $\beta$ outside the cluster region, by binning the $\beta(\bm{m})$ determined as described above. In the first case, we weigh objects by their probability of being near the cluster redshift, $P_{\rm cl}(\bm{m})$. In the second case we limit the analysis to the part of the field sufficiently separated from the cluster core and use all objects with equal weight. Fig.~\ref{fig:pbeta} shows $p_{\rm cl}(\beta)$ and $p_{\rm f}(\beta)$ for an exemplary cluster redshift of $z_{\rm cl}=0.4$ as calculated in one of our fields on \emph{RI}, \emph{BRI} and \emph{BVRI} colour information. Fortunately, the two distributions are distinct in all these cases, such that a decomposition is possible.

\begin{figure}
\centering
\includegraphics[width=0.48\textwidth]{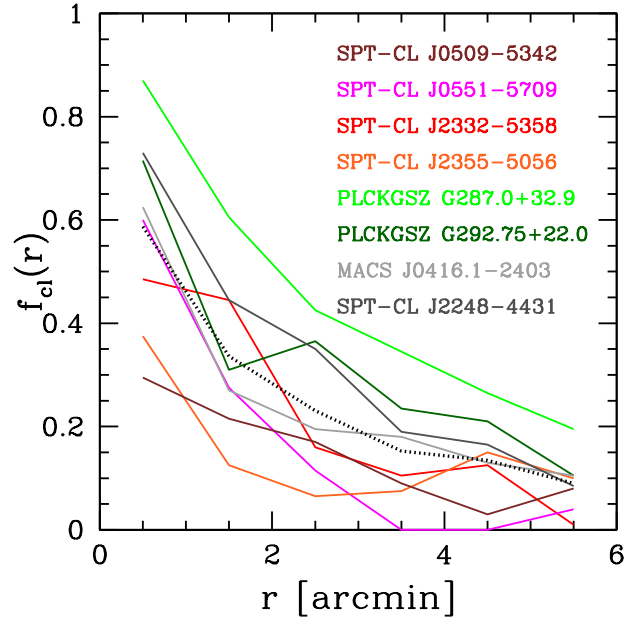}
\caption{Cluster member fraction $f_{\rm cl}$ in the photometric catalogue as a function of radius for the cluster field without photometric redshift information. Colour coding for individual clusters as indicated, with the mean value given by the dotted black line. The richest among these clusters, PLCKESZ G287.0+32.9 (light green), is visually confirmed to contain numerous cluster member galaxies spread over a large region.}
\label{fig:fclr}
\end{figure}

As a function of radius from the cluster centre, we fit a parameter $0\leq f_{\rm cl}(r)\leq 1$ for the fraction of cluster member galaxies in our shape catalogue at that separation. This is done by demanding that
\begin{equation}
 p(\beta,r)=f_{\rm cl}(r)\times p_{\rm cl}(\beta)+(1-f_{\rm cl}(r))\times p_{\rm f}(\beta) \; ,
\end{equation}
which can be optimized in terms of $f_{\rm cl}(r)$ using minimum $\chi^2$ with Poissonian errors. In practice, we do this in bins of 1~arcmin width out to a radius of 6~arcmin. For application to individual objects, we linearly interpolate $f_{\rm cl}(r)$ to the galaxy position.

Results for $f_{\rm cl}(r)$ are shown in Fig.~\ref{fig:fclr}. The overall profile of cluster member galaxies is of the order of 50 per cent in the central region and drops smoothly towards the outskirts. We note that the data are fitted best not by a single power law but by a broken power -aw profile with logarithmic slope near 0 at small and -2 at large radii. This is consistent with the expected NFW profile of cluster member number density \citep[cf.][]{1997ApJ...485L..13C,2004ApJ...610..745L,2011ApJ...734....3Z,2012MNRAS.423..104B} rather than a single power law (but cf. \citealt{2012MNRAS.427.1298H}, who find a global $r^{-1}$ dependence).

Knowing $f_{\rm cl}(r)$, we can correct the probabilities for a galaxy of belonging to the foreground, the cluster redshift region and the background as
\begin{equation}
(P_{\rm fg},P_{\rm cl},P_{\rm bg})\rightarrow(P_{\rm fg}',P_{\rm cl}',P_{\rm bg}')=\frac{\left(P_{\rm fg},P_{\rm cl} b_{\rm cl},P_{\rm bg}\right)}{1+(b_{\rm cl}-1) P_{\rm cl}} \; ,
\label{eqn:pp}
\end{equation}
where we have defined the cluster member bias $b_{\rm cl}$ as
\begin{equation}
 b_{\rm cl}-1=f_{\rm cl}(r)\times\left(1/\langle P_{\rm cl}\rangle-1\right) \;,
\end{equation}
with $\langle P_{\rm cl}\rangle$ the fraction of galaxies in the field which lie in the cluster redshift slice. 

The true field population appears reduced to a fraction $(1+(b_{\rm cl}-1) P_{\rm cl})^{-1}$ due to the excess cluster members. For any galaxy, the probability of actually belonging to the background population and with it $\beta$ therefore decrease by that same factor, 
\begin{equation}
 \beta\rightarrow\beta'=\beta\times(1+(b_{\rm cl}-1) P_{\rm cl})^{-1} \; .
\label{eqn:betap}
\end{equation}
Note that for galaxies in regions of magnitude space unlikely to be populated with galaxies at the cluster redshift ($P_{\rm cl}\approx0$), no correction is necessary. Indeed, we find that the changes to the best-fitting mass when not applying the cluster member correction are typically below $5\%$ in our sample.

In the case where $\beta$ is to be calculated for a secondary lens at a redshift different from the cluster, a small generalization to equation (\ref{eqn:betap}) must be made. The relevant mutually exclusive and collectively exhaustive cases here are $P_{\rm cl}$, the probability of a galaxy belonging to the main cluster redshift slice, and $P_{\rm fg\neg cl}$ and $P_{\rm bg\neg cl}$, the probabilities for a galaxy to be in the foreground or background of the secondary lens, but now explicitly excluding the cluster redshift slice in both. These three probabilities transform exactly as in equation (\ref{eqn:pp}). However, the initial estimate for $\beta$ is composed of
\begin{equation}
 \beta=P_{\rm cl}\beta(z_{\rm cl})+P_{\rm bg\neg cl}\beta_{\rm bg\neg cl} \; .
\end{equation}
Note that cluster member galaxies could have $\beta(z_{\rm cl})>0$ if the cluster is in the background of the secondary lens, in which case equation (\ref{eqn:betap}) is no longer correct. Rather then,
\begin{eqnarray}
\beta\rightarrow\beta'&=&\beta(z_{\rm cl}) P'_{\rm cl}+P'_{\rm bg\neg cl}\beta_{\rm bg\neg cl} \nonumber \\
&=&\frac{\beta(z_{\rm cl}) P_{\rm cl}(b_{\rm cl}-1)+\beta}{1+(b_{\rm cl}-1) P_{\rm cl}} \; .
\label{eqn:betapp}
\end{eqnarray}
For $\beta(z_{\rm cl})=0$, this reduces to equation (\ref{eqn:betap}).

\subsubsection{Optimised $\beta$ threshold}

\label{sec:minbeta}

One can finally make optimised cuts by selecting objects whose $\beta$ is above some threshold. A threshold too high would remove too many actual background objects and increase the shape noise. A threshold too low would increase the noise by including objects for which the lensing effect is small compared to the intrinsic ellipticity scatter. The optimum can be found as follows.

Let the density of objects in $\beta$ space be given by $p(\beta)$, such that $\int_0^{\beta(z_s\rightarrow\infty)} p(\beta)\; d\beta=1$. If we assume a constant shear $\gamma_t$ for objects at a hypothetical $\beta=1$, a constant shape noise of $\sigma_{\epsilon}$ and a $\beta$ threshold $\beta_{\rm min}$, then the S/N of the measurement will be 
\begin{figure}
\centering
\includegraphics[width=0.48\textwidth]{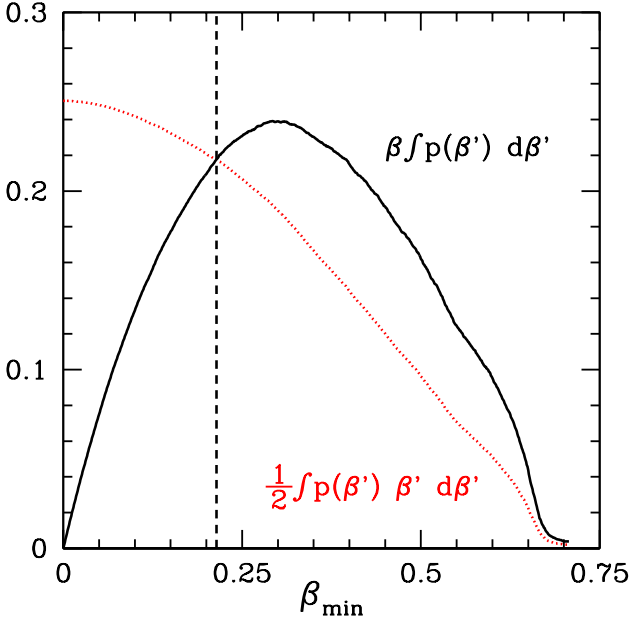}
\caption{Both sides of equation (\ref{eqn:betamin}) plotted against the $\beta$ threshold $\beta_{\rm min}$, illustrated for a hypothetical lens redshift $z_{\rm cl}=0.4$ based on \emph{BVRI} photometry in the field of SPT-CL J0509--5342. The optimal threshold is in this case $\beta_{\rm min,opt}\approx0.214$ as indicated by the dashed vertical line.}
\label{fig:betamin}
\end{figure}

\begin{equation}
 \mathrm{S/N}=\frac{\sqrt{N_{\rm gal}}\gamma_t}{\sigma}\frac{\int_{\beta_{\rm min}}^{\beta(z_s\rightarrow\infty)} p(\beta) \beta\; d\beta}{\sqrt{\int_{\beta_{\rm min}}^{\beta(z_s\rightarrow\infty)} p(\beta) \; d\beta}} \; ,
\end{equation}
where $N_{\rm gal}$ is the number of galaxies in the sample.

A maximum of the S/N can be found at some $\beta_{\rm min,opt}$ where $\frac{\mathrm{d}{\rm S/N}}{\mathrm{d}\beta_{\rm min}}=0$. This entails
\begin{equation}
 \beta_{\rm min,opt} \int_{\beta_{\rm min,opt}}^{\beta(z_s\rightarrow\infty)} p(\beta) \mathrm{d}\beta = \frac{1}{2} \int_{\beta_{\rm min,opt}}^{\beta(z_s\rightarrow\infty)} p(\beta) \beta \mathrm{d}\beta \; .
\label{eqn:betamin}
\end{equation}
The above equation can be solved numerically for any ensemble of galaxies to which $\beta$ values have been previously assigned, as is illustrated in Fig.~\ref{fig:betamin}.

\subsubsection{Uncertainty in $\beta$ and Reduced Shear}

The observed shear signal is not the gravitational shear $\gamma$ itself, but rather the reduced shear 
\begin{equation}
 g=\frac{\gamma}{1-\kappa} \; .
\label{eqn:greduced}
\end{equation}
Note that, unlike $\gamma$ or convergence $\kappa$, this is not linear in $\beta$. When a model predicts some value $\gamma_1$ and $\kappa_1$ for a hypothetical $\beta=1$, the expectation value for a galaxy whose $\beta$ is determined with non-zero uncertainty is (\citealt{1997A&A...318..687S}, cf. also \citealt{2012arXiv1208.0605A})
\begin{equation}
 \langle g\rangle=\frac{\langle\beta\rangle\gamma_1}{1-\frac{\langle\beta^2\rangle}{\langle\beta\rangle}\kappa_1} \; ,
\label{eqn:ss}
\end{equation}
which includes a correction for the non-linear response of equation (\ref{eqn:greduced}) to the dispersion of $\beta$.

While the $\beta(\bm{m})$ determined above equals $\langle\beta\rangle$ in equation (\ref{eqn:ss}), $\langle\beta^2\rangle$ can be determined in an equal fashion by taking the mean of $\beta^2$ in a magnitude space volume around the position of the source galaxy. Likewise, equation (\ref{eqn:betapp}) can be used to correct the estimate for cluster members by substituting $\beta^2(z_{\rm cl})$ for $\beta(z_{\rm cl})$. Compared to the direct application of equation (\ref{eqn:greduced}), this correction yields a 1\% decrease in best-fitting mass for our clusters, where the effect is, as expected, strongest for the most massive systems, for which the approximation $\kappa\ll1$ does not hold.

For the analysis of the two CFHTLS WISCy clusters with photometric redshifts, we also account for the non-singular probability distribution \citep[cf.][]{2012arXiv1208.0605A}. In this case we approximate the uncertainty by a scatter of photometric redshifts of $\sigma_{\Delta z/(1+z)}=0.03$ and an outlier rate $\eta=0.04$, as defined and measured against spectroscopic samples in \citet{fabrice}. From the simulated distribution, $\langle\beta\rangle$ and $\langle\beta^2\rangle$ are calculated for the individual galaxy and then used in equation (\ref{eqn:ss}).

\subsubsection{Comparison to Photometric Redshifts}
\label{sec:zcomp}
We finally make a comparison between the $\beta(\bm{m})$ determined by colour-magnitude matching as described above and the alternative method of calculating $\beta(z_{\rm phot})$ directly from best-fitting photometric redshifts. In particular, we are interested in how biased $\langle\beta(z_{\rm phot})\rangle$ becomes when estimated from photometric redshifts in various regimes.

To this end, we use the galaxies from the DPS reference catalogue with \emph{UBVRIJK} magnitude information. We generate photometric redshifts for these galaxies, limiting the bands used to either \emph{BVRI} or \emph{UBVRI} only. We select objects either by their $\beta(BRI)>0.22$ estimated as described above from \emph{BRI} magnitudes only or by these \emph{BVRI} or \emph{UBVRI} photometric redshifts. In the latter case, we select the background by the condition 
\begin{equation}
 z_s>1.1\times z_l + 0.15 \; .
\label{eqn:zsel}
\end{equation}
The conservative offset of $z_s-z_l>0.1\times z_l+0.15$ is meant to prevent significant scatter of foreground galaxies into the background \citep{fabrice}. In all cases, we compare the mean $\beta$ of the sample selected and estimated with the respective method to the mean $\beta$ of the sample based on the \emph{UBVRIJK} photometric redshifts. This corresponds to a comparison to the truth under the assumption that the latter are accurate redshifts, an approximation justified by their excellent quality with respect to the spectroscopic control sample (cf. Fig~\ref{fig:photspec}).

\begin{figure}
\centering
\includegraphics[width=0.48\textwidth]{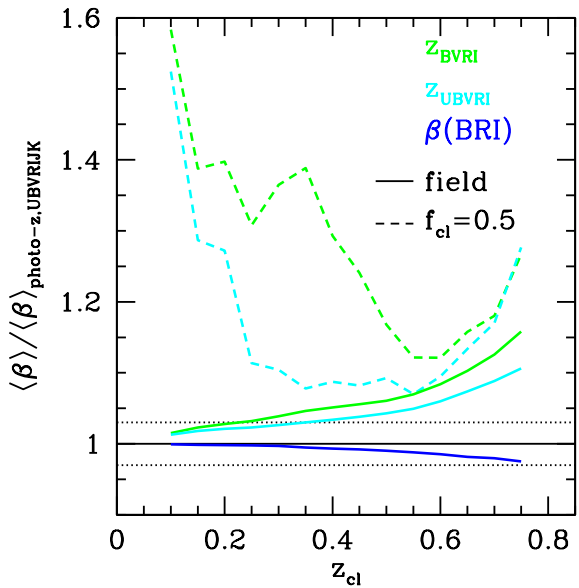}
\caption{Bias in estimated mean $\beta$ for background galaxy samples selected by various methods compared to the $\beta$ estimated from \emph{UBVRIJK} photometric redshifts.}
\label{fig:ztest}
\end{figure}

Fig.~\ref{fig:ztest} shows these comparisons both in an empty field (solid lines) and in a region heavily populated with cluster galaxies (half of the galaxies at the cluster redshift in excess of the field distribution, i.e. $f_{\rm cl}=0.5$, typical for the central region of our cluster pointings, cf. Fig.~\ref{fig:fclr}). We note the following points.
\begin{itemize}
 \item Photometric redshifts from a limited number of bands show little bias, in particular at low to medium cluster redshifts, in the field. For \emph{UBVRI}, the bias in $\langle\beta\rangle$ is below 3\% up to $z_{\rm cl}\approx0.35$.
 \item The ability of photometric redshifts with a limited number of bands to yield an unbiased background selection and $\beta$ estimate is strongly degraded by the presence of excess cluster members. This is also a strong function of the available bands, as can be seen in the case of \emph{BVRI} and \emph{UBVRI} redshifts for $z_{\rm cl}\approx0.4$ and $f_{\rm cl}=0.5$: here the \emph{BVRI} background sample has a $\approx35\%$ bias in $\beta$, while \emph{UBVRI} is closer to a $\approx10\%$ bias.
 \item The $\beta(\mathrm{m})$ method described in this section is always biased below the 3\% level. The small systematic effect is caused by our symmetric weighting over the magnitude space sphere despite the non-uniform magnitude space density of galaxies. Note that the curve of $\beta(\mathrm{BRI})$ for $f_{\rm cl}=0.5$ and in the field are identical by definition in this plot due to our cluster member correction (cf. Section~\ref{sec:fcl}).
\end{itemize}

These observations are consistent with our data: for six of our clusters with sufficiently deep WFI multi band photometry (namely SPT-CL J2332--5358, SPT-CL J0551--5709 and SPT-CL J2355--5056 with \emph{BRI}, SPT-CL J0509--5342 and MACS J0416.1--2403 with \emph{BVRI} and SPT-CL J2248--4431 with \emph{UBVRIZ}), we perform single halo NFW fits on the same shape catalogue, supplying $\beta$ with either method. The best-fitting masses measured with these photo-$z$ values are, in all cases, significantly lower than the ones we find with $\beta(\bm{m})$, on average by 30\% (cf. also the similar effect noted in \citealt{2012arXiv1208.0605A}). The difference in $\Delta\Sigma$ profiles is primarily in the central region, in line with the observation that the bias of photometric redshift based $\beta$ is a strong function of excess cluster member density. For the photometric redshifts calculated with the better spectral coverage and greater depth of the field of SPT-CL J2248--4431, best-fitting masses agree between both methods at the level of shape noise expected from the different object selection.

\subsubsection{Background samples}

For the background samples determined as above, we present some fundamental metrics in Table~\ref{tbl:background}. 

\begin{table}
\begin{center}
\begin{tabular}{|l|r|r|r|r|r|}
\hline
\# & $\beta_{\rm min}$ & $\langle\beta\rangle$ & $z(\beta_{\rm min})$  & $z(\langle\beta\rangle)$ & $N_{\rm gal, bg}$ \\ \hline
1  & 0.20            & 0.40        & 0.60        & 0.85          & 8316  \\ \hline
2  & 0.21            & 0.42        & 0.55        & 0.80          & 4307  \\ \hline
3  & 0.23            & 0.46        & 0.54        & 0.83          & 9152  \\ \hline 
4  & 0.25            & 0.51        & 0.44        & 0.72          & 6259  \\ \hline
5  & 0.19            & 0.38        & 0.49        & 0.67          & 5763  \\ \hline
6  & 0.23            & 0.46        & 0.40        & 0.60          & 7467  \\ \hline
7  & 0.23            & 0.46        & 0.54        & 0.82          & 8663  \\ \hline
8  & -               & 0.59        & -           & -             & 8045  \\ \hline
9  & -               & 0.77        & -           & -             & 10639 \\ \hline
10 & -               & 0.60        & -           & -             & 8067 \\ \hline
11 & -               & 0.34        & -           & -             & 2628 \\ \hline
12 & -               & 0.83        & -           & -             & 10053 \\ \hline
\end{tabular}
\end{center}
\caption{Background sample statistics for WISCy clusters. IDs are taken from Table~\ref{tbl:list}. $\beta$ is determined as described in Section~\ref{sec:beta}. The threshold $\beta_{\rm min}$ is found by means of equation (\ref{eqn:betamin}). The remaining columns give the mean of $\beta$ in the selected background, the corresponding source redshifts and the number of background galaxies with successful shape measurements selected inside the $\approx30\times30$~arcmin$^2$ WFI pointing. For the five clusters where photometric redshifts are used, only $\langle\beta\rangle$ and $N_{\rm gal, bg}$ of the photo-$z$ selected background sample are given. For the four CFHT clusters 9-12, source counts are within a WFI pointing centred on the cluster for direct comparison.}
\label{tbl:background}
\end{table}

The remaining SPT-CL J2248--4431 WFI data and the CFHT fields have sufficient coverage in five or more bands for the determination of photometric redshifts. In these cases, we select sources by equation (\ref{eqn:zsel}).

\section{Weak Lensing Analysis}
\label{sec:shapemeas}

Weak gravitational lensing changes the positions, sizes and shapes of background galaxy images. While the effect on position is indeterminable and changes in size induce changes in the observed surface density of galaxies which depend on the less than fully known intrinsic distribution of galaxy sizes and magnitudes, the weak assumption that the orientations of galaxies are intrinsically random allows an unbiased measurement of the mass density field.

One important difficulty is that this requires the accurate measurement of pre-seeing galaxy shapes in the presence of a point spread function (PSF) with size of the order of or exceeding that of small background galaxies and ellipticity similar to the gravitational shear signal. This step can be divided into making a model of the PSF at the galaxy position in the image and measuring the pre-seeing galaxy shape using this model. The pipeline used in this work performs these tasks with the publicly available software \textsc{\mbox{psfex}}\footnote{\texttt{http://www.astromatic.net/software/psfex}} \citep{2011ASPC..442..435B} and an implementation of the shape measurement method of \citet[][hereafter KSB]{1995ApJ...449..460K}. It is described also in \citet{2013MNRAS.tmp.1221G}, and we only give a summary here, with emphasis on aspects particular to the WISCy sample.

\subsection{Model of the Point-Spread Function}

From bright but unsaturated stars without significant blending selected in a radius-magnitude diagram, we generate a model of the PSF using \textsc{\mbox{psfex}} \citep{2011ASPC..442..435B}. The model assigns a value to each pixel in a centred vignet image of the star that is a polynomial of the $x$ and $y$ coordinate in the field of view.

It is necessary to ensure that the model represents the PSF accurately. The most important metrics for the purpose of WL are residuals in size and shape. Both are based on second moments of the PSF profile (cf. \citealt{2013ApJS..205...12K}), which we calculate from actual star images and the model vignets at star positions. 

\begin{figure*}
\centering
\includegraphics[width=0.48\textwidth]{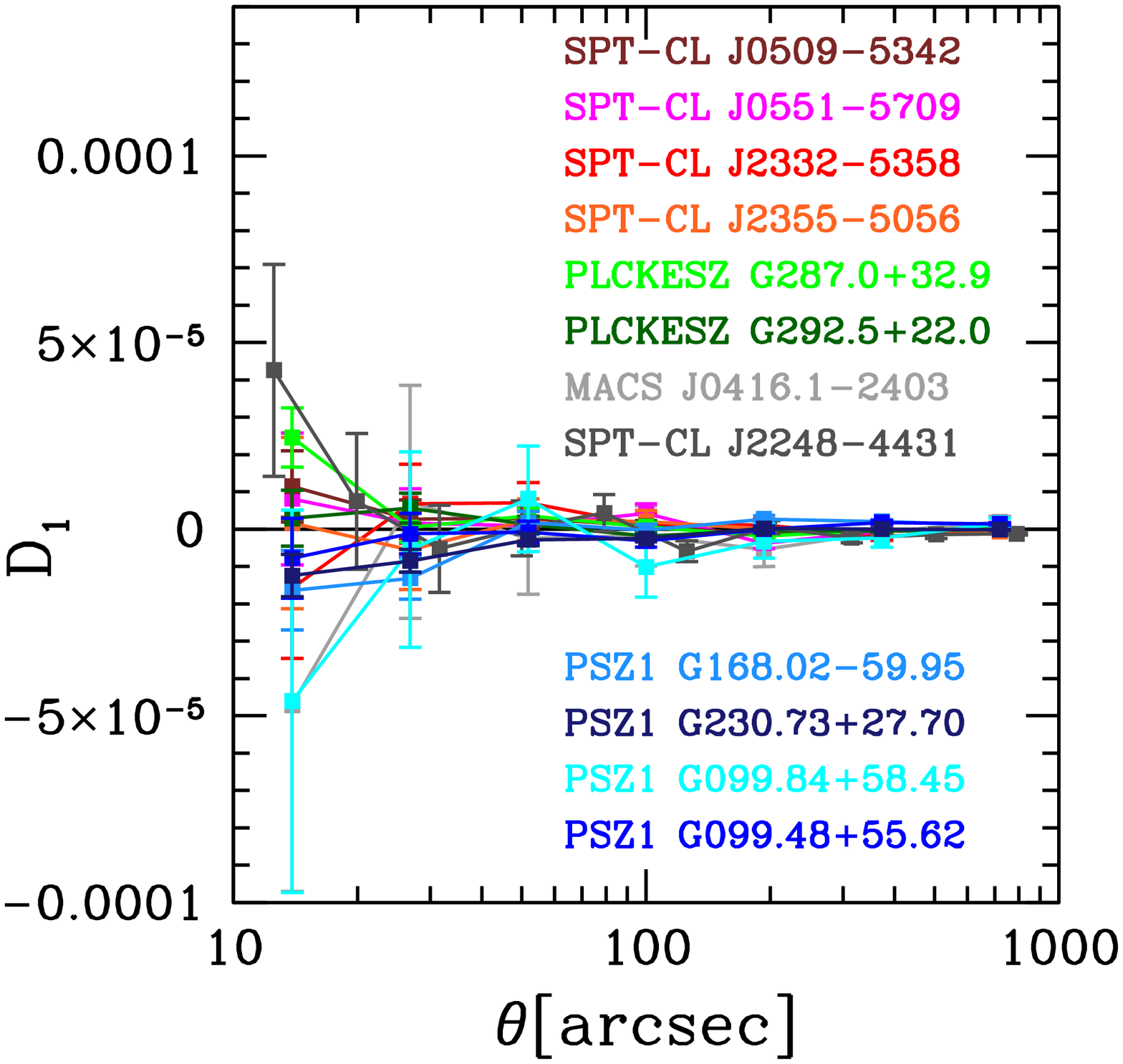}
\includegraphics[width=0.48\textwidth]{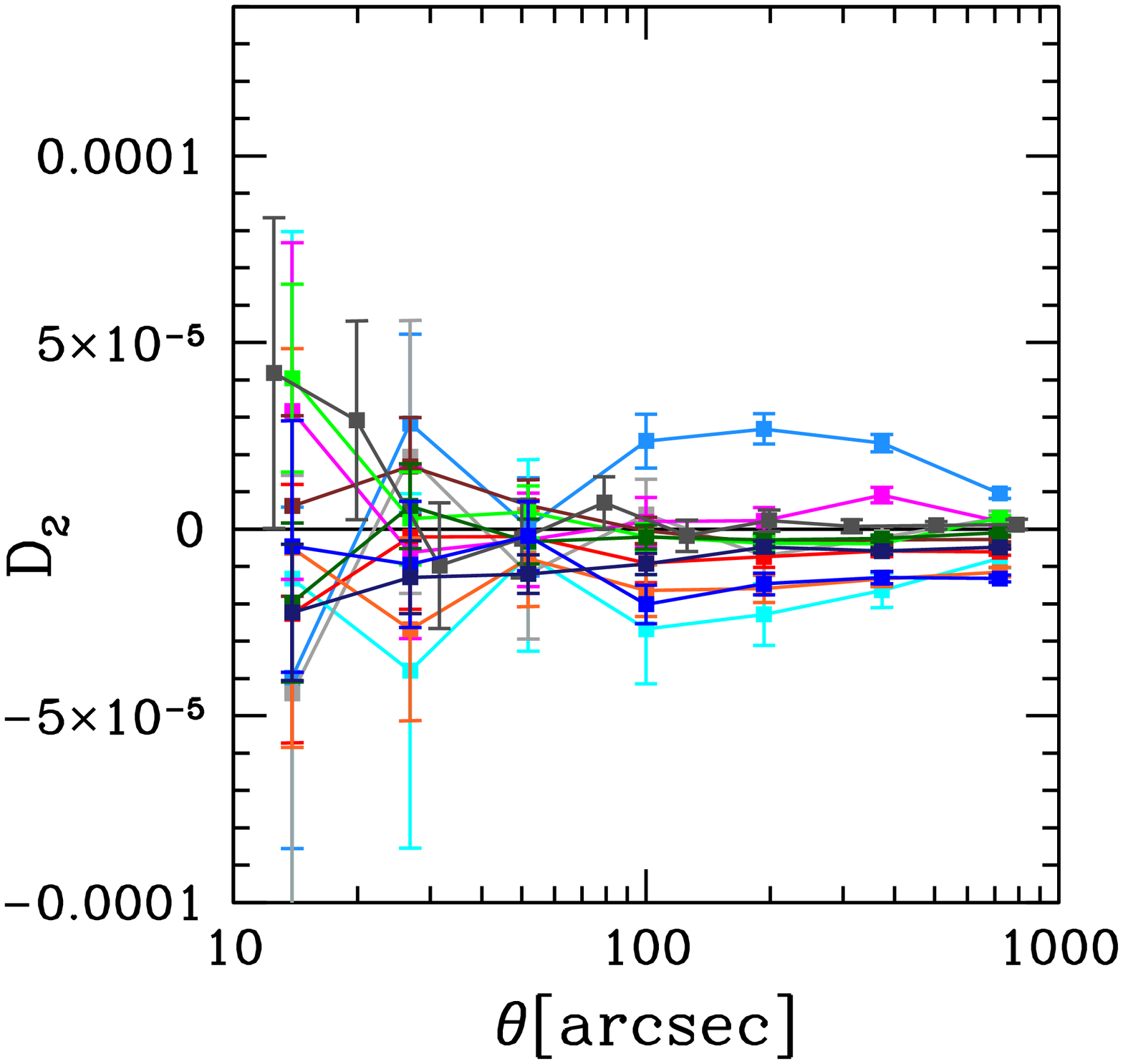}
\caption{\citet{2010MNRAS.404..350R} statistics for the main lensing bands of the WISCy sample. The left- (right)-hand panel shows the autocorrelation of stellar ellipticity residuals (cross-correlation of residual and measured ellipticity). We note that the requirement $|D_i(\theta)|<10^{-5}$ is always fulfiled for $D_1$, the component that directly connects to spurious E/B modes due to PSF mismodelling. For $D_2$, it is violated moderately for three of the four CFHT fields (PSZ1 clusters) and SPT-CL J2355--5056, yet to a level that can be tolerated in cluster analyses.}
\label{fig:rowe}
\end{figure*}

The main requirement for WL is that PSF sizes and shapes are modelled accurately with low statistical uncertainty and without spatial coherence in the residuals. The latter could cause a systematic signal, since it induces a spatially coherent additive and/or multiplicative component on the measured shear field. We therefore accept a PSF model only if the residual between measured stars and the model at the same position are small. Our qualitative and quantitative criteria are as follows:
\begin{itemize}
 \item no offset of the residual histogram from zero mean for both size and shape,
 \item no regions of visibly problematic fit for either size or shape,
 \item root-mean-squared residual ellipticity below 0.004 in any single component, and
 \item auto-correlation of ellipticity residuals $D_1(\theta)$ and cross-correlation of residuals and measured stellar ellipticity $D_2(\theta)$, i.e. the \citet{2010MNRAS.404..350R} statistics (defined as in their equations 13 and 14), are consistent with $|D_i(\theta)|<10^{-5}$ at a separation angle of $\theta\geq100$~arcsec, the scales relevant to our analysis.
\end{itemize}
The last criterion is a useful test of having selected the correct polynomial order of PSF interpolation, since it is sensitive to both over- and underfitting \citep{2010MNRAS.404..350R}. Fig.~\ref{fig:rowe} shows $D_1$ and $D_2$ for the main lensing band of all our clusters. In each frame, we use the lowest order at which the $D_i$ statistics are either consistent with zero or do not significantly improve when increasing the orders further.

For the subsample of eight clusters from the WISCy sample imaged with WFI, it is always possible to fulfill the PSF model quality criteria for the R-band stack and in several cases also in other bands (cf. Table~\ref{tbl:list} for the list of bands with successful shape measurement). The negative $D_2$ of SPT-CL J2355--5056 (cf. Fig.~\ref{fig:rowe}) is a minor exception. Since this is in principle a sign of overfitting despite the use of a polynomial of only 4th order, we conclude that it is likely a statistical effect and accept the model regardless. In some of the \emph{I} band frames  it is not possible to control $D_2$ to an acceptable level, and we reject these frames for shape measurement. For the four clusters imaged by CFHTLS, the criteria of root-mean-square (rms) residual and $|D_i(\theta)|<10^{-5}$ prove to be too strict to be fulfilled for the models tested. Modelling the PSF on individual chips while excluding the chip gaps and regions of the field where the PSF model is visibly problematic, we accept the frames despite a slightly higher $D_2$. We note that this higher level of PSF complexity is typical for the CFHTLS frames (cf. also \citealt{fabrice}, their section~3.4). 

\subsection{Shape Measurement}
\label{sec:shapes}

We run an implementation of KSB+ (KSB; \citealt{1997ApJ...475...20L} and \citealt{1998ApJ...504..636H}) which uses the \textsc{psfex} PSF model (\textsc{ksbpsfex}). Galaxies are prepared for shape measurement by the following steps:
\begin{itemize}
 \item unsaturated, reasonably isolated sources with \textsc{SExtractor} flags $\leq3$ are selected,
 \item postage stamps of 64$\times$64pix size are extracted and neighbouring objects are masked according to the \textsc{SExtractor} segmentation map,
 \item the \textsc{SExtractor} photometric background estimate at the object position is subtracted from the image in order to compensate for small-scale background variations insufficiently subtracted by the data reduction pipeline,
 \item bad and masked pixels are interpolated using a Gauss Laguerre model of the galaxy \citep{2002AJ....123..583B}, discarding objects with more than 20 per cent of postage stamp area or 5 per cent of model flux in bad or masked pixels.
\end{itemize}
For the actual shape measurement, KSB+ is run on the prepared postage stamp of the galaxy and the sub-pixel resolution PSF model at the same position. Here we only give a brief summary of this step, referring the reader to the original papers for further details.

KSB measure polarizations
\begin{equation}
\bm{e} = \frac{1}{Q_{11}+Q_{22}} \left(
\begin{array}{c}
Q_{11}-Q_{22}\\
2Q_{12}
\end{array} \right)\\
\end{equation}
defined by second moments calculated inside an aperture with Gaussian weight,
\begin{equation}
Q_{ij}=\int \mathrm{d}^2\theta \, I(\bm{\theta}) \, w(|\theta|) \, \theta_i \, \theta_j \; ,
\end{equation}
of the surface brightness distribution of the galaxy $I(\bm{\theta})$ with a weight function $w(|\theta|)$ centred on the galaxy centroid. Our implementation allows measuring the galaxy and PSF moments with the identical weight function, scaled with the observed half-light radius of the galaxy.

In the presence of an elliptical PSF, the linear approximation of how observed post-seeing polarization $\bm{e_o}$ reacts to a reduced shear $\bm{g}$ (cf. \citealt{2001PhR...340..291B}, p. 60) can be expressed as
\begin{equation}
 \bm{e_o}=\bm{e}_i + \hat{P}^{\mathrm{sm}}\bm{p}+\hat{P}^{\gamma}\bm{g} \; ,
\end{equation}
where $\bm{e}_i$ is the intrinsic post-seeing ellipticity of the galaxy, $\hat{P}^{\mathrm{sm}}$ is a $2\times2$ tensor quantifying the response of observed shear to PSF polarization $\bm{p}$ and $\hat{P}^{\gamma}$ is the shear responsivity tensor. We invert $(\hat{P}^{\gamma})^{-1}\approx\frac{2}{\tr P^{\gamma}}$ and assume that $(\hat{P}^{\gamma})^{-1}\bm{e}_i$ is zero on average (because of the random intrinsic orientation of galaxies) to find the ensemble shear estimate
\begin{equation}
 \langle\bm{g}\rangle=\langle\bm{\epsilon}\rangle=\left\langle \frac{2}{\tr P^{\gamma}} \left(\bm{e_o}-\hat{P}^{\mathrm{sm}}\bm{p}\right) \right\rangle \; .
\label{eqn:gfrome}
\end{equation}

We apply some final filtering and corrections to our catalogue, namely
\begin{itemize}
 \item removing objects with any of $\tr P^{\gamma}<0.1$, half-light radius less than 5\% larger than the PSF half-light radius or S/N as defined by \citet{2001AuA...366..717E} smaller than 10, since these give extremely noisy shape estimates prone to selection effects and noise biases,
 \item for objects with measured $|\bm{\epsilon}|>1+\sqrt{\Delta\epsilon_1^2+\Delta\epsilon_2^2}$, which can happen due to noise, re-scaling ellipticity to $|\bm{\epsilon}|=1+\sqrt{\Delta\epsilon_1^2+\Delta\epsilon_2^2}$ where $\Delta\epsilon_i$ is the Gaussian error of measured pre-seeing ellipticity from pixel noise; note that a strict clipping at $|\bm{\epsilon}|=1$ could potentially induce biases since it asymmetrically limits the measurement errors on objects with non-zero ellipticity,
 \item correcting for shape measurement biases (see Section~\ref{sec:bias} for details), and
 \item merging shape catalogues measured on different bands (or, in the case of the CFHTLS, overlapping pointings), where they exist, into one; since due to differences in depth and colour the statistical uncertainty of these shape estimates can greatly differ between bands, we use for any galaxy the inverse-variance ($\Delta\epsilon^{-2}$) weighted average of all shape estimates.
\end{itemize}

These catalogues are then matched against photometry extracted as described in Section~\ref{sec:photometry}.

\subsection{Correction of shape measurement bias}
\label{sec:bias}

Biases of shape estimators are commonly expressed as a multiplicative and an additive term \citep[see, e.g.,][]{2006MNRAS.368.1323H},
\begin{equation}
 \epsilon_{\rm obs}-\epsilon_{\rm true} = m\times\epsilon_{\rm true} + c \; .
\end{equation}
It is known from programmes testing shape measurement methods on simulated images with known shear \citep[cf.][Young et al., in preparation]{2006MNRAS.368.1323H,2007MNRAS.376...13M,2010MNRAS.405.2044B,2012MNRAS.423.3163K} that even under these well-controlled conditions shape measurement biases exist in virtually all pipelines, at least at the typical low levels of S/N that one must deal with in WL studies.

Multiplicative biases $m\neq0$ translate directly to a bias in the amplitude of the shear signal and, consequently, the mass. Additive biases $c\neq0$, while severely hindering cosmic shear analyses where the two-point correlation of shapes is measured, average out in spherically symmetric shear analyses such as the ones presented here. They can, however, still influence our analysis in the case of significant masked regions or near the image borders, where spherical averaging is incomplete. We therefore decide to test for and, where necessary, calibrate both effects, using a combination of simulated and real galaxy images. 

For calibrating the multiplicative bias, we use a set of simulations of galaxy images with realistic distributions of size, ellipticity and S\'{e}rsic parameter convolved with circular and elliptical Gaussian PSFs (Young et al., in preparation). The images are resampled using the same Lanczos kernel as in our stacks, applying a random sub-pixel shift, before galaxies are selected and their shapes measured with our pipeline. S/N is a parameter known to relate to several effects relevant for multiplicative shape bias \citep{2000ApJ...537..555K,2002AJ....123..583B,2012MNRAS.425.1951R,2012arXiv1203.5049K,2012MNRAS.424.2757M}. We therefore determine the multiplicative bias as the deviation of the ratio of mean measured shapes to known true shears from one as a function of S/N. We note that the magnitude of multiplicative bias and also its observed increase towards lower S/N is typical for present shape measurement pipelines \citep[e.g.][]{2010AuA...516A..63S,2012arXiv1208.0597V}. The fitted functional form for multiplicative bias is shown in \citet[][equation 9]{2013MNRAS.tmp.1221G} and the corresponding correction is applied to our shape catalogues.

No significant additive bias is detected in our simulations, yet could potentially be caused in real data by more intricate observational effects not present in the former. The fact that the mean ellipticity of galaxies in an unbiased catalogue should be zero can be used to empirically calibrate constant additive biases directly from the data \citep[cf. e.g.,][]{2012MNRAS.427..146H}. In our initial catalogues, we found a significantly negative $\epsilon_1$ component, indicating preferential orientation of objects along the vertical direction. We trace this back to a selection effect in the latest public version of \textsc{SExtractor}.\footnote{Version 2.8.6, cf. \texttt{http://www.astromatic.net/software/sextractor}} Related to buffering with insufficiently large memory settings (Melchior, Bertin, private communication), horizontally elongated objects are preferentially deselected (\texttt{FLAG}=16). An increase of \texttt{MEMORY\_BUFSIZE} fixes this and yields shape catalogues without significant additive bias.

\subsection{Mass mapping}

The WL shear field can be used to estimate the surface mass density as a function of position. We create such maps for the eight WFI clusters in the WISCy sample. They are shown in the respective subsections of Section~\ref{sec:individual} and used for the purpose of illustrating the cluster density fields only.

For the surface density reconstruction we use the finite field reconstruction technique described in \citet{1996A&A...305..383S}. From the observed galaxy ellipticity we estimate the reduced shear using equations 5.4 to 5.7 in \citet{1996A&A...305..383S}. We choose a smoothing length of 1.5~arcmin for the spatial averaging of the galaxy ellipticities (their equation 5.7), accounting for the relatively low galaxy density (of order 10 galaxies per square arcminute) of background objects used for the reconstruction. The $\kappa$-maps are obtained on a $100\times100$ grid for the FOV of shear data of typically 28$\times$28~arcmin$^2$. Spatial resolution is limited by the required large smoothing length and not by the grid on which $\kappa$ is calculated.

As was pointed out by \citet{1995A&A...294..411S} and \citet{1995A&A...297..287S,1997A&A...318..687S}, as long as one uses shear data (at one effective redshift) only, the surface density maps can only be obtained up to the mass sheet degeneracy. We thus arbitrarily fix this constant such that within the reconstructed field the mean density is $\kappa=0.01$ (accounting for the fact that the field is not empty but contains a massive cluster with approximately this mean surface density). If we would have chosen this constant differently the contour-pattern of course would not change, but the values would be slightly altered as described by the mass sheet degeneracy.

For the four clusters imaged by the CFHTLS, masking of the chip border areas is required in our shape catalogue, since the complex and discontinuous behaviour of the stack PSF in this region cannot be controlled well enough. As a result, two-dimensional density mapping is extremely noisy, and consequently we do not provide density maps for them.

\subsection{Mass measurement}

For the interpretation of the WL signal in each of our cluster fields, we follow the scheme described in this section. We follow two different approaches, one ignoring secondary structures along the line of sight (as is commonly done in similar WL studies) and one explicitly modelling these. The following sections detail the components of this procedure.

\subsubsection{Density profile}
\label{sec:nfw}
It is supported by a range of observational and simulation studies that dark matter haloes of clusters of galaxies on average follow the profile described first by NFW. We also adopt this mass profile, with the three-dimensional density $\rho(r)$ at radius $r$ given as
\begin{equation}
\rho(r)=\frac{\rho_0}{(r/r_{\rm s})(1+r/r_{\rm s})^2} \; .
\label{eqn:nfwrho}
\end{equation}
The profile can be rewritten in terms of two other parameters such as mass $M_{200m}$ and concentration $c_{200m}=r_{200m}/r_{\rm s}$ instead of the central density $\rho_0$ and scale radius $r_{\rm s}$. Expressions for the projected density and shear of the NFW profile are given by \citet{1996AuA...313..697B} and \citet{2000ApJ...534...34W}.

The two parameters of the NFW profile are not independent, but rather connected through a concentration-mass relation. This can be used, for instance, in cases where the data are not sufficient for fitting the two parameters simultaneously. It also allows for proper marginalization over concentration from the two-dimensional likelihood in cases where one is only interested in mass. In our analysis, we assume the concentration-mass relation of \citet{2008MNRAS.390L..64D} with a lognormal prior \citep{2001MNRAS.321..559B} with $\sigma_{\log c}=0.18$. Offsets of the assumed concentration-mass relation from the truth, which can be due to differences between assumed and true cosmology or imperfect simulations from which the relation is drawn, impact the mass measurement, however only mildly (see, for example, the discussion in \citealt{2012MNRAS.427.1298H}, their section~4.3).

Unless otherwise noted, we use the brightest cluster galaxy (BCG) position as the centre of the halo (cf. also \citealt{2012MNRAS.427.1298H} and section~\ref{sec:centralshear} for a discussion for choosing the centre for lensing analyses). In order to be less sensitive to miscentring, heavy contamination with cluster members, very strong shears and possible deviations from the NFW profile, a central region of 2~arcmin radius around the BCG is not used in our likelihood analysis (see following section). This is in line with previous studies (cf. \citealt{2012arXiv1208.0605A} and \citealt{2012MNRAS.427.1298H}, who exclude the central projected 750 and 500~kpc, respectively, or \citet{2010MNRAS.405.2078M}, who propose an aperture mass measurement insensitive to surface mass density inside $r_{200c}/5$ and discuss, in more detail, the reasons why this is beneficial for the mass determination). Note, however, that we a posteriori find the measured central shears to be consistent with the prediction from our fit (cf. Section~\ref{sec:centralshear}).

\subsubsection{Likelihood analysis}

Assuming Gaussian errors of the shape estimates, the likelihood of any model can be calculated from the shear catalogue by means of the $\chi^2$ statistics. Given model predictions $\hat{g}_{i,j}$ for the $i$ component of the shear on background galaxy $j$ (which depend on the masses and concentrations of one or more haloes in the model and the estimates of $\beta$ and $\beta^2$ of the background galaxy, cf. Section~\ref{sec:beta}), the likelihood $L$ can be written as
\begin{equation}
 -2\ln \mathcal{L}=\sum_{i,j}\frac{(\hat{g}_{i,j}-\epsilon_{i,j})^2}{\sigma^2_{i,j}+\sigma^2_{\mathrm{int}}} + \mathrm{const}\; .
\label{eqn:likelihood}
\end{equation}
Here $\epsilon_{i,j}$ is the corresponding shape estimate with uncertainty $\sigma_{i,j}$ and the intrinsic dispersion of shapes is $\sigma_{\mathrm{int}}\approx0.25$. The shape measurement uncertainty is calculated by the \textsc{ksbpsfex} pipeline based on linear error propagation and then multiplied with an empirically calibrated factor of $1.4$ to approximate non-linear effects. The latter can be estimated by assuming that the observed variance of shape estimates around the true mean of 0 be equal to the sum of measurement and intrinsic variance, which, unlike the measurement error, is constant under different observing conditions. In addition, the factor is confirmed by a comparison of uncertainties measured on CFHT data with our pipeline with the uncertainties given in the CFHTLenS shape catalogues (cf. \citealt{2013MNRAS.429.2858M} and \citealt{2012MNRAS.427..146H}).

Equation (\ref{eqn:likelihood}) readily allows finding maximum-likelihood solutions. Marginalisation over the concentration parameter can easily be done if a sufficiently large range in concentration around the concentration prior is probed (cf. Section~\ref{sec:nfw}). By means of the $\Delta\chi^2$ method \citep{1976ApJ...210..642A}, we can also determine projected or combined confidence regions in the parameter space of our model.

\subsubsection{Single halo versus multiple halo analysis}
\label{sec:central}
\label{sec:redz}

Our measurement of masses is done with two different fitting procedures, which we denote in the following as the \emph{single halo} and \emph{multiple halo} analysis.

For the single halo analysis, we fit only the central halo of the SZ cluster, placing its centre at the brightest cluster galaxy.

However, no cluster of galaxies is completely isolated in the field. Rather, both uncorrelated structures along the line of sight and correlated structures in the vicinity of the cluster contain additional matter and therefore cause a lensing signal of their own.

It is in this important sense that while lensing accurately weights the \emph{matter} along the line of sight (up to the mass sheet degeneracy), a \emph{mass} estimate for a cluster is the result of our interpretation of the signal. In the same way as in the case of full knowledge (as it is available only in a high-resolution simulation), the mass we determine therefore differs by method, yet in the case of lensing with the additional difficulty of projection.

Several studies in the past have shown that the accuracy of WL analyses depends on proper modelling of such projected structures (see, for example, \citealt{2004PhRvD..70b3008D} and \citealt{2005AuA...442..851M} for minimizing the impact of uncorrelated structures, \citealt{2011MNRAS.412.2095H} for the effect of modelling massive projected structures and \citealt{2011MNRAS.416.1392G} for the impact of correlated structures). Projected structures, depending on their position, can bias the mass estimate in either direction, requiring a modelling of the particular configuration rather than a global calibration.

For each of the systems considered here, our multiple halo analysis is therefore done as follows. From known clusters and density peaks identified visually and in photometric redshift catalogues we compile a list of candidate haloes within the WFI field of view. Details of this procedure are given in the respective subsection of Section~\ref{sec:individual} and Table~\ref{tbl:companions} in the appendix.

We determine the redshift of each of the candidate haloes by three different strategies, depending on the field. In the case of PSZ1 G168.02--59.95 and PSZ1 G099.84+58.45, coverage with spectroscopy from SDSS DR 10 \citep{2014ApJS..211...17A} is used where candidate structure members have been observed. In the fields with $\geq5$ bands, the median photometric redshift of visually selected cluster members is taken. For the remaining seven WFI fields without accurate photometric redshifts, a different scheme is applied.

In these cases, from a list of visually identified red member galaxies, we find the median colours with respect to the R-band magnitude. These could be compared to a red galaxy template, yet it is known that the SED of red galaxies indeed changes with redshift and shows significant variability even at fixed redshift \citep{2013ApJ...768..117G}. We therefore determine the mean \emph{B}-\emph{R}, \emph{V}-\emph{R} and \emph{R}-\emph{I} colours of red galaxies in redshift bins from the DPS catalogues (cf. Section~\ref{sec:refcat}). Fitting a parabola to the squared deviations between these colours and the median colour of red member galaxies of the candidate clusters, we find its minimum $\chi^2$ redshift. Excluding the outlier of SPT-CL J0551--5709 (where our estimate of $z\approx0.30$ deviates from the spectroscopic value of $0.423$, potentially due to residual star-forming activity in cluster member galaxies), this estimate yields a root-mean-squared error of $\sigma_z=0.02$ w.r.t. known redshifts in our WFI cluster fields.

In order to remove the effect of projected structures on the mass estimate of the central halo, we then determine the maximum-likelihood masses for each of the candidate haloes in a combined fit, assuming a fixed concentration-mass relation. As a first-order approximation, in this analysis we assume simple addition of reduced shears due to multiple lenses. Where the best-fitting masses of candidate off-centre haloes deviate from zero, we subtract the signal of the best-fitting haloes from the shear catalogue, on which we then perform the same two-parameter likelihood analysis as in Section~\ref{sec:central}.

\section{Sunyaev-Zel'dovich Catalogues}

\label{sec:psz}
The comparison of our WL results to the SZ effect of the clusters is performed purely on the basis of publicly available catalogues. This section lists the data and our methods of using the catalogues.

\subsection{South Pole Telescope}

The SPT is presently performing multiple surveys of 2500~deg$^2$ of the southern sky in three microwave bands, one of its goals being a cosmological analysis of the SZ signal of clusters of galaxies \citep{2009ApJ...701...32S,2010ApJ...722.1180V,2011ApJ...738..139W,2013ApJ...763..127R}.  Five of our clusters have been observed in SPT surveys and four of them are indeed SPT discovered. 

SPT mass estimates are based on the detection significance $\zeta$ in the band where it is maximal. In this work, we take \citet{2010ApJ...722.1180V} as the primary reference for mass calibration. They calibrate the mass-significance relation in SPT empirically as
\begin{equation}
 \zeta = A\times\left(\frac{M_{200m}}{5\times10^{14}h^{-1}\Msol }\right)^B\times\left(\frac{1+z}{1.6}\right)^C \; ,
\label{eqn:sptmor}
\end{equation}
with parameters calibrated from simulations as $A=5.62$, $B=1.43$ and $C=1.40$ \citep{2010ApJ...722.1180V}. Here $\zeta$ is the SPT detection significance corrected from the observed significance $\xi$ for several noise dependent biases, namely first the preferential selection of objects for which intrinsic and measurement noise have a positive contribution to the signal and secondly the preferential up-scatter of intrinsically less massive objects because of the steepness of the mass function. These effects are sometimes called the Malmquist and Eddington bias, and are indeed related to the works by Malmquist (although only widely and with varying definition) and \citet{1913MNRAS..73..359E}. A third effect particular to SPT is an additional bias on measured significances, caused by the fact that these are measured at the position where they are highest (instead of the unknown true position of the cluster). While the first and third effect can be corrected directly, the second requires to assume a mass function, which in turn depends on cosmology. Where we use $\zeta$ in our analysis, we therefore calculate it from the mass estimate published by the SPT collaboration (\citealt{2010ApJ...722.1180V} for two of our clusters, \citealt{2013ApJ...763..127R} for two of the clusters where new spectroscopic redshifts had become available in the meantime and \citealt{2011ApJ...738..139W} for SPT-CL J2248--4431, all of which have made corrections for all three effects), which we insert into equation (\ref{eqn:sptmor}) or (\ref{eqn:sptmorr}).

As an alternative calibration based on a very similar scheme, we also compare to the mass estimates of \citet{2013ApJ...763..127R}. Using a different mass definition and redshift scaling, \citet{2013ApJ...763..127R} and \citet{2013ApJ...763..147B} define the MOR as
\begin{equation}
\zeta=A\times\left(\frac{M_{500c}}{3\times10^{14}h^{-1}\Msol}\right)^{B}\times\left(\frac{E(z)}{E(0.6)}\right)^C \; ,
\label{eqn:sptmorr}
\end{equation}
with $E(z)=\sqrt{\Omega_{\rm m}(1+z)^3+\Omega_{\Lambda}}$. The parameters are determined from survey simulations as $A=6.24\pm1.87$, $B=1.33\pm0.27$ and $C=0.83\pm0.42$ \citep{2013ApJ...763..127R}, taking into account modelling uncertainty. Posterior scaling parameters from the combined cosmological analysis including X-ray cluster measurements and CMB are reported in \citet{2013ApJ...763..147B} as $A=4.91\pm0.71$, $B=1.40\pm0.15$ and $C=0.83\pm0.30$ with an intrinsic scatter of $\zeta$ at fixed mass corresponding to lognormal $\sigma_{\mathrm{int},\log_{10}}=0.09\pm0.04$.

\subsection{\emph{Planck}}

\label{sec:planckmor}

\emph{Planck} is a mission complementary to SPT in its SZ applications. Unlike SPT, which has imaged a relatively small area, \emph{Planck} detects an all-sky catalogue of the SZ-brightest objects with sufficient angular extent. However, while the SPT produces arcminute-resolution images, the Planck SZ analyses are complicated by the much lower resolution (above 4~arcmin full width at half-maximum even for the bands with smallest beam size), which leaves most clusters unresolved and high redshift clusters hard to detect, and the inhomogeneity of noise over the observed area.

For this reason, Compton parameters $Y$ are provided by \citet{2013arXiv1303.5089P} only in the form of a two-dimensional likelihood grid in terms of $Y=Y_{5\theta_{500c}}^{\rm cyl}$, integrated inside a cylindrical volume of $5\times\theta_{500c}$ diameter, and the angular size $\theta_{\rm s}=r_{\rm s}/D_A(z)$, where $r_{\rm s}$ is the scale radius of the generalized NFW profile \citep[cf.][]{2010AuA...517A..92A} of the intracluster gas.\footnote{For a detailed description of the format, please refer to \texttt{http://www.sciops.esa.int/wikiSI/planckpla/index.php?\\title=Catalogues\&instance=Planck\_Public\_PLA}} Due to the large beam size, the Compton parameter can only be confined well if prior information on $\theta_{\rm s}$ is available. 

\subsubsection{De-biasing}

As a first step, Compton parameters $Y$ measured near the detection limit must be (Malmquist) noise-bias corrected. At the detection limit, there exists a selection effect based on the preferential inclusion (exclusion) of objects whose signal scatters up (down) from the fiducial value. Consequently, a majority of objects near the detection limit have a positive noise and intrinsic scatter contribution.\footnote{This is true even without the Eddington bias (preferential up-scatter of less massive haloes due to the steepness of the mass function), an effect ignored here for consistency with the \citet{2013arXiv1303.5089P} calibration.} The resulting multiplicative bias in 
\begin{equation}
Y_{\rm obs}=b_m\times Y_{\rm true} 
\end{equation}
can be estimated as \citep{2009ApJ...692.1033V}
\begin{equation}
 b_m=\exp\left[\sigma\times \frac{\exp(-x^2/2\sigma^2)}{\sqrt{\pi/2}\mathrm{erfc}(x/\sqrt{2}\sigma)} \right] \; ,
\label{eqn:malm}
\end{equation}
where 
\begin{eqnarray}
x&=&-\ln[(\mathrm{S/N})_0/(\mathrm{S/N})_{\rm cut}] \; , \\
\label{eqn:xsigma}
\sigma&=&\sqrt{\ln^2\left((\mathrm{S/N}+1)/(\mathrm{S/N})\right)+\ln^2(10\sigma_{\rm int, \log10})} \; , 
\end{eqnarray}
with the nominal S/N according to the MOR at the mass of the cluster $(\mathrm{S/N})_0$, the threshold of the catalogue $(\mathrm{S/N})_{\rm cut}=4.5$ and the ln-normal scatter of the MOR $\sigma$. We adopt a value of the lognormal intrinsic scatter of $\sigma_{\rm int,\log10}\approx0.07$ \citep{2013arXiv1303.5089P}. As done by \citet{2013arXiv1303.5089P}, we simply divide the Compton decrement by $b_m$.\footnote{We consistently work with the natural logarithm here. \citet{2013arXiv1303.5089P} appear to use $\log$ and $\ln$ interchangeably in their equation 8 and the following description.}

For the seven \emph{Planck}-detected clusters in the WISCy sample, $b_m\in(1.00,1.10)$. Note that for a higher $\sigma_{\mathrm{int},\log_{10}}$ as suggested by previous studies \citep{2009ApJ...701L.114M,2012MNRAS.427.1298H,2012ApJ...754..119M}, $b_m$ and with it $Y_{\rm true}$ and estimated mass for our clusters change significantly, in particular for the low S/N detections.

\subsubsection{$\theta_{\rm s}$-$Y$ degeneracy}

\begin{figure}
\centering
\includegraphics[width=0.48\textwidth]{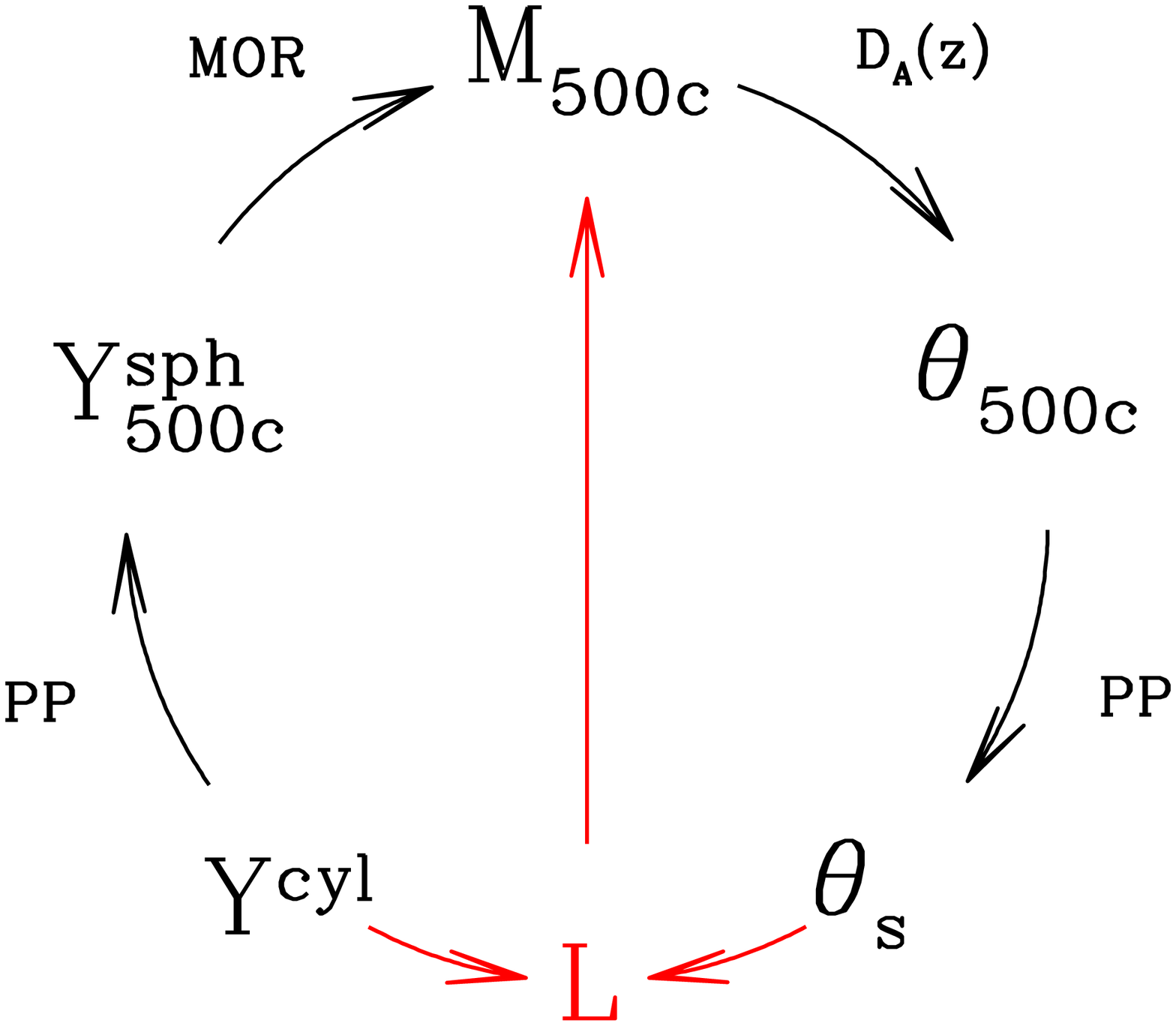}
\caption{Self-consistent mass estimation from \emph{Planck} likelihood in $\theta_{\rm s}$-$Y^{\rm cyl}$ space. The diagram illustrates how, assuming a pressure profile (PP), a mass-observable relation (MOR) and the angular diameter distance of the cluster $D_{\rm A}(z)$, $\theta_{\rm s}$ is a function of $Y^{\rm cyl}$ (and vice versa). The likelihood for mass need then be evaluated only for consistent tuples of $Y^{\rm cyl}$ and $\theta_{\rm s}$.}
\label{fig:sc}
\end{figure}

How can the degeneracy between angular size and integrated Compton parameter be broken? In this work we use two different schemes of calculating mass estimates from the Planck SZ signal. For their own calibration, the Planck Collaboration uses subsets of clusters with X-ray mass estimates. From these and assumptions about the pressure profile (PP; see below) they derive $\theta_{\rm s}$, thereby breaking the $\theta_{\rm s}$-$Y$ degeneracy. We perform a similar analysis, using X-ray mass estimates for our sample from various sources. The resulting SZ mass estimates, however, are not made truly independently from the SZ information, since in fact prior information on the mass itself is used. In addition, we therefore develop a scheme of self-consistent SZ masses, similar to the one suggested by \citet[][their section~7.2]{2013arXiv1303.5089P}. It is illustrated in Fig.~\ref{fig:sc}.

\begin{figure}
\centering
\includegraphics[width=0.48\textwidth]{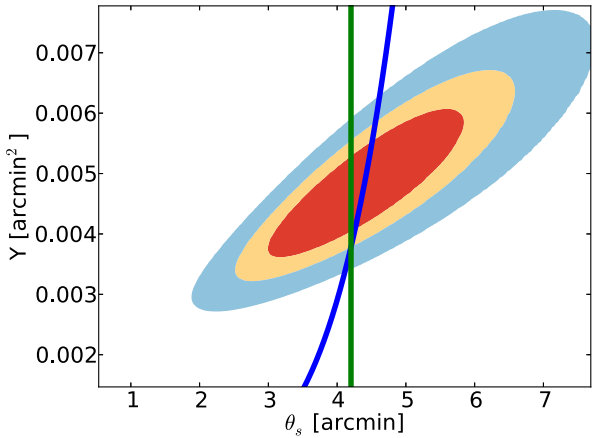}
\caption{Mass estimation from \emph{Planck} likelihood in $Y^{\rm cyl}$-$\theta_{\rm s}$ space for SPT-CL J2248--4431. From the tabulated likelihood \citep{2013arXiv1303.5089P} we determine the confidence interval in $Y$ using a fixed prior for the scale angle from X-ray measurements (green line) and using self-consistency with the MOR (blue line). Colour contours indicate the 68, 90 and 99 per cent confidence levels for two degrees of freedom.}
\label{fig:psz2248}
\end{figure}

Likelihoods are given by \citet{2013arXiv1303.5089P} in terms of cylindrically integrated Compton parameters $Y$ and angular scale radii $\theta_{\rm s}$.\footnote{These likelihoods are provided by \citet{2013arXiv1303.5089P} for different extraction methods. We will use the Matched Multi-filter method (\textsc{MMF3}) catalogues of \citet{2006A&A...459..341M} for consistency with \citet{p2013cosmology} in version R1.11.} A value for $\theta_{500c}$ corresponding to an external X-ray mass estimate can be converted to $\theta_{\rm s}$ by means of the PP concentration parameter $c_{P,500c}$=1.1733 \citep{2010AuA...517A..92A,2013arXiv1303.5089P}. At the fixed value of $\theta_{s}$, we obtain a confidence interval of $Y$ (see Fig.~\ref{fig:psz2248} for an illustration), which can be converted to the spherical estimate $Y_{500c}$ and to $M_{500c}$ with the prescriptions of \citet{2011AuA...536A..11P,2013arXiv1303.5089P}, namely
\begin{equation}
 Y_{500c} = 0.5567\times Y
\label{eqn:yy500}
\end{equation}
(cf. also \citealt{2010AuA...517A..92A}, their section~6.3.1) and
\begin{equation}
 E(z)^{-2/3}\times\left(\frac{D_A^2\times Y_{500c}}{\mathrm{Mpc}^2}\right)=10^A\times\left(\frac{M_{500c}\times(1-b)}{6\times10^{14}h_{70}^{-1}\Msol}\right)^B
\label{eqn:pszm}
\end{equation}
with $E(z)=\sqrt{\Omega_{\rm m}(1+z)^3+\Omega_{\Lambda}}$ and fiducial calibration parameters $A=-4.19\pm0.02$, $B=1.79\pm0.08$ and $b=0.2^{+0.1}_{-0.2}$ \citep[cf.][their equation A.15]{p2013cosmology}. We find that the value of $Y_{500c}^{R_X}=(2.36\pm0.20)\times10^{-4}$~Mpc$^{-2}$ determined such from the \citet{2013arXiv1303.5089P} likelihoods for SPT-CL J2248--4431 is consistent with the value given by \citet{2011AuA...536A..11P} from earlier \emph{Planck} data, also using X-ray priors on size, of $Y_{500c}^{R_X}=(2.21\pm0.16)\times10^{-4}$~Mpc$^{-2}$. For PLCKESZ G287.0+32.9 and PLCKESZ G292.5+22.0, \citet{2011AuA...536A...8P} give Compton decrements as $Y=(6.1\pm0.6)\times10^{-3}$arcmin$^2$ and $Y=(3.7\pm0.6)\times10^{-3}$arcmin$^2$. These are again consistent with the values we extract of $Y=(6.16\pm0.52)\times10^{-3}$ and $Y=(3.16\pm0.52)\times10^{-3}$arcmin$^2$, respectively.

The second option we use is to calculate, by way of equations \ref{eqn:yy500} and \ref{eqn:pszm}, for any $Y$ the corresponding 
\begin{equation}
\theta_{\rm s}=\theta_{500c}(M_{500c}(Y))/c_{P,500c} \;.
\label{eqn:selfcons}
\end{equation}
For the PP concentration we again use the fixed value of $c_{P,500c}=1.1733$. 

Equation (\ref{eqn:selfcons}) is a criterion of self-consistency, namely that the analysis result in a mass to which corresponds the virial radius used in the extraction of the signal. The likelihood is then evaluated at the combination of $Y$ and $\theta_{\rm s}$ matched by equation (\ref{eqn:selfcons}). Thereby we reduce the dimensionality of the likelihood analysis by one, applying only knowledge about the angular diameter distance to the system and no mass information external to the Planck SZ signal. An illustration of the method is given in Fig.~\ref{fig:psz2248} for one of our clusters.

\begin{table*}
\begin{center}
\begin{tabular}{|l|r|r|r|r|r|r|}
\hline
SZ name             & $\theta_{500c}$ & S/N & $Y_{500c}^{R_X}$ & $M^{\mathrm{SZ},R_X}_{500c}$ & $M^{\rm SZ,sc}_{500c}$ \\ \hline

PLCKESZ G287.0+32.9 & 4.85            & 17.2 & $3.43\pm0.29$     & $17.7^{+0.8}_{-0.9}$ $^{+1.7}_{-1.5}$ $^{+2.7}_{-3.6}$ & $18.2^{+1.0}_{-0.9}$ $^{+1.7}_{-1.6}$ $^{+2.8}_{-3.7}$ \\ \hline

PLCKESZ G292.5+22.0 & 5.00            &  8.9 & $1.76\pm0.29$     & $10.3^{+0.9}_{-1.0}$ $^{+1.0}_{-0.9}$ $^{+1.5}_{-2.1}$ & $10.6^{+1.1}_{-1.1}$ $^{+1.0}_{-0.9}$ $^{+1.5}_{-2.1}$ \\ \hline

SPT-CL J2248--4431  & 4.93            & 16.7 & $2.71\pm0.23$     & $14.5^{+0.7}_{-0.7}$ $^{+1.4}_{-1.2}$ $^{+2.1}_{-2.9}$ & $14.8^{+0.8}_{-0.8}$ $^{+1.4}_{-1.3}$ $^{+2.2}_{-3.0}$ \\ \hline

PSZ1 G168.02--59.95 & 5.97            &  5.4 & $1.21\pm0.33$     & $4.6^{+0.6}_{-0.8}$ $^{+0.4}_{-0.4}$ $^{+0.7}_{-0.9}$  & $5.0^{+0.9}_{-0.9}$ $^{+0.5}_{-0.4}$ $^{+0.7}_{-1.0}$  \\ \hline

PSZ1 G230.73+27.70  & 3.71            &  5.4 & $0.69\pm0.20$     & $6.0^{+0.9}_{-1.0}$ $^{+0.6}_{-0.5}$ $^{+0.9}_{-1.2}$  & $6.2^{+1.2}_{-1.3}$ $^{+0.6}_{-0.5}$ $^{+0.9}_{-1.3}$  \\ \hline

PSZ1 G099.84+58.45  & 2.98            &  5.9 & $0.58\pm0.14$     & $8.6^{+1.1}_{-1.2}$ $^{+0.8}_{-0.7}$ $^{+1.2}_{-1.7}$  & $8.3^{+1.2}_{-1.3}$ $^{+0.8}_{-0.7}$ $^{+1.2}_{-1.7}$  \\ \hline

PSZ1 G099.48+55.62  & 6.27            &  7.0 & $1.08\pm0.23$     & $3.2^{+0.4}_{-0.4}$ $^{+0.3}_{-0.3}$ $^{+0.5}_{-0.7}$  & $3.8^{+0.5}_{-0.5}$ $^{+0.4}_{-0.3}$ $^{+0.6}_{-0.8}$  \\ \hline
\end{tabular}
\end{center}
\caption{Mass estimates from the \citet{2013arXiv1303.5089P} SZ catalogue. $M^{\mathrm{SZ},R_X}_{500c}$ is calculated using X-ray mass estimates for determining the aperture size $\theta_{500c}$. $M^{\rm SZ,sc}_{500c}$ uses the self-consistent mass determination algorithm described herein. All Compton parameters have been corrected for Malmquist bias by means of equation (\ref{eqn:malm}) according to their S/N in the MMF3 catalogue.  Masses are given in units of $10^{14}h_{70}^{-1}\Msol$. The Compton parameter integrated inside a sphere of the size of the X-ray $r_{500c}$, $Y_{500c}^{R_X}$, is given in units of $10^{-3}$~arcmin$^2$. The $M^{\rm SZ}$ estimates are listed with statistical, intrinsic and systematic uncertainties.}
\label{tbl:psz}
\end{table*}

\subsubsection{Mass estimates}

We give confidence intervals for the mass estimates based on the \citet{2013arXiv1303.5089P} catalogues in Table~\ref{tbl:psz}, listing three separate uncertainties: 
\begin{itemize}
 \item the statistical uncertainty of $Y_{500c}$, extracted from the likelihood as described above,
 \item the intrinsic scatter of $Y_{500c}$ at fixed mass (and, conversely, mass at fixed $Y_{500c}$, which we approximate as $\sigma_{\log Y_{500c}|M_{500c}}=0.07$, and
 \item the systematic uncertainty based on the confidence regions of parameters $A$, $B$ and $b$ in equation (\ref{eqn:pszm}), where we assume all deviations from the fiducial values to be uncorrelated.
\end{itemize}

\section{Individual Cluster Analysis}
\label{sec:individual}

This section presents the individual analyses of the WISCy sample. The subsections are ordered by cluster ID (cf. Table~\ref{tbl:list}) and contain information on the following:
\begin{enumerate}
 \item \textit{Visual appearance.} Including a colour image of the central parts and a map of projected galaxy density,
 \item \textit{Previous work.} With a review of published optical, spectroscopic, X-ray, radio and SZ observations and known neighbouring projected structures,
 \item \textit{Weak lensing analysis.} Including shear profile, individual mass estimates and remarks on particularities of the analysis, especially in terms of the modelling of nearby structures, and
 \item \textit{Strong lensing.} With a review of previous strong lensing analyses and strong lensing candidates discovered in our cluster fields.
\end{enumerate}

\subsection{SPT-CL J0509--5342}

\subsubsection{Visual Appearance}

\begin{figure}
\centering
\includegraphics[width=0.41\textwidth]{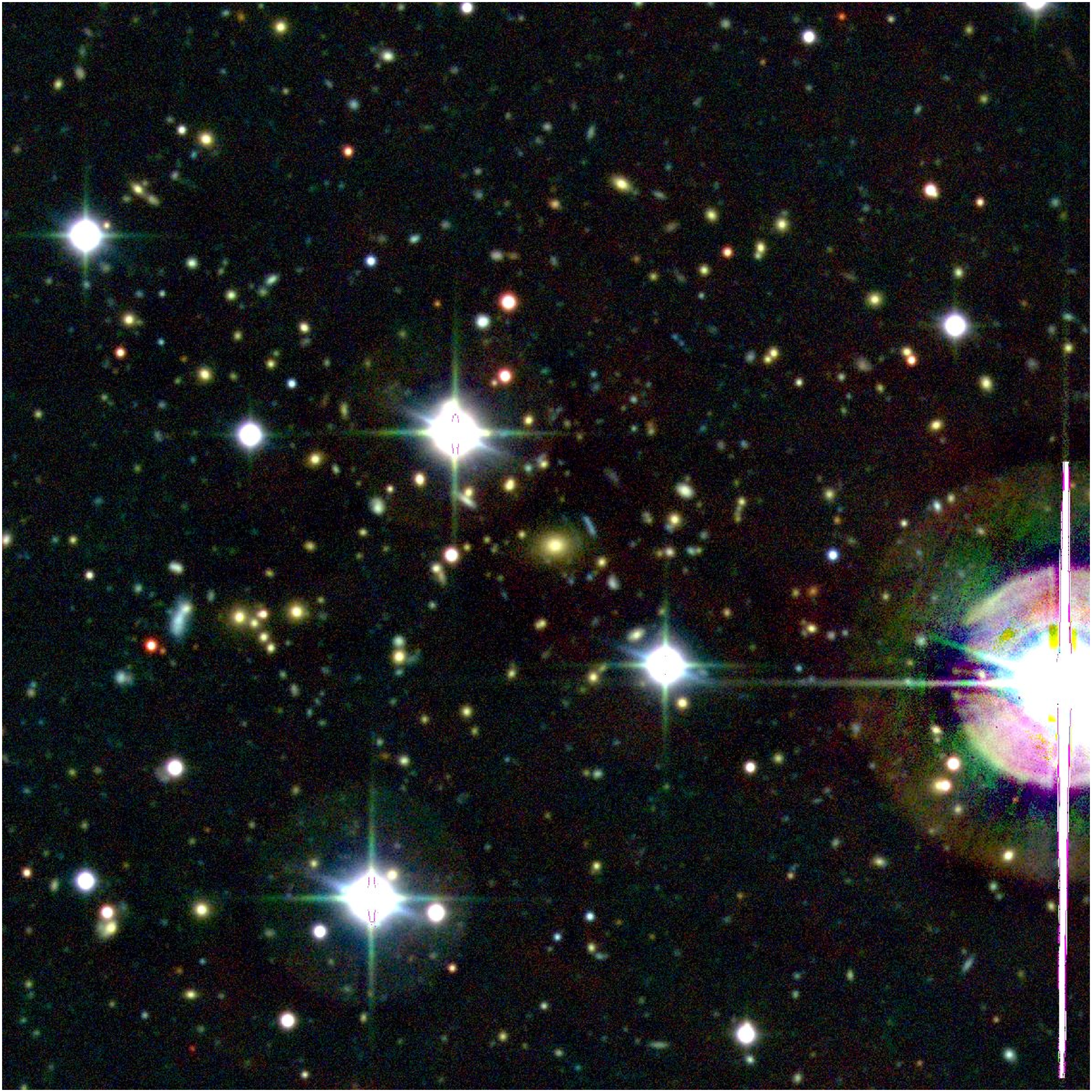}
\includegraphics[width=0.47\textwidth]{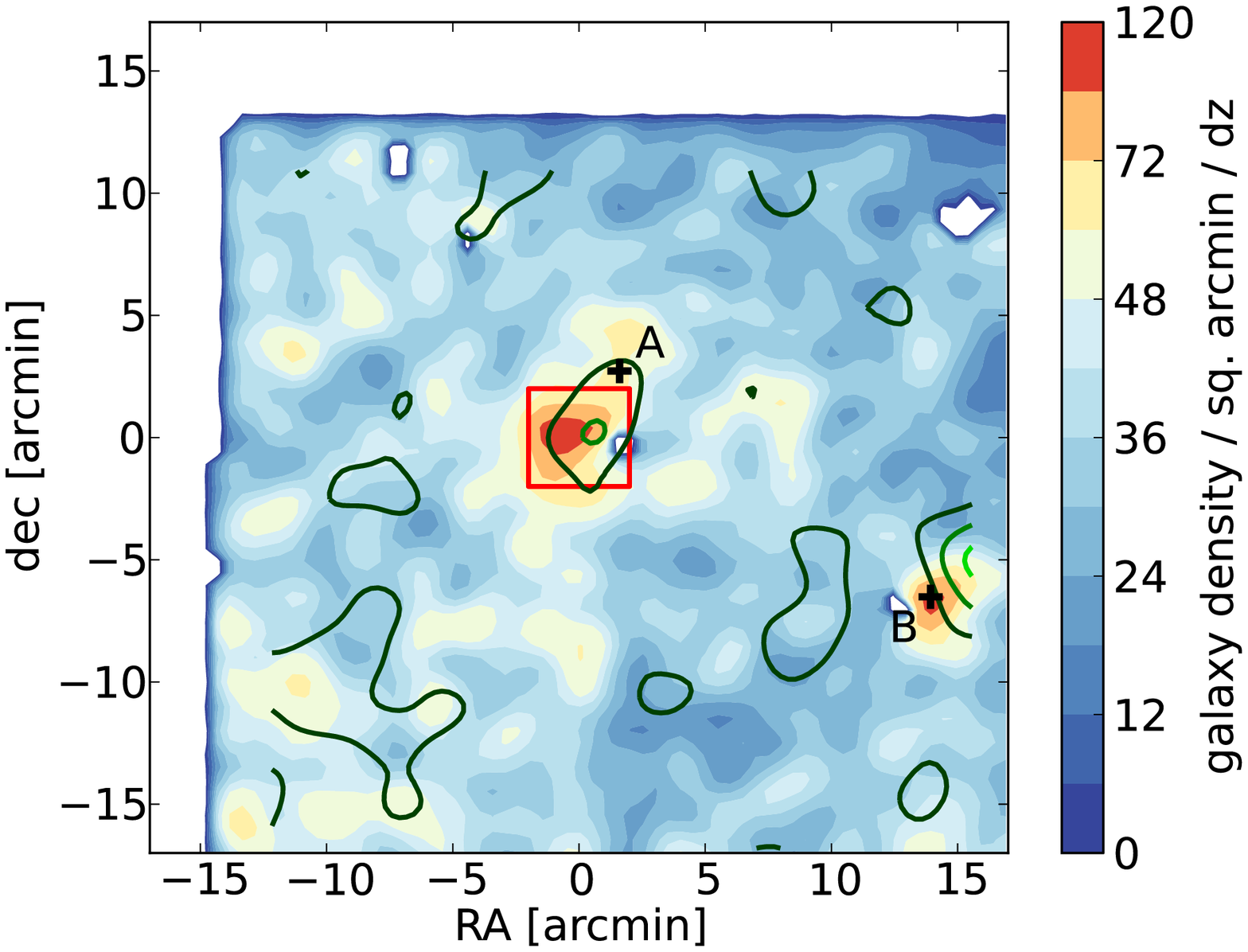}
\includegraphics[width=0.43\textwidth]{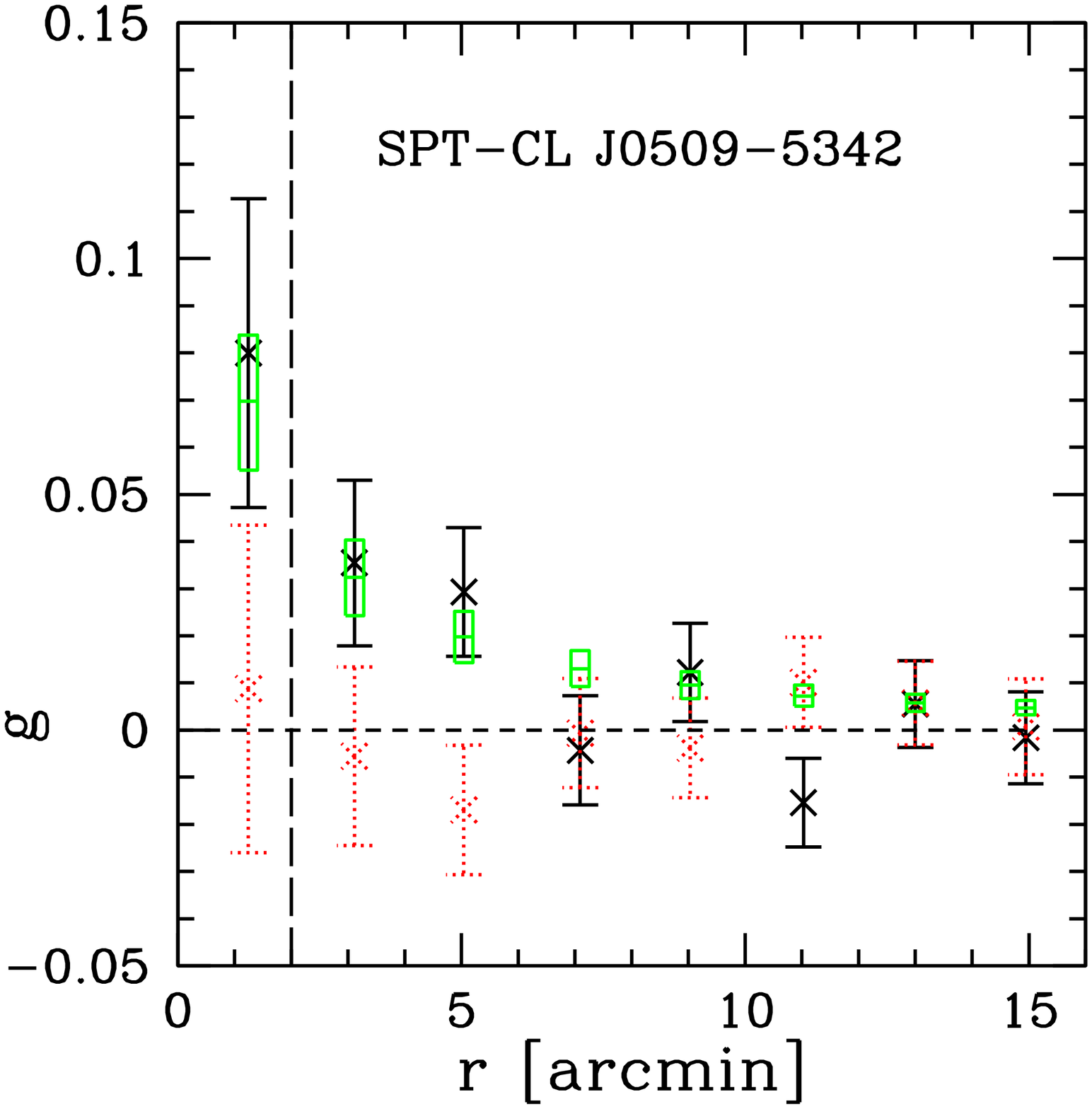}
\caption{Colour image (top panel) from \emph{VRI} frames of the central $4\times4$~arcmin region of SPT-CL J0509--5342, corresponding to the red square in the three-dimensional galaxy density map (central panel) with overlaid $\kappa$ contours at $\kappa=0.05,\,0.10,\,0.15$. The bottom panel shows shear profile with tangential shear (black), cross-shear (red, dotted) and NFW fit (green boxes), using only data $>2$~arcmin from the centre (long dashed line).}
\label{fig:picspt4}
\end{figure}

\label{sec:spt4visual}

The centre of SPT-CL J0509--5342 (cf. Fig.~\ref{fig:picspt4}, top panel, showing a $4\times4$~arcmin$^2$ region) is marked by a diffuse BCG, embedded in a number of relatively bright foreground stars and neighboured by at least two giant gravitational arcs of background galaxies strongly lensed by the system.

The central panel of Fig.~\ref{fig:picspt4} shows the three-dimensional density of galaxies in a 30$\times$30~arcmin map around the cluster, smoothed with a minimum-variance (Epanechnikov) kernel of radius 1~arcmin and corrected for masked area.  Galaxies are selected by their photometric redshift within $|z-z_{\rm cl}|\leq 0.06\times(1+z_{\rm cl})$. Note that the photometric redshifts used here are very noisy due to the small number of bands and are therefore provided for illustrative purposes only. The colour image cutout, as in the respective figures of all following clusters, corresponds to the red frame in the overview map (central panel). 

\subsubsection{Previous Work}

\citet{2009ApJ...701...32S} describe the system as one of the first clusters discovered in an SZ survey. Its detection significance is quoted by \citet{2010ApJ...722.1180V} as $\xi=6.61\sigma$. \citet{2010ApJ...723.1523M} independently discover the SZ signal of the cluster and label it as ACT-CL J0509-5341. The cluster coincides with the source 1RXS J050921.2-534159 from the \emph{ROSAT} faint source catalogue \citep{2000IAUC.7432R...1V}. 

\citet{2010ApJ...723.1736H} determine a spectroscopic redshift of the cluster of $z=0.4626$, which we adopt in this analysis. The independent spectroscopy of \citet{2010ApJ...723.1523M} yields $z=0.461$.

Using \emph{griz} data from the Blanco Cosmology Survey and Magellan, \citet{2010ApJ...723.1736H} provide an optical richness of $N_{\mathrm{gal}}=41\pm8$, corresponding to their mass estimate of $M_{200m}(N_{\mathrm{gal}})=(3.3\pm2.0\pm1.0)\times10^{14}h^{-1}\Msun$. \citet{2011ApJ...734....3Z} fit the luminosity function and the number density profile of cluster member galaxies, noting that there is no indication that this and the other clusters in their SZ-selected sample differ in their properties from samples selected by other kinds of observations.

\citet{2013ApJ...772...25S} use spectroscopic redshifts of 76 cluster members to determine a velocity dispersion of $(846\pm111)$km s$^{-1}$ and a corresponding dynamical mass estimate of $M_{200c}=(5.5\pm2.1)\times10^{14} h_{70}^{-1}\Msol$.

The SZ mass of the system is estimated by \citet{2010ApJ...722.1180V} based on the significance-mass relation, yielding $M_{200m}=(5.09\pm_{\rm stat}1.02\pm_{\rm sys}0.69)\times10^{14}h^{-1}\Msun$. The unbiased significance according to equation (\ref{eqn:sptmor}) is $\zeta=5.08$.

The independent discovery of the cluster by ACT is listed as ACT-CL J0509--5341 \citep{2011ApJ...737...61M}. The SZ mass is given by \citet{2013arXiv1301.0816H} as $M_{500c}=(4.0\pm0.8)\times10^{14} h_{70}^{-1} \Msun$.

\citet{2011ApJ...738...48A} present the system in their X-ray measurements of SPT clusters with \emph{Chandra}, finding $T=7.0^{+1.4}_{-1.1}$keV, $M_{g,500}=5.6^{+0.2}_{-0.2}\times10^{13}h_{70.2}^{-1}\Msol$ and a corresponding $Y_{X,500c}=(4.3\pm0.8)\times10^{14}h_{70.2}^{-1}\Msol$keV. Using the MORs of \citet{2009ApJ...692.1033V} with mean values calibrated from local clusters under the assumption of HSE, they estimate $M_{500c}=(5.43\pm0.60)\times10^{14} h_{70.2}^{-1}\Msun$ from $Y_{X,500c}$ and $M_{500c}=(6.71\pm1.69)\times10^{14} h_{70.2}^{-1}\Msun$ from the temperature. In this work, we will adopt the weighted mean value $M_{500c}=(5.57\pm0.56)\times10^{14} h_{70.2}^{-1}\Msun$.

\citet{2011ApJ...738...48A} classify SPT-CL J0509--5342 as a merger at an early stage based on the double-peaked X-ray centre. In contrast, \citet{2012ApJ...761..183S} find the system to have a moderate cool core.

\citet{2013ApJ...763..127R} quote a combined SZ+X-ray mass of $M_{500c}=(5.36\pm0.71)\times 10^{14} h_{70}^{-1} \Msun$.

\citet{2009MNRAS.399L..84M} perform a WL analysis of the system as one of the first SZ detected clusters. They calculate KSB shapes on \emph{i} band and photometric redshifts on \emph{g}-, \emph{r}-, \emph{i}-, and \emph{z}-band public imaging data from the Blanco Cosmology Survey. Their mass estimate from an NFW fit is $M_{200m}=3.54^{+2.07}_{-1.68}h_{71}^{-1}\times10^{14}\Msol$.

\subsubsection{Weak Lensing Analysis}

The $\kappa$ map (Fig.~\ref{fig:picspt4}, central panel) contains peaks at the centre of SPT-CL J0509--5342 and at a secondary location near the western edge of the field (B). The latter is the location of the candidate strong lensing feature discussed in the following section and shown in Fig.~\ref{fig:spt4sl}.

A single halo fit, fixing the centre at the BCG and marginalizing over concentration, yields $M_{200m}=(8.4^{+3.4}_{-2.9})\times h_{70}^{-1}10^{14}\Msol$ ($M_{500c}=(4.7^{+1.9}_{-1.7})\times h_{70}^{-1} 10^{14} \Msol$). The observed shear profile and the confidence interval of the NFW fit are shown in Fig.~\ref{fig:picspt4} (bottom panel). 

Given the correlated shape noise, this result is inconsistent with the earlier analysis of \citet{2009MNRAS.399L..84M}, although without a cross-match to their photometric redshift catalogue the significance of inconsistency cannot be estimated confidently. We hypothesize that the photometric redshifts calculated by \citet{2009MNRAS.399L..84M} on four bands only, the bluest being the \emph{g} band, could be a cause for a low bias in their mass estimate (cf. Section~\ref{sec:zcomp} and \citealt{2012arXiv1208.0605A}). We note that there is additional evidence from the strong lensing feature (cf. Section~\ref{sec:spt4sl}) and the consistency of our result with all SZ and X-ray based measurements supporting our higher mass estimate.

After subtracting the maximum likelihood signal of surrounding structures fitted simultaneously, the resulting mass estimate for the central halo is slightly lower at $M_{200m}=(6.6^{+3.1}_{-2.6})\times h_{70}^{-1}10^{14}\Msol$ ($M_{500c}=(3.8^{+1.7}_{-1.5})\times h_{70}^{-1} 10^{14} \Msol$).

Both our mass estimates are in agreement with the SZ masses of \citet{2010ApJ...722.1180V} and \citet{2013arXiv1301.0816H}, the X-ray masses of \citet{2011ApJ...738...48A}, the combined SZ+X-ray mass of \citet{2013ApJ...763..127R} and the dynamical mass of \citet{2011ApJ...738...48A}.

\subsubsection{Strong Lensing}
\label{sec:spt4sl}
Two giant arcs are visible in Fig.~\ref{fig:picspt4} towards the north-eastern direction from the BCG, as previously reported by \citet{2010ApJ...723.1523M}. At a separation of $\theta_1=4.8$~arcsec and $\theta_2=8.7$~arcsec from the BCG they correspond to the Einstein radii of the best-fitting NFW profile at source redshifts $z_{s,1}=1.75$ and $z_{s,2}=5.7$. For the 68\% upper limit of our NFW fit, the corresponding redshifts are at more plausible values of $z_{s,1}=1.37$ and $z_{s,2}=2.6$. 

We note that these strong lensing features pose additional evidence for our mass estimate that significantly exceeds the previous one by \citet{2009MNRAS.399L..84M}. An NFW profile following the concentration-mass relation \citep{2008MNRAS.390L..64D} at their best-fitting mass (their upper limit), $M_{200m}=3.59\times10^{14}h_{70}^{-1}\Msol$ ($5.69\times10^{14}h_{70}^{-1}\Msol$), has a limiting Einstein radius of 4.7~arcsec (7.6~arcsec) for $z_s\rightarrow\infty$. Likewise, at their best-fitting mass, a high concentration of $c_{200m}=8.6$ (an outlier from the mean by $\approx1.5\sigma$) is required to reproduce the outer arc for a hypothetical source redshift of $z_s=2$. Since the BCG is visibly elliptical at an axis ratio of $A/B=1.27$, a strongly increased projected concentration is, however, unlikely.

\begin{figure}
\centering
\includegraphics[width=0.28\textwidth]{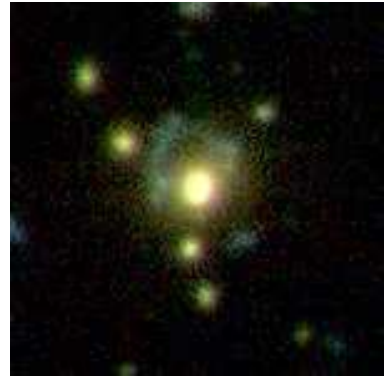}
\caption{Colour image of strong lensing candidate in the field of SPT-CL J0509--5342. The cutout is centred at $(\alpha,\delta)=(05^{\rm h}07^{\rm m}47.1^{\rm s},-53^{\circ}48'40.9'')$ and $0.5\times0.5$~arcmin$^2$ in size.}
\label{fig:spt4sl}
\end{figure}

In addition, the field contains a strangely shaped blue structure around a red galaxy at the centre of a group in the west of the main cluster (Fig.~\ref{fig:spt4sl}). At a radius of $r\approx4$~arcsec and for the 68\% upper limit of the group mass of $M_{200m}=1.8\times10^{14}\Msol$ (although not constrained well due to its vicinity to the edge of the field) and red sequence colour redshift (cf. Section~\ref{sec:redz}) of $z_l=0.41$, the Einstein radius for plausible source redshifts is smaller by a factor of at least $1/2$. Considering this, the inhomogeneity of the surface brightness of the feature and its unusual geometry, we conclude that the candidate is more likely to be a pair of interacting galaxies, yet higher resolution imaging would be needed to unambiguously explore its nature.

\subsection{SPT-CL J0551--5709}

\subsubsection{Visual Appearance}

\begin{figure}
\centering
\includegraphics[width=0.44\textwidth]{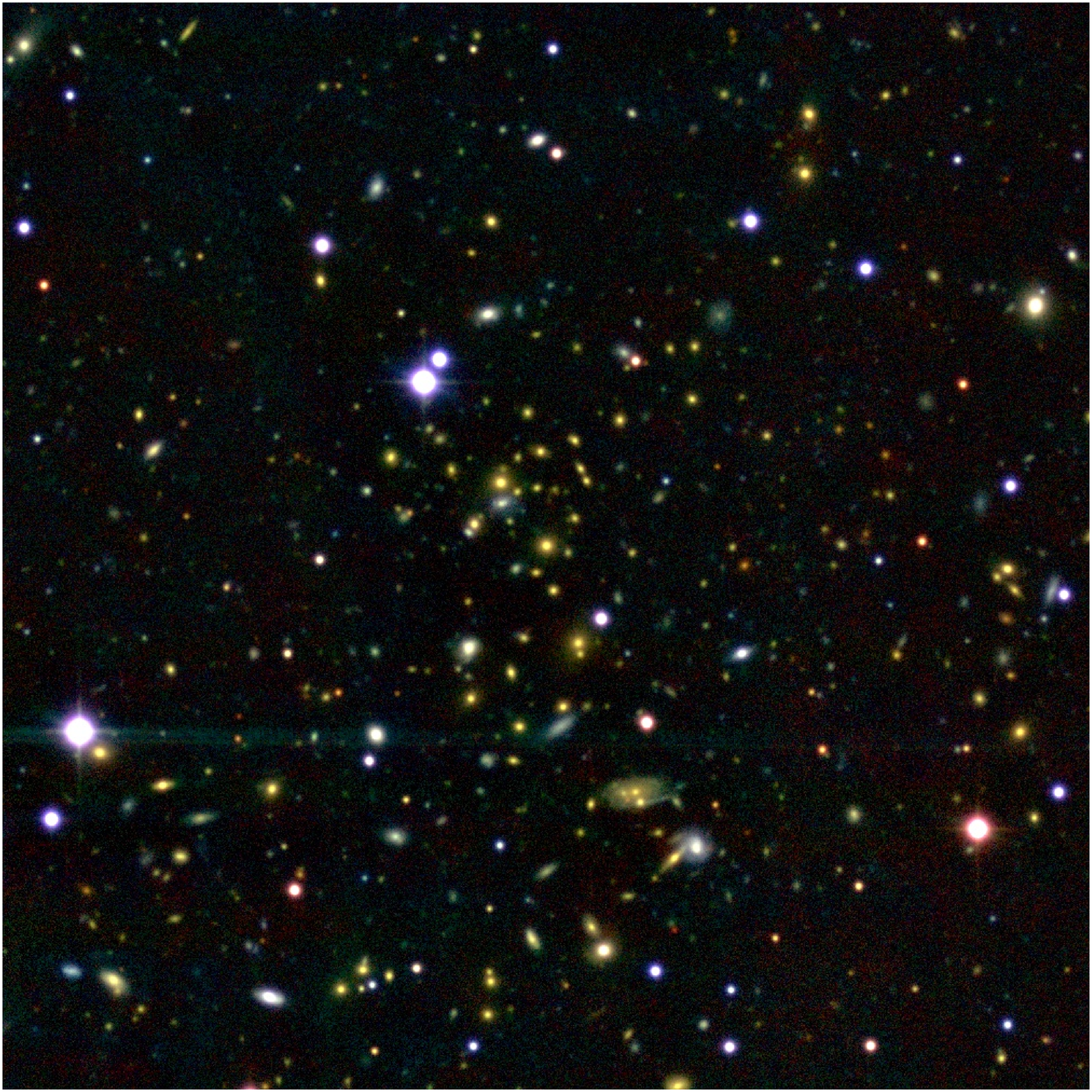}
\includegraphics[width=0.47\textwidth]{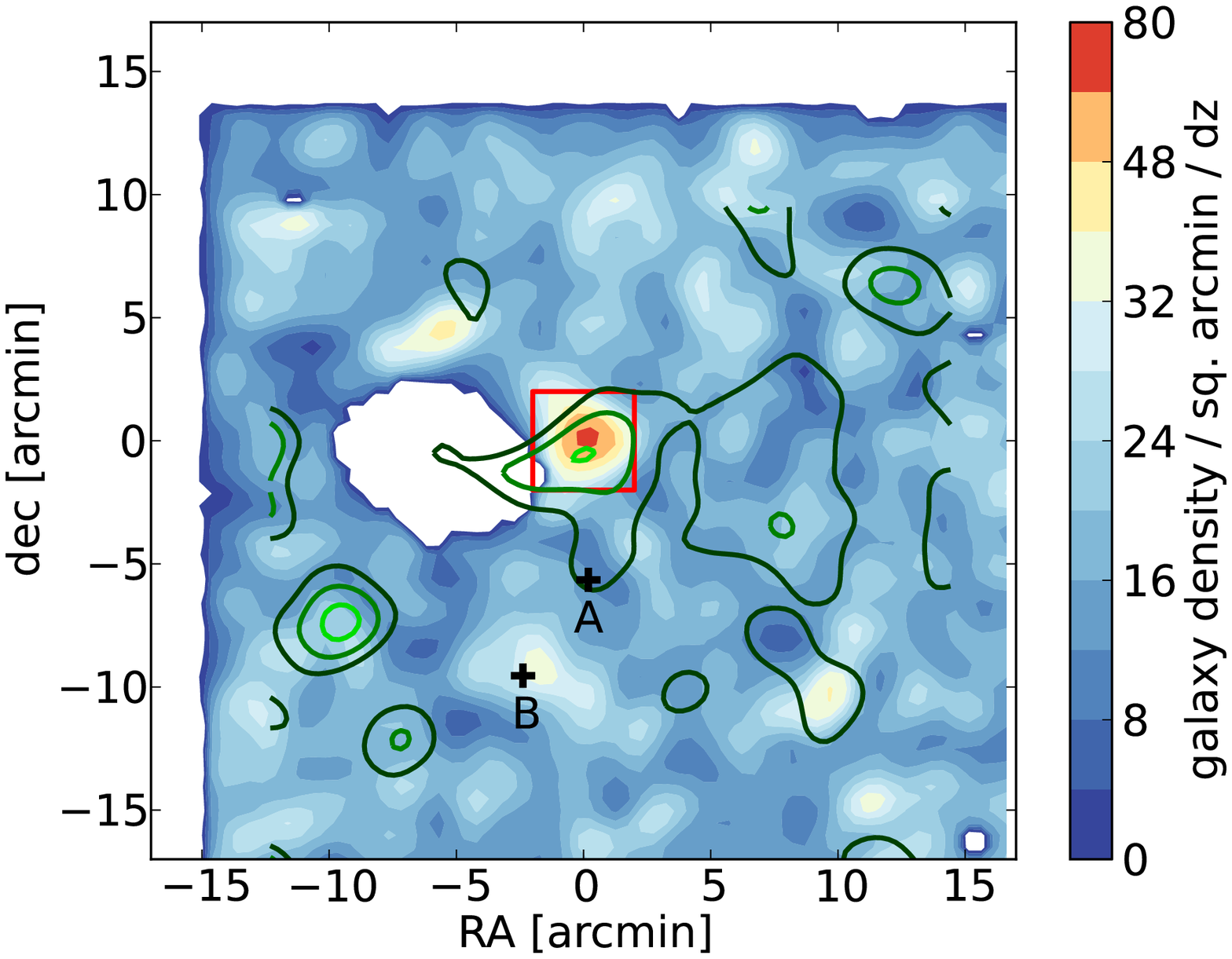}
\includegraphics[width=0.46\textwidth]{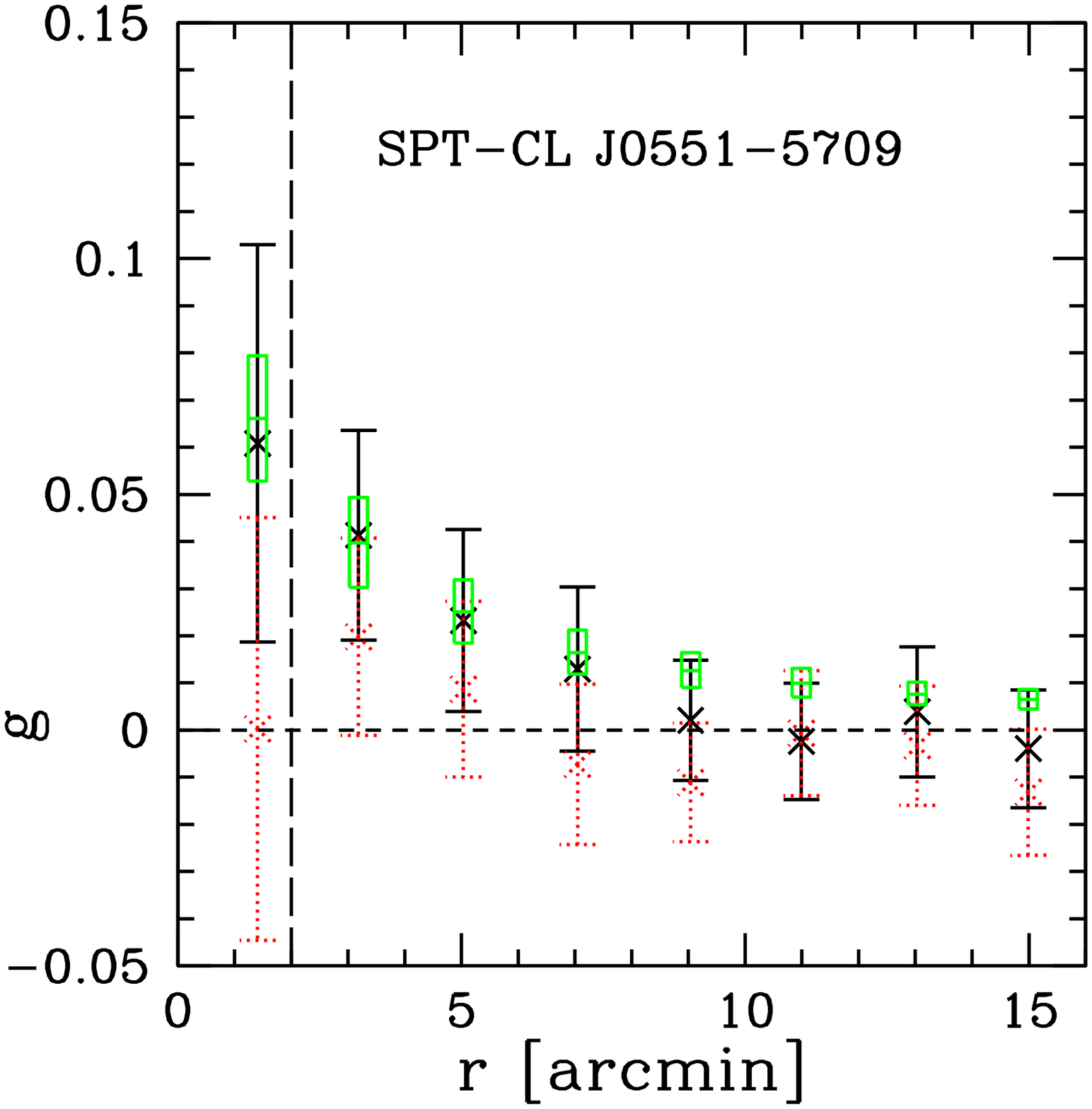}
\caption{Colour image from \emph{BRI} frames (top panel), three-dimensional galaxy density and $\kappa$ contours (central panel) and shear profile (bottom panel) of SPT-CL J0551--5709. See Fig.~\ref{fig:picspt4} and Section~\ref{sec:spt4visual} for details.}
\label{fig:picspt2}
\end{figure}

The central part of SPT-CL J0551--5709 (cf. Fig.~\ref{fig:picspt2}, top panel, showing a $4\times4$~arcmin$^2$ region) is formed by three galaxies of similar brightness aligned in almost north-south direction. Clean imaging of the cluster is impaired by the nearby bright star HR~2072 of magnitude $m_{V}\approx6$ at approximately 6~arcmin separation, and a large area has to be masked for the purpose of photometry and shape measurement.

\subsubsection{Previous Work}

The cluster was detected in SZ by SPT \citep{2010ApJ...722.1180V} at a significance of $\xi=6.13\sigma$. 

Its spectroscopic redshift is given as $z=0.423$ \citep{2010ApJ...722.1180V}.

Using \emph{griz} data from the Blanco Cosmology Survey and Magellan, \citet{2010ApJ...723.1736H} provide an optical richness of $N_{\mathrm{gal}}=54\pm15$, which they relate to a mass of $M_{200m}(N_{\mathrm{gal}})=(5.4\pm_{\rm stat}3.8\pm_{\rm sys}1.6)\times10^{14}h^{-1}\Msun$.

\citet{2010ApJ...722.1180V} give the SZ mass estimate at $M_{200m}=(4.84\pm_{\rm stat}1.06\pm_{\rm sys}0.68)\times10^{14}h^{-1}\Msun$. The de-biased significance according to equation (\ref{eqn:sptmor}) is $\zeta=4.55$.

\citet{2011ApJ...738...48A} present the first X-ray measurements SPT-CL J0551--5709 with \emph{Chandra}. Their analysis is updated in \citet{2013ApJ...763..127R} with deeper data, finding $T=4.0^{+0.6}_{-0.6}$keV, $M_{g,500c}=5.1^{+0.6}_{-0.6}\times10^{13}h_{70.2}^{-1}\Msol$ and a corresponding $Y_{X,500c}=(1.9\pm0.4)\times10^{14} h_{70.2}^{-1}\Msol$keV. Using the MORs of \citet{2009ApJ...692.1033V} with mean values calibrated from local clusters under the assumption of HSE, they convert the X-ray measurements to a mass estimate of $M_{500c}=(3.4\pm0.4)\times10^{14}h_{70.2}^{-1}\Msol$.

\citet{2011ApJ...738...48A} identify the cluster as a merger based on its disturbed X-ray morphology. This is consistent with the analysis of the X-ray surface brightness concentration by \citet{2012ApJ...761..183S}, who classify the system as a non cool-core cluster with a relatively large offset of $82\pm3$kpc between BCG and X-ray centroid.

\citet{2013ApJ...763..127R} quote a combined SZ+X-ray mass of $M_{500c}=(3.82\pm0.54)\times10^{14}h_{70}^{-1}\Msol$.

The foreground cluster Abell S 552 (with lowest Abell richness class, \citealt{1989ApJS...70....1A}) lies at 5~arcmin separation from SPT-CL J0551-5709. \citet{2010ApJ...723.1736H} give its photometric redshift as $z=0.09$, the value we adopt in this analysis. Our red galaxy method (cf. Section~\ref{sec:redz}) yields a consistent value of $z=0.08$.

\subsubsection{Weak Lensing Analysis}

The $\kappa$ mapping (Fig.~\ref{fig:picspt2}, central panel) of SPT-CL J0551--5709 is made difficult by the large masked area around the bright star towards the east from the cluster. Nevertheless, a density peak is identified in the central region of the cluster, its odd shape likely being a result of the masking. The secondary $\kappa$ peak in the south-eastern direction from SPT-CL J0551--5709 is associated only with a diffuse galaxy overdensity at $z\approx0.2\ldots0.3$ and also noisified by a number of moderately bright stars that mask out background galaxies near the peak position.

A single halo fit, fixing the centre at the BCG and marginalizing over concentration, yields $M_{200m}=(11.7^{+5.1}_{-4.2})\times10^{14}h_{70}^{-1}\Msol$ ($M_{500c}=(6.6^{+2.6}_{-2.4})\times10^{14}h_{70}^{-1}\Msol$). The observed shear profile and the confidence interval of the NFW fit are shown in Fig.~\ref{fig:picspt2} (bottom panel).

When including nearby structures in a combined fit and subtracting their best-fitting NFW signal from the shear catalogue, this result is unchanged. The WL mass measurement is consistent with the combined X-ray and SZ estimate of \citet{2013ApJ...763..127R} and the SZ mass estimate at the $1\sigma$ level and with the X-ray estimate of \citet{2011ApJ...738...48A} at $\approx1.2\sigma$.

For the purpose of our WL analysis, we have set the centre to be the central cluster galaxy, even though the northern one is the most luminous of the central trio, brighter by 0.13 magnitudes in \emph{R}. We test the impact of this and find that the best-fit mass at the position of the northern, brighter galaxy is lower by less than 10 per cent of the statistical error (less than 4 per cent of the mass), indicating that WL does not yield significant information on the true centring of the system, but that our off-BCG centring approach following the overall light of the brightest cluster galaxies is reasonable. 

\subsubsection{Strong Lensing}

We find no compelling evidence for strong lensing features in the field of SPT-CL J0551--5709.

\subsection{SPT-CL J2332--5358}

\subsubsection{Visual Appearance}
\begin{figure}
\centering
\includegraphics[width=0.44\textwidth]{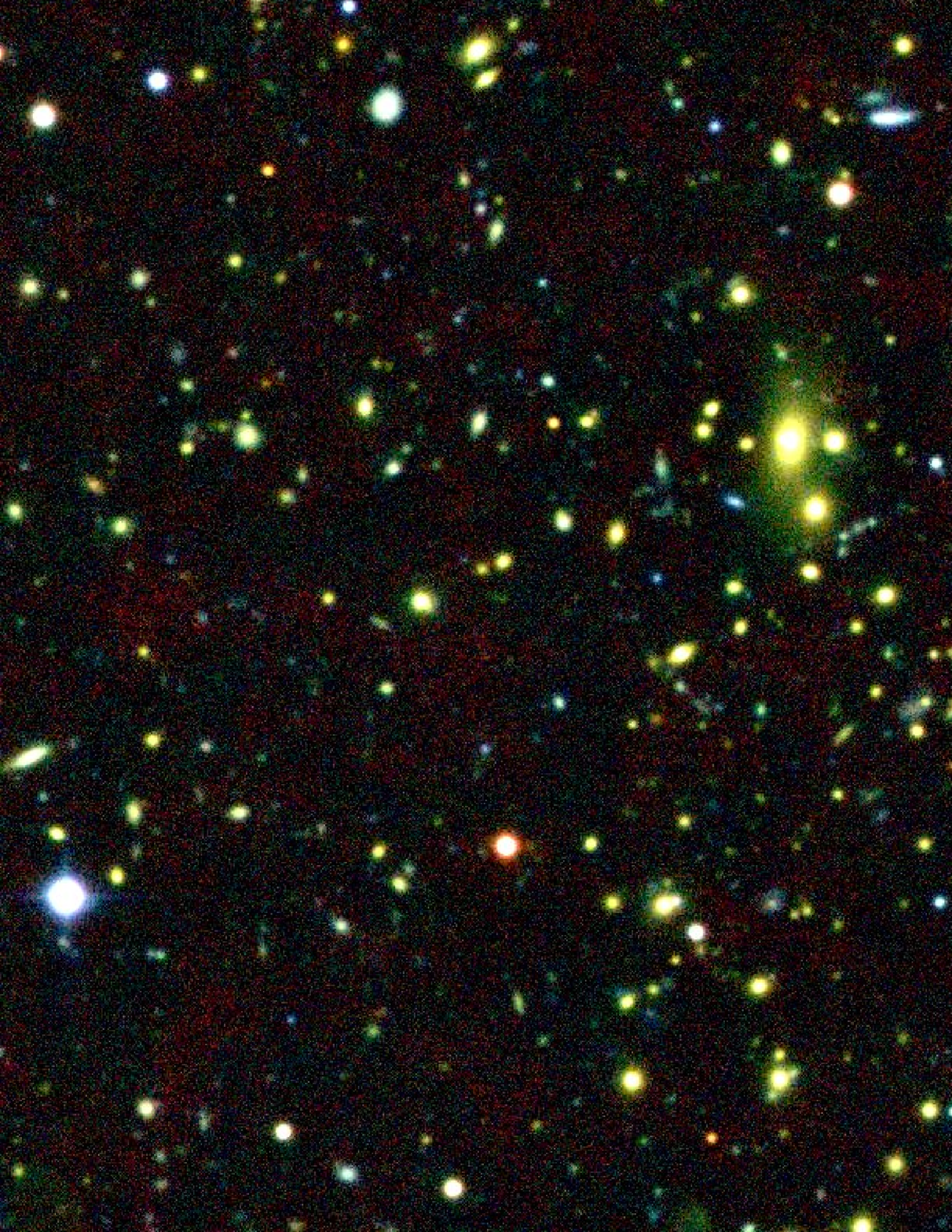}
\includegraphics[width=0.47\textwidth]{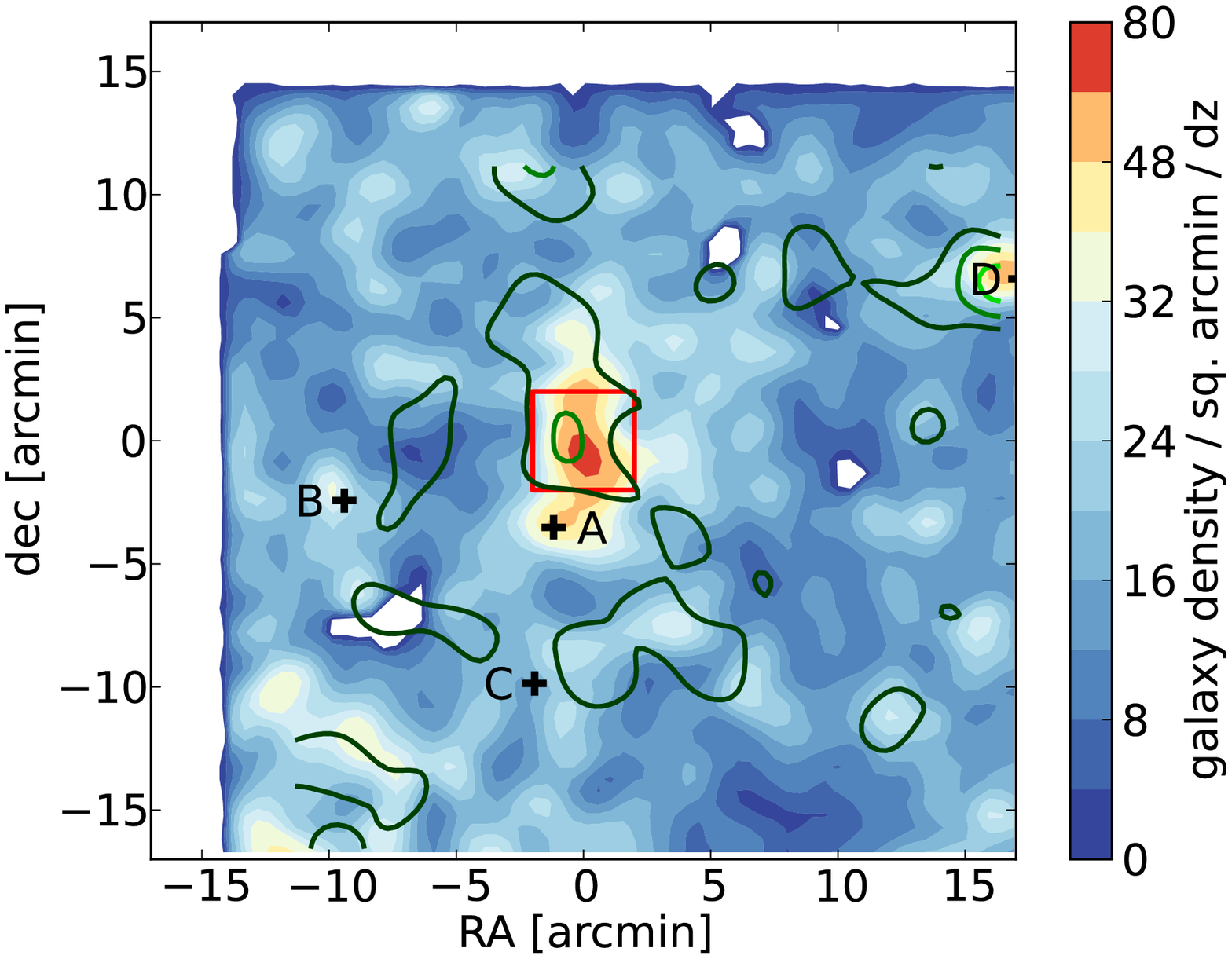}
\includegraphics[width=0.46\textwidth]{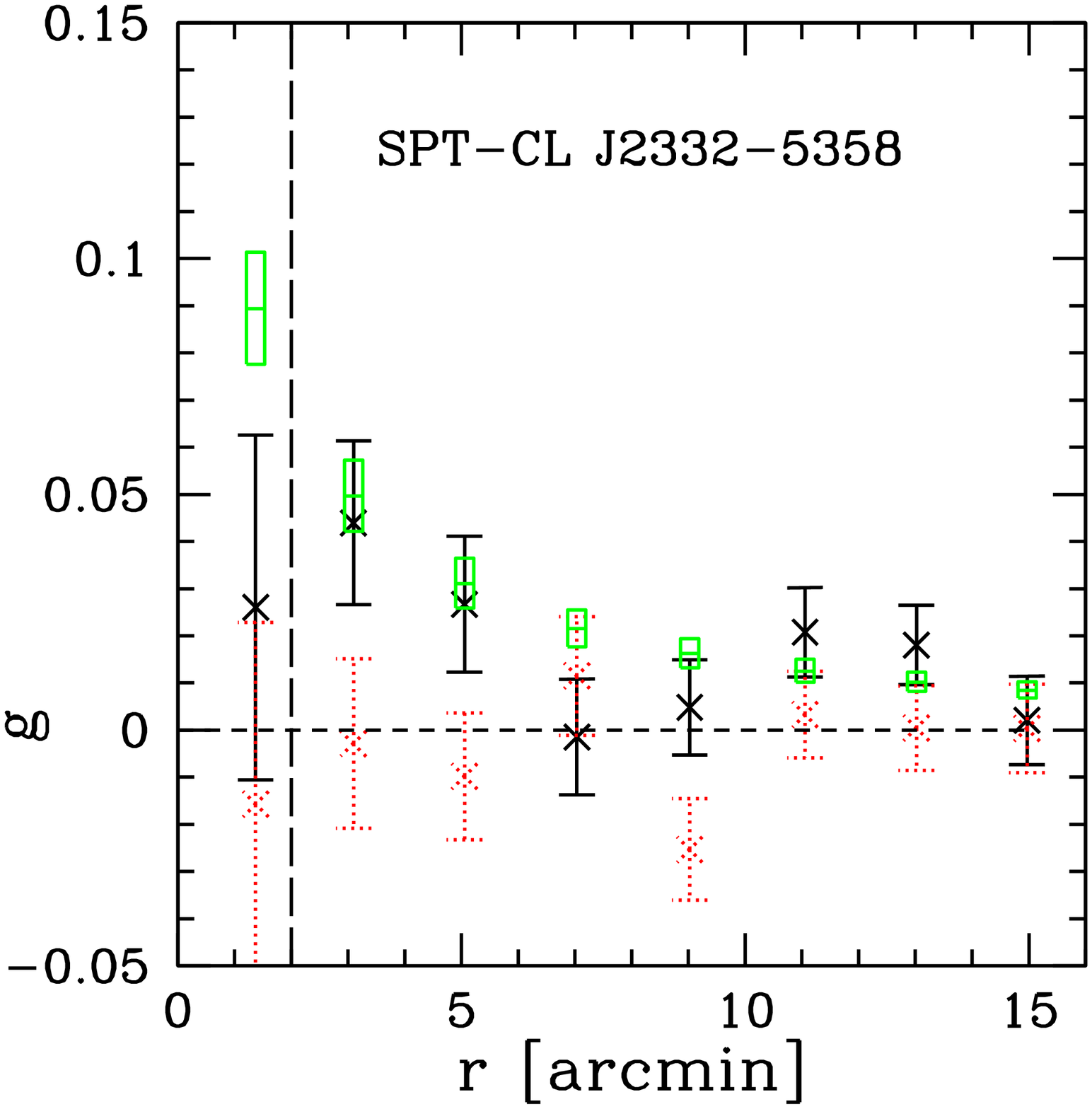}
\caption{Colour image from \emph{BRI} frames (top panel), three-dimensional galaxy density and $\kappa$ contours (central panel) and shear profile (bottom panel) of SPT-CL J2332--5358. See Fig.~\ref{fig:picspt4} and Section~\ref{sec:spt4visual} for details.}
\label{fig:picspt1}
\end{figure}

SPT-CL J2332--5358 (cf. Fig.~\ref{fig:picspt1}, top panel, showing a $4\times4$~arcmin$^2$ region) appears regular with a single dominant brightest cluster galaxy (BCG) and no obvious massive structures in the vicinity. The elongation of the galaxy distribution along the north-south direction is clearly visible, in accordance with the BCG orientation.

\subsubsection{Previous Work}

SPT-CL J2332--5358 was discovered by SPT \citep{2010ApJ...722.1180V} at an SZ significance of $\xi=7.3\sigma$. It is also identified as SCSO J233227-535827 after its optical detection by the Southern Cosmology Survey \citep{2010ApJS..191..340M} and associated with 1RXS J233224.3-535840, a source in the \emph{ROSAT} bright source catalogue \citep{1999A&A...349..389V}. \citet{2010AuA...514L...3S} independently discover the cluster in X-ray observations as part of the \emph{XMM-Newton} Blanco Cosmology Survey, making the system one of the first clusters to be detected independently both by X-ray and SZ surveys.

Its redshift was initially estimated photometrically as $z=0.32$ \citep{2010ApJ...723.1736H}, but has been determined more recently spectroscopically to be $z=0.402$ \citep{2012ApJ...761...22S}. 

Using \emph{griz} data from the Blanco Cosmology Survey and Magellan, \citet{2010ApJ...723.1736H} provide an optical richness of $N_{\mathrm{gal}}=42\pm8$, which they translate to a mass of $M_{200m}(N_{\mathrm{gal}})=(3.5\pm2.1\pm1.1)\times10^{14}h^{-1}\Msun$ including statistical and systematic uncertainty, although using the erroneous redshift $z=0.32$.

An SZ mass estimate based on a significance-mass relation is given by \citet{2010ApJ...722.1180V} as $M_{200m}=(6.21\pm_{\rm stat}1.15\pm_{\rm sys}0.94)\times10^{14}h^{-1}\Msun$, yet calculated at the erroneous redshift of $z=0.32$ and with the caveat of potential blending with a radio point source. The more recent measurement of \citet{2013ApJ...763..127R} reports a higher $M_{500c}=(6.5\pm0.79)\times10^{14}h_{70}^{-1}\Msun$ at the spectroscopic redshift, which is the value we will compare to in this work. The de-biased significance according to equation (\ref{eqn:sptmorr}) is $\zeta=9.91$, which yields $M_{200m}=(12.1\pm1.4)\times10^{14}h_{70}^{-1}\Msun$ by means of equation (\ref{eqn:sptmor}).

X-ray imaging of the cluster exists from \emph{XMM-Newton}, analysed by \citet{2010AuA...514L...3S} and \citet{2011ApJ...738...48A} at the erroneous photometric redshift of $z=0.32$. The analysis is repeated by \citet{2013ApJ...763..147B} for the correct redshift. They find a temperature of $T=7.8^{+1.0}_{-0.9}$keV, a gas mass of $M_{g,500c}=7.6^{+0.2}_{-0.3}\times 10^{13}h_{70.2}^{-1}\Msol$ and a corresponding $Y_{X,500c}=(6.1\pm0.8)\times10^{14}h_{70.2}^{-1}\Msol$keV. Using the MORs of \citet{2009ApJ...692.1033V} with mean values calibrated from local clusters under the assumption of HSE, they convert the X-ray measurements to a mass estimate of $M_{500c}=(6.7\pm0.5)\times10^{14}h_{70.2}^{-1}\Msol$.

\citet{2011ApJ...738...48A} characterize the cluster as relaxed based on its X-ray morphology.

The combined SZ+X-ray mass estimate of \citet{2013ApJ...763..127R} is $M_{500c}=(6.54\pm0.82)\times 10^{14}h_{70}^{-1}\Msun$.

\citet{2012ApJ...756..101G} describe a candidate strongly lensed sub-mm galaxy with multiple images at approximately 0.5~arcmin separation from the cluster centre with a spectroscopic redshift of $z=2.7$.

The nearby system SCSO J233231.4-540135.8 at $\theta\approx4$~arcmin separation is a cluster of galaxies optically discovered by \citet{2009ApJ...698.1221M}, at a photometric redshift of $z=0.33$. Using the relation of \citet{2008MNRAS.390.1157R}, they calculate mass estimates based on a combination of BCG luminosity and either richness or total luminosity of $M_{200m}=4.1\times10^{14}h_{70}^{-1}\Msun$ and $M_{200m}=1.7\times10^{14}h_{70}^{-1}\Msun$, respectively.

\subsubsection{Weak Lensing Analysis}

The $\kappa$ map (Fig.~\ref{fig:picspt1}, central panel) shows that the cluster is elongated along the north-south axis, aligned with the BCG ellipticity. An overdensity of galaxies at an estimated $z\approx0.24$, 18~arcmin from the BCG at the western edge of the field, is associated with a clear peak in $\kappa$ and included in the multi-halo fit (see below).

A single halo fit, fixing the centre at the BCG and marginalizing over concentration, yields $M_{200m}=(14.5^{+4.0}_{-3.5})\times10^{14}h_{70}^{-1}\Msol$ ($M_{500c}=(7.5^{+1.8}_{-1.7})\times10^{14}h_{70}^{-1}\Msol$). The observed shear profile and the confidence interval of the NFW fit are shown in Fig.~\ref{fig:picspt1} (bottom panel).

When including nearby structures in a combined fit and subtracting their best-fitting NFW signal from the shear catalogue, a repeated fit of the central halo yields a slightly higher $M_{200m}=(15.3^{+4.2}_{-3.6})\times10^{14}h_{70}^{-1}\Msol$ ($M_{500c}=(7.9^{+1.8}_{-1.7})\times10^{14}h_{70}^{-1}\Msol$).

Both results are mildly higher than the SPT SZ mass estimate of \citet{2010ApJ...722.1180V} at $\approx1.5\sigma$, yet in good agreement with the X-ray, SZ and combined mass estimate of \citet{2013ApJ...763..127R}.

\subsubsection{Strong Lensing}

We find several small blue background sources in the south-east and south-west from the BCG. It is unclear from our data whether any of these are multiply imaged or correspond to the strongly lensed system discovered by \citet{2012ApJ...756..101G}.

\subsection{SPT-CL J2355--5056}

\subsubsection{Visual Appearance}

\begin{figure*}
\centering
\includegraphics[width=\textwidth]{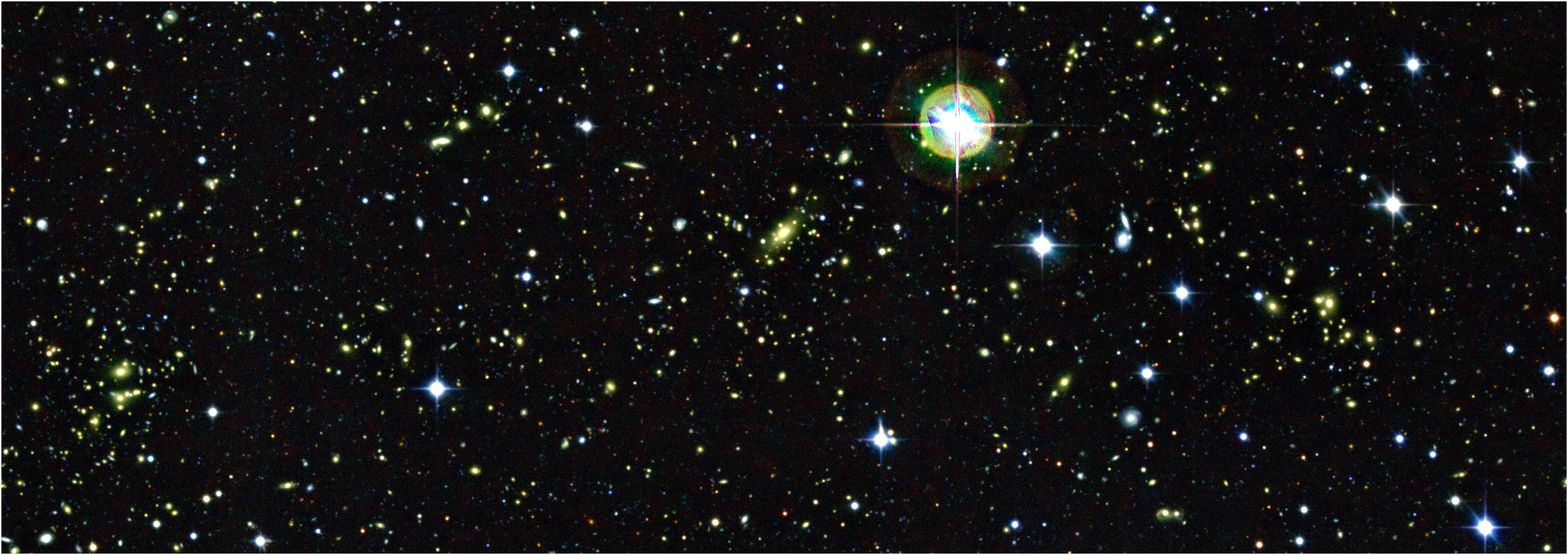}
\includegraphics[width=0.52\textwidth]{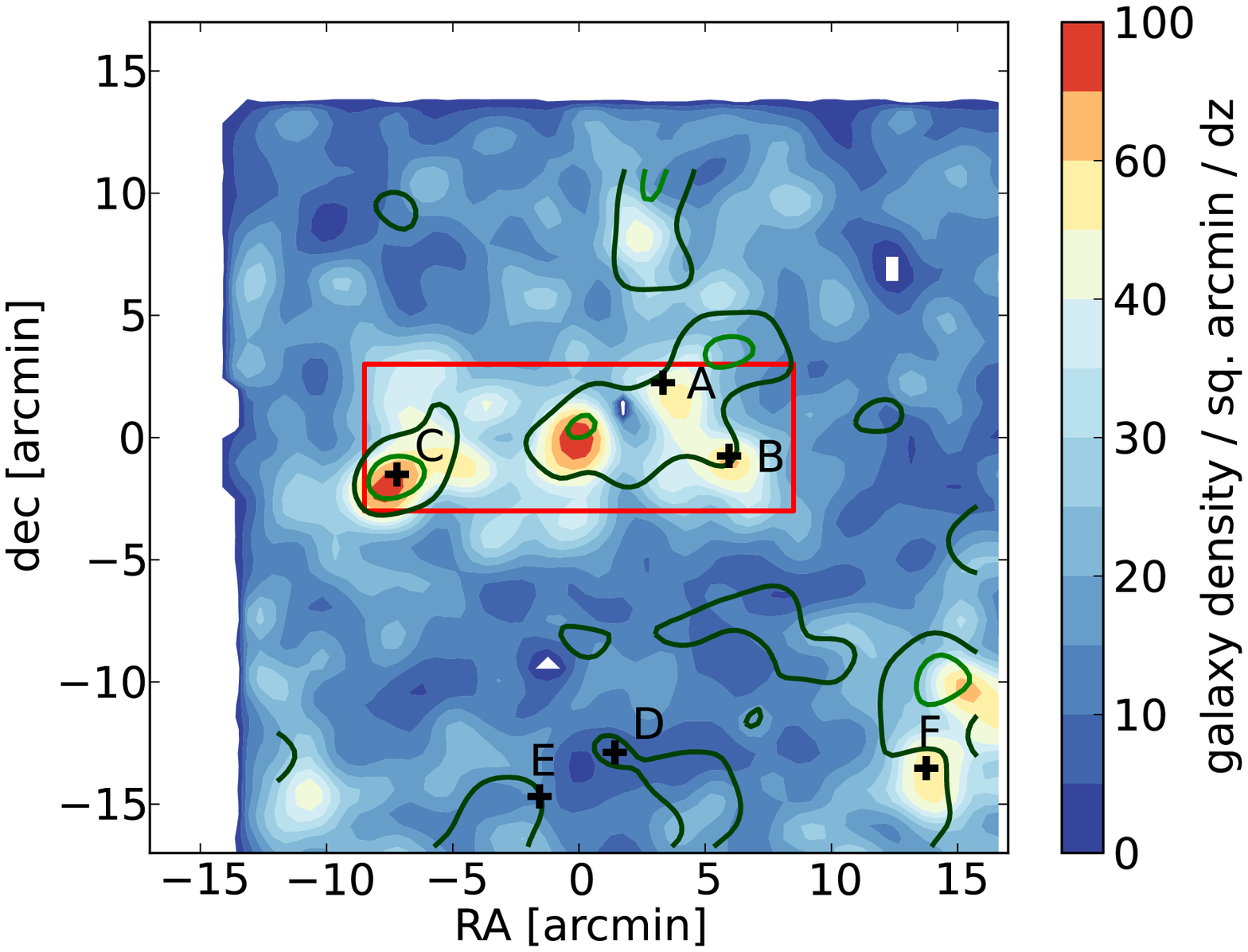}
\includegraphics[width=0.44\textwidth]{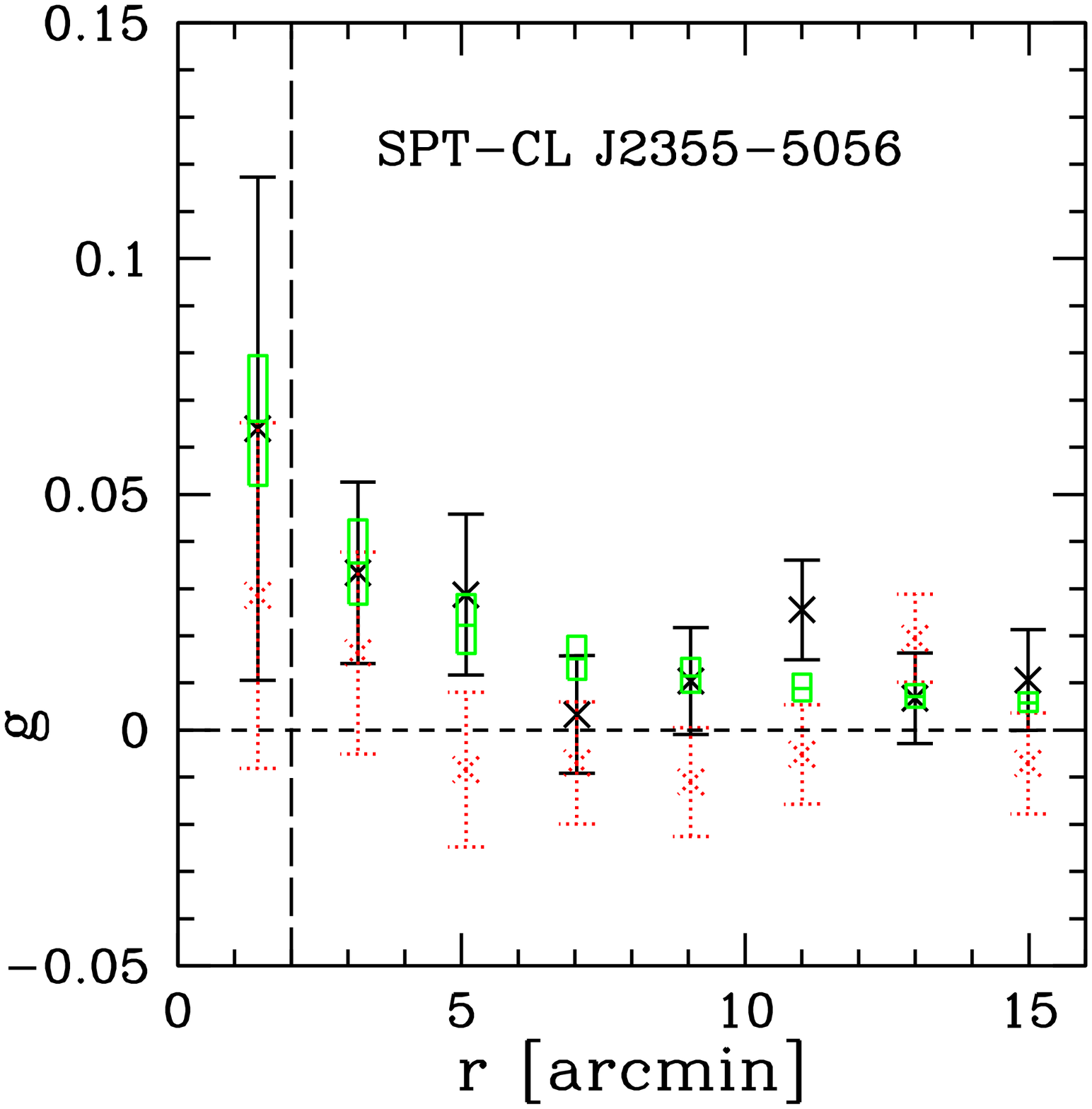}
\caption{Colour image from \emph{BRI} frames of the central $17\times6$~arcmin region (top panel), three-dimensional galaxy density and $\kappa$ contours (central panel) and shear profile (bottom panel) of SPT-CL J2355--5056. See Fig.~\ref{fig:picspt4} and Section~\ref{sec:spt4visual} for details.}
\label{fig:picspt3}
\end{figure*}

The central part of SPT-CL J2355--5056 (cf. Fig.~\ref{fig:picspt3}, upper panel, showing a $17\times6$~arcmin$^2$ region) features a single large and diffuse BCG. However, the cluster is neighboured by structures towards the East and est at separations of 3-4~arcmin, which from visual inspection seem to have similar red galaxy colours and thus redshifts. There is no indication for the neighbouring structures from the archival \emph{Chandra} X-ray data, although the field of view is not large enough for a conclusive statement.

\subsubsection{Previous Work}

SPT-CL J2355--5056 is discovered by SPT \citep{2010ApJ...722.1180V} at a significance of $5.89\sigma$.

\citet{2012ApJ...761...22S} and \citet{2013ApJ...763..127R} provide a spectroscopic redshift of 0.3196, correcting the previous photometric estimate of 0.35 \citep{2010ApJ...723.1736H}.

Using \emph{griz} data from the Blanco Cosmology Survey and Magellan, \citet{2010ApJ...723.1736H} provide an optical richness of $N_{\mathrm{gal}}=55\pm5$, on which they base a mass estimate of $M_{200m}(N_{\mathrm{gal}})=(5.6\pm3.0\pm1.7)\times10^{14}h^{-1}\Msun$ when assuming the slightly erroneous redshift of $z=0.35$.

An SZ mass estimate is given by \citet{2010ApJ...722.1180V}, yet at the erroneous $z=0.35$. The more recent measurement of \citet{2013ApJ...763..127R} reports  $M_{500c}=(4.07\pm0.57)\times10^{14}h_{70}^{-1}\Msun$ at the spectroscopic redshift, which is the value we will compare to in this work. The de-biased significance according to equation (\ref{eqn:sptmorr}) is $\zeta=5.12$, which yields $M_{200m}=(7.6\pm1.0)\times10^{14}h_{70}^{-1}\Msun$ by means of equation (\ref{eqn:sptmor}).

\citet{2011ApJ...738...48A} present the first X-ray measurements SPT-CL J2355--5056 with \emph{Chandra}. Their analysis is updated in \citet{2013ApJ...763..127R} for the corrected redshift, finding $T=5.3^{+0.9}_{-0.7}$keV, $M_{g,500c}=3.9^{+0.2}_{-0.1}\times10^{13}h_{70.2}^{-1}\Msol$ and a corresponding $Y_{X,500c}=(2.2\pm0.4)\times10^{14} h_{70.2}^{-1}\Msol$keV. Using the MORs of \citet{2009ApJ...692.1033V} with mean values calibrated from local clusters under the assumption of HSE, they convert the X-ray measurements to a mass estimate of $M_{500c}=(3.8\pm0.4)\times10^{14}h_{70.2}^{-1}\Msol$.

\citet{2011ApJ...738...48A} report the cluster to be relaxed based on its X-ray morphology, likely with a cool core. \citet{2012ApJ...761..183S} consistently classify the system as a cool core cluster, with a relatively small offset of $6.7\pm2.3$kpc between BCG and X-ray centroid.

\citet{2013ApJ...763..127R} quote, using the spectroscopic redshift, a combined SZ+X-ray mass of $M_{500c}=(4.11\pm0.54)\times 10^{14} h_{70}^{-1} \Msun$. 

The APM Galaxy Survey \citep{1997MNRAS.289..263D} lists the cluster APMCC 936 at $(\alpha,\delta)=(23^{\rm h}53^{\rm m}21.5^{\rm s}, -51^{\circ}26'57'')$, $\approx14$~arcmin separated from SPT-CL J2355--5056. They estimate its redshift at $z=0.118$ with a richness of $R=61.6$.

\subsubsection{Weak Lensing Analysis}

The $\kappa$ map (Fig.~\ref{fig:picspt3}, central panel) nicely maps the complex structure of the multiple components in the field of SPT-CL J2355--5056. Aside the central region, the galaxy peak near the south-western corner of the field is associated with a $\kappa$ overdensity and red galaxy population at $z\approx0.42$, included as a component in the multi-halo fit (see below).

A single halo fit, fixing the centre at the BCG and marginalizing over concentration, yields $M_{200m}=(8.0^{+3.7}_{-3.1})\times10^{14}h_{70}^{-1}\Msol$ ($M_{500c}=(4.1^{+1.7}_{-1.5})\times10^{14}h_{70}^{-1}\Msol$). The observed shear profile and the confidence interval of the NFW fit are shown in Fig.~\ref{fig:picspt3} (lower right panel).

When including nearby structures in a combined fit and subtracting their best-fitting NFW signal from the shear catalogue, a repeated fit of the central halo yields a significantly lower $M_{200m}=(3.8^{+2.4}_{-2.0})\times10^{14}h_{70}^{-1}\Msol$ ($M_{500c}=(2.1^{+1.4}_{-1.1})\times 10^{14}h_{70}^{-1}\Msol$). In return, significant mass is attributed to the eastern peak $\approx$~7~arcmin from the BCG ($M_{200m}=(4.5^{+1.7}_{-1.7})\times10^{14}h_{70}^{-1}\Msol$ at $z\approx0.24$) and a northern and southern component of the western peak ($M_{200m}=(1.2^{+1.7}_{-1.2})\times 10^{14}h_{70}^{-1}\Msol$ and $M_{200m}=(6.7^{+3.0}_{-2.1})\times 10^{14}h_{70}^{-1}\Msol$ at redshifts consistent with the central cluster, respectively).

The single-peak analysis is consistent with the X-ray, SZ and combined result of \citet{2013ApJ...763..127R}, while the WL result with neighbouring structures subtracted is $1\sigma$ below. The large uncertainty in the WL analysis of this complex system allows for no significant conclusions on the effect of blending on the SZ analysis. Yet we note that the structure of the cluster warrants deeper follow-up, since the interplay of cluster gas and dark matter might be interesting.

\subsubsection{Strong Lensing}

\begin{figure}
\centering
\includegraphics[width=0.23\textwidth]{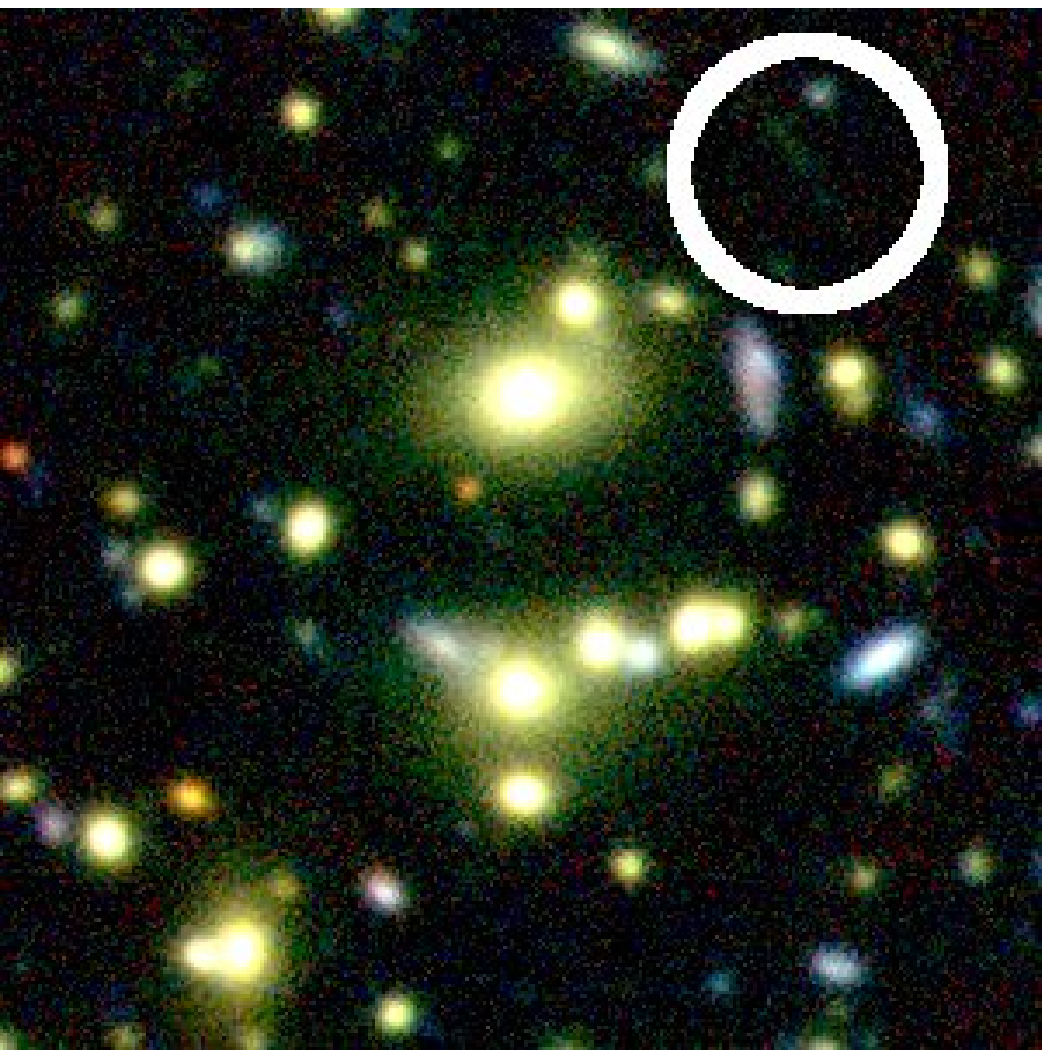}
\includegraphics[width=0.23\textwidth]{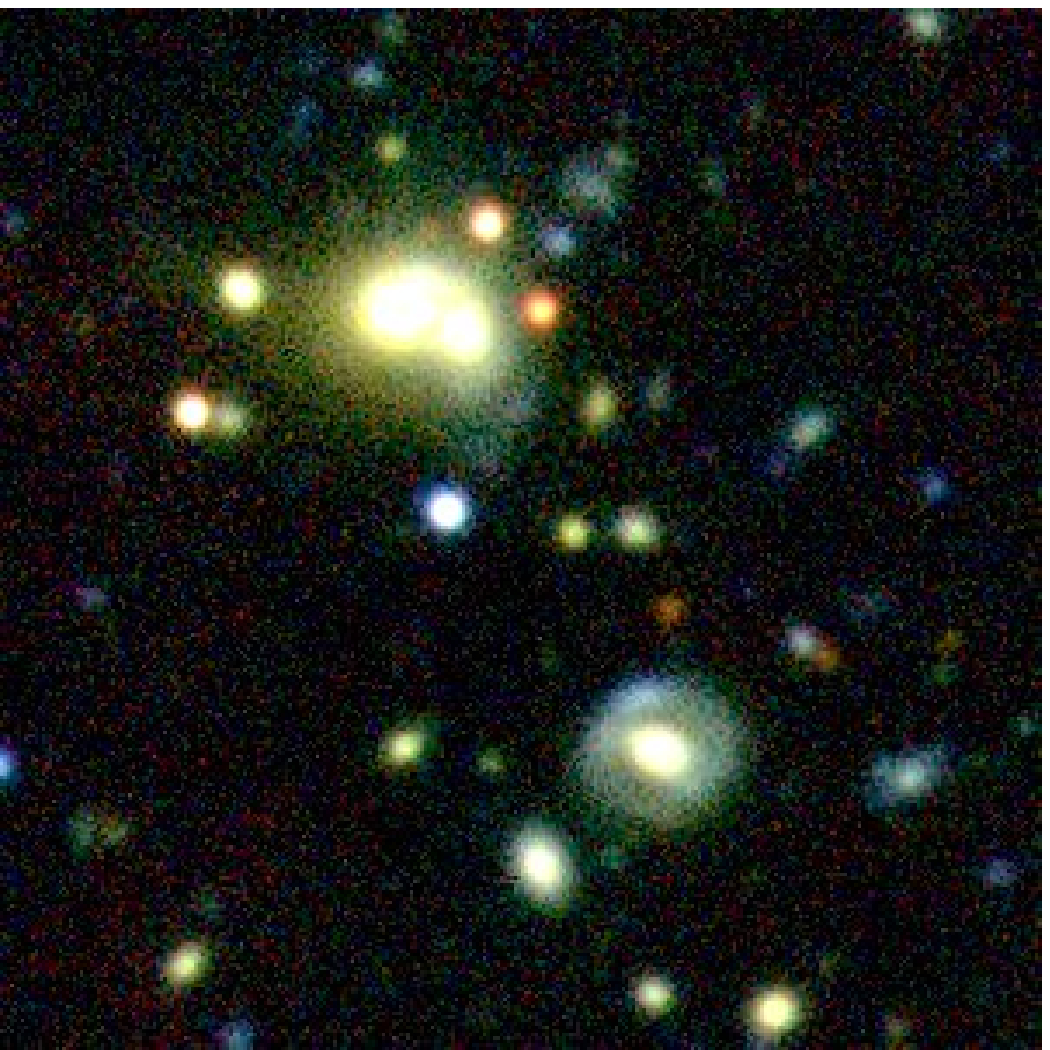}
\caption{Colour images of strong lensing candidates in the field of SPT-CL J2355--5056. Left panel is centred at $(\alpha,\delta)=(23^{\rm h}56^{\rm m}33.1^{\rm s},-50^{\circ}57'20'')$, right panel is centred at $(\alpha,\delta)=(23^{\rm h}56^{\rm m}30.4^{\rm s},-50^{\circ}53'08'')$, both are $1\times1$~arcmin$^2$ in size.}
\label{fig:spt3sl}
\end{figure}

Apart from inconclusive arc-like features around the BCG, two candidate strong lensing features can be seen towards the east. One is a potential arc at a separation of $\theta\approx20.6$~arcsec from the main galaxy of the eastern structure $\approx$~7~arcmin from the centre of the field (left panel of Fig.~\ref{fig:spt3sl}). The other is composed of a blue symmetric ring with radius $\theta\approx3.3$~arcsec around a red galaxy and a bluish structure blended with two red foreground galaxies close in projection (right panel of Fig.~\ref{fig:spt3sl}). Both features are at unexpectedly large radii given the likely mass of the lenses, and deeper follow-up observations at higher resolution would be required to confirm their strong lensing nature.

\subsection{PLCKESZ G287.0+32.9}

\subsubsection{Visual Appearance}

\begin{figure}
\centering
\includegraphics[width=0.44\textwidth]{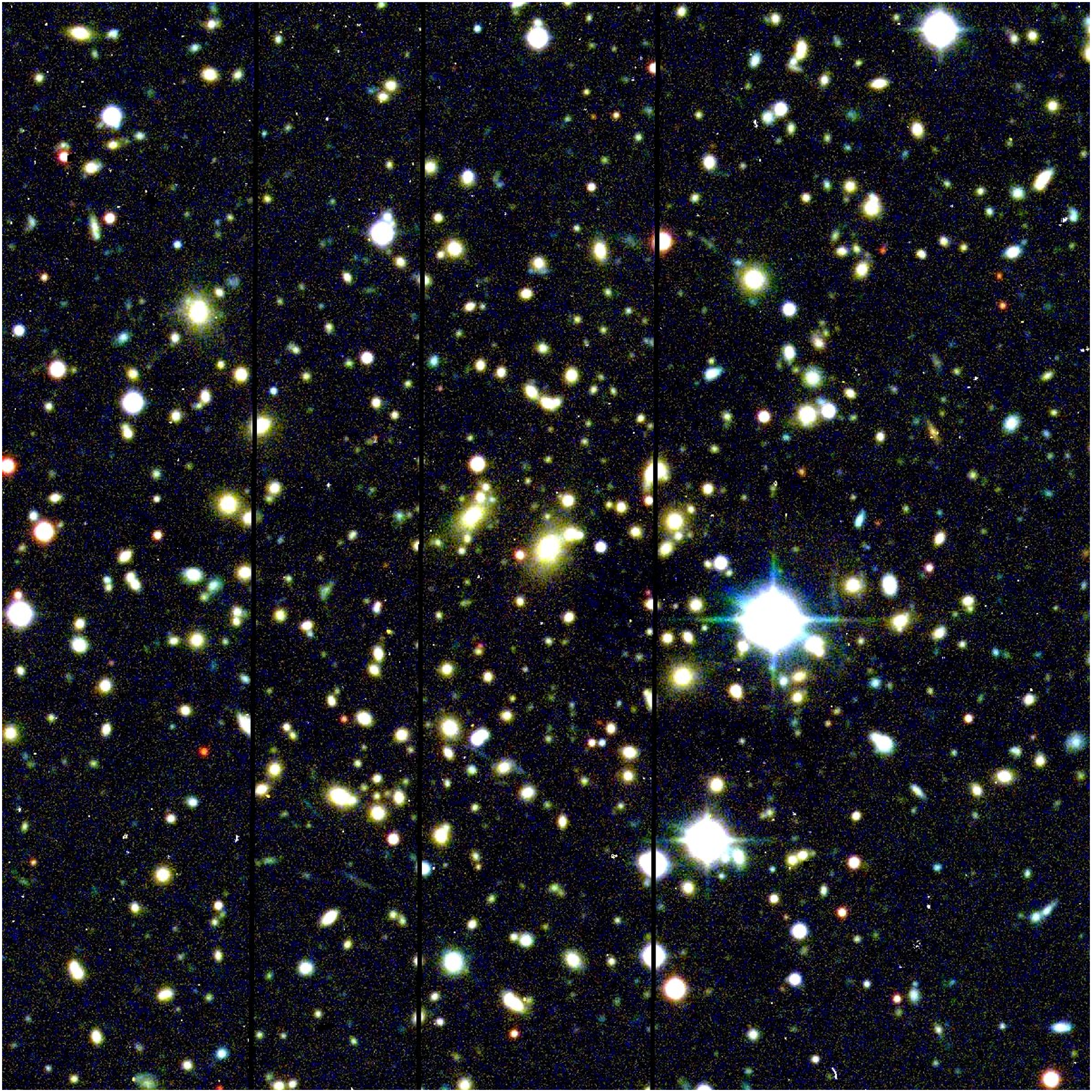}
\includegraphics[width=0.47\textwidth]{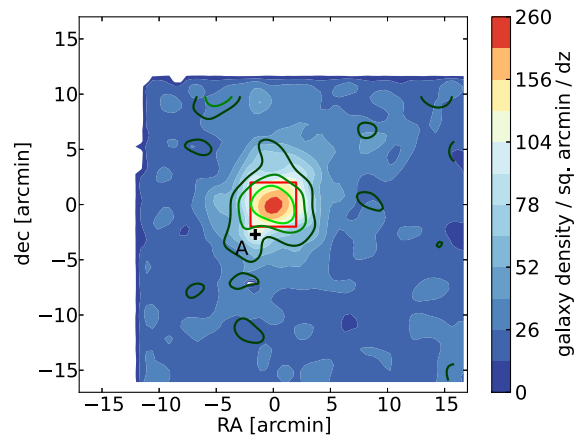}
\includegraphics[width=0.46\textwidth]{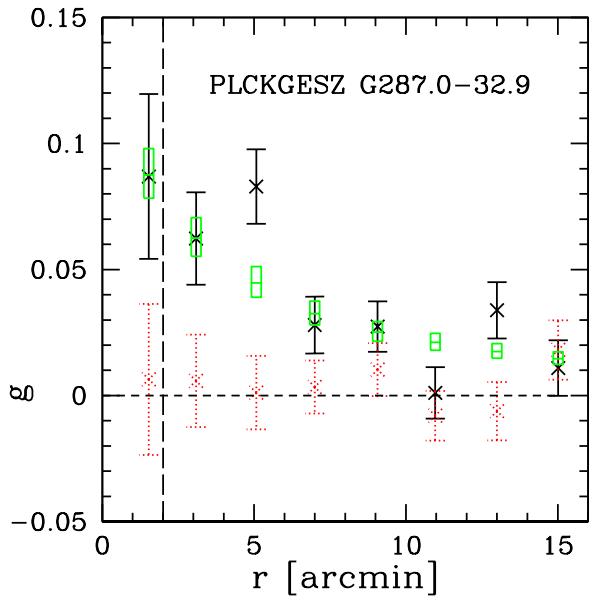}
\caption{Colour image (top panel), galaxy density weighted with $P_{\rm cl}$ (cf. Section~\ref{sec:fcl}) and $\kappa$ contours (central panel) and shear profile (bottom panel) of PLCKESZ G287.0+32.9. See Fig.~\ref{fig:picspt4} and Section~\ref{sec:spt4visual} for details.}
\label{fig:picplckg287}
\end{figure}

PLCKESZ G287.0+32.9 features a large number of cluster member galaxies, with a clear brightest galaxy in the centre and numerous gravitational arcs visible near cluster members at arcminute separations from the core (cf. Section~\ref{sec:plckg287sl}). Fig.~\ref{fig:picplckg287} shows a \emph{VRI} colour image, using the single \emph{V} band frame available.

\subsubsection{Previous Work}

The cluster was discovered by its SZ signal, detected by Planck \citep{2011AuA...536A...8P} at a significance of $10.62\sigma$.

A redshift estimate is given by \citet{2011AuA...536A...9P} at $z=0.39$, although based on X-ray emission lines and therefore not of optimal confidence.

The cluster is followed-up with X-ray observations with \emph{XMM-Newton}. The \citet{2011AuA...536A...9P} find from these data a temperature of $T=(12.86\pm0.42)$keV, $M_{g,500c}=(2.39\pm0.03)\times10^{14}h_{70}^{-1}\Msol$ and a corresponding $Y_{X,500c}=(30.69\pm0.36)\times10^{14}h_{70}^{-1}\Msol$keV. From a $Y_{X}$ MOR calibrated with relaxed objects under the assumption of HSE \citep{2010AuA...517A..92A}, they calculate a mass of $M_{500c}=(15.72\pm0.27)\times10^{14}h_{70}^{-1}\Msun$.

\citet{2011ApJ...736L...8B} investigate non-thermal radio emission from the cluster. They find a double radio relic, which is an indication of a major merger. Notably, the double relic has a projected separation of over 4~Mpc, the largest found to date at a redshift as high as $z=0.39$ \citep{2011ApJ...736L...8B}. They characterize the X-ray morphology based on the same \emph{XMM-Newton} that is used by \citet{2011AuA...536A...9P} as disturbed, consistent with the merger hypothesis. They find the X-ray peak to be separated from the BCG by a large distance of 410~kpc.

From the catalogued SZ likelihood of \citet{2013arXiv1303.5089P} we calculate mass estimates using the X-ray $\theta_{500}$ as a fixed size and the self-consistency method as $M_{500c}=17.7^{+0.8}_{-0.9}\times10^{14}$ and $18.2^{+0.9}_{-1.0}\times10^{14}h_{70}^{-1}\Msol$, respectively.

\subsubsection{Weak Lensing Analysis}

The $\kappa$ map (Fig.~\ref{fig:picplckg287}, central panel) is dominated by the central peak, which has symmetric appearance and a central value of $\kappa=0.22$.

A single halo fit, fixing the centre at the BCG and marginalizing over concentration, yields $M_{200m}=(37.7^{+9.5}_{-7.6})\times 10^{14}h_{70}^{-1}\Msol$ ($M_{500c}=(19.5^{+3.3}_{-3.2})\times 10^{14}h_{70}^{-1}\Msol$). The observed shear profile and the confidence interval of the NFW fit are shown in Fig.~\ref{fig:picplckg287} (bottom panel). 

We note that the excess shear at 5~arcmin radius is potentially due to the added mass from the structure A in Fig.~\ref{fig:picplckg287} at that projected separation.
Subtracting the maximum likelihood signal of structure A from a combined fit, the resulting mass is slightly lower at $M_{200m}=(35.4^{+8.9}_{-7.1})\times 10^{14}h_{70}^{-1}\Msol$ ($M_{500c}=(18.7^{+3.2}_{-3.1})\times 10^{14}h_{70}^{-1}\Msol$).

We note that both results are consistent with previous X-ray observations \citep{2011AuA...536A...9P} and the Planck-based SZ mass estimate (\citealt{2013arXiv1303.5089P}; this work).

\subsubsection{Strong Lensing}
\label{sec:plckg287sl}

\begin{figure}
\centering
\includegraphics[width=0.48\textwidth]{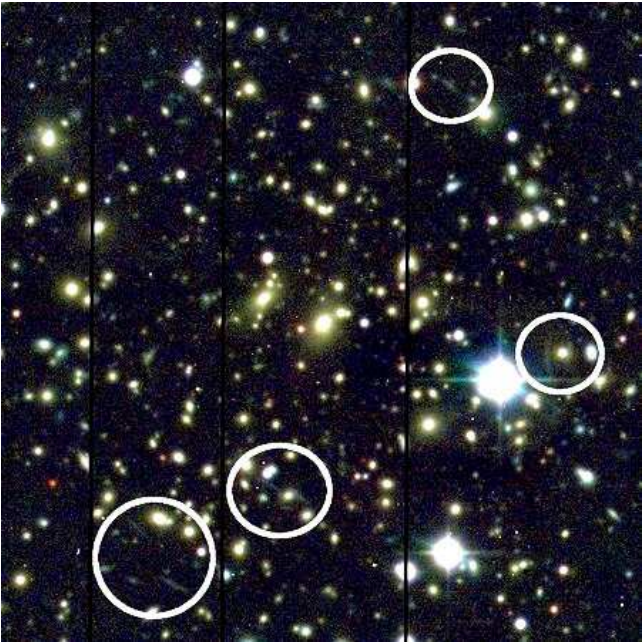}
\caption{Colour image of centre of PLCKESZ G287.0+32.9, arc candidates marked with circles. The cutout size is $3\times3$~arcmin$^2$, making this an exceedingly large region for strong lensing features even for a massive cluster.}
\label{fig:plckg287sl}
\end{figure}

The central region of PLCKESZ G287.0+32.9 shows several arc-like features, marked in the enlarged view in Fig.~\ref{fig:plckg287sl}. Candidate features can be found over a large region, the two most separated ones $2\theta\approx165$~arcsec apart. Even at the 68\% upper limit of the NFW fit and for a hypothetical source redshift of $z_s=5$, the predicted Einstein radius is only $\theta_E\approx37$~arcsec, although a larger than mean concentration can increase this value considerably. It is the most likely explanation that the features seen are therefore either not actual multiple images or caused by the particular geometry of the system and its subcomponents (e.g. a major merger as indicated by the radio signature, cf. \citealt{2011ApJ...736L...8B}). Despite this, the unusual constellation merits follow-up with spectroscopy and higher-quality imaging, since it would, if confirmed, make PLCKESZ G287.0+32.9 potentially the largest strong lens known to date and a candidate for the largest strong lens in the Universe, with some tension with $\Lambda$CDM (cf. \citealt{2013arXiv1304.1223M}, who find MACS J0717.5+3745 at $\theta_E\approx55$~arcsec for $z_s\approx3$ and \citealt{2009MNRAS.392..930O}, \citealt{2012A&A...547A..67W}, and \citealt{2012A&A...547A..66R}).

\subsection{PLCKESZ G292.5+22.0}
\label{sec:p292}
\subsubsection{Visual Appearance}

\begin{figure}
\centering
\includegraphics[width=0.44\textwidth]{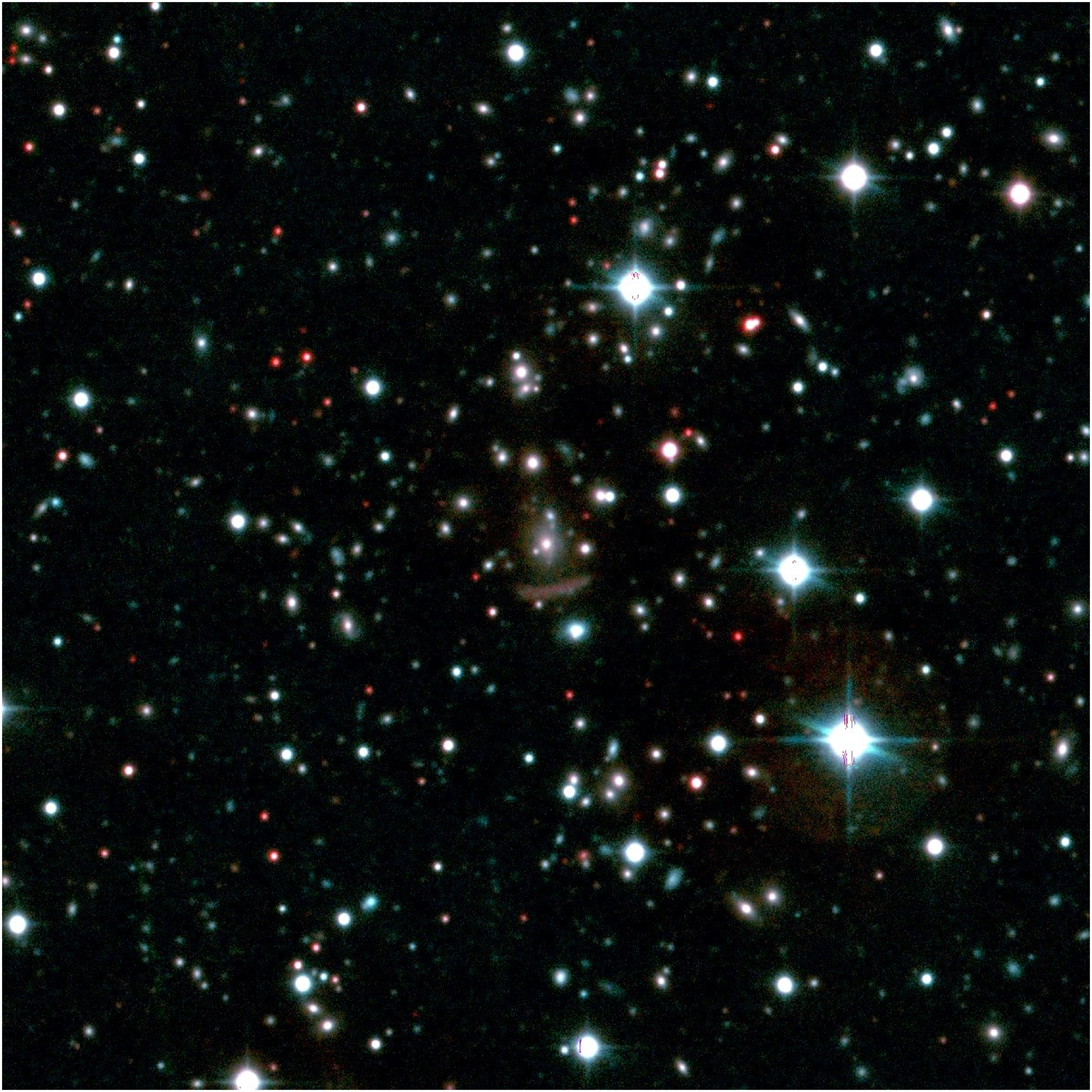}
\includegraphics[width=0.47\textwidth]{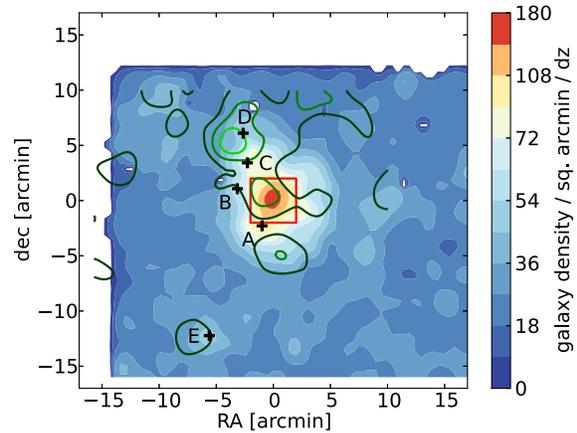}
\includegraphics[width=0.46\textwidth]{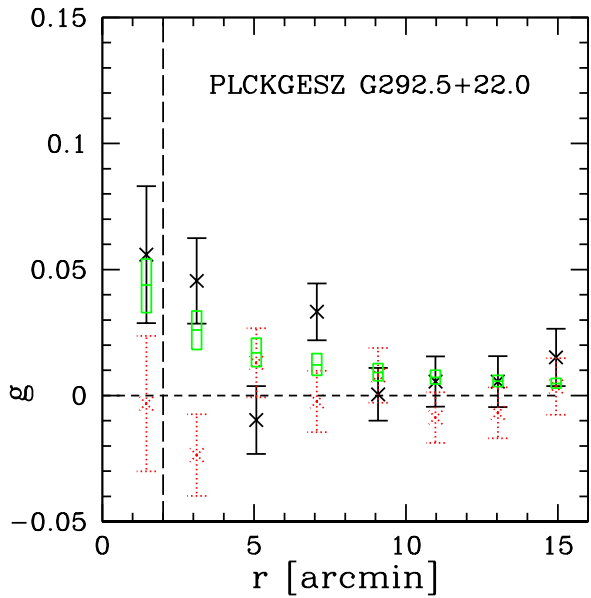}
\caption{Single colour image from \emph{RI} frames (top panel), galaxy density weighted with $P_{\rm cl}$ (cf. Section~\ref{sec:fcl}) and $\kappa$ contours (central panel) and shear profile (bottom panel) of PLCKESZ G292.5+22.0. See Fig.~\ref{fig:picspt4} and Section~\ref{sec:spt4visual} for details.}
\label{fig:picplckg292}
\end{figure}

PLCKESZ G292.5+22.0 has a central region dominated by the brightest cluster galaxy (cf. Fig~\ref{fig:picplckg292}), with a highly magnified strongly gravitationally lensed image of a background galaxy between it and a bright galaxy towards the southern direction.

\subsubsection{Previous Work}

This cluster was discovered by its SZ signal, detected by \citet{2011AuA...536A...8P} at a significance of $6.88\sigma$. 

\citet{2011AuA...536A...9P,2012AuA...543A.102P} provide a redshift of $z=0.31$ using X-ray spectral fitting and of $z=0.29$ from \emph{V}-, \emph{R}- and \emph{I}-band optical photometry. In this paper we adopt $z=0.30$, noting that spectroscopic confirmation would benefit the analysis.

The cluster is followed-up with X-ray observations with \emph{XMM-Newton}. The \citet{2011AuA...536A...9P} find from these data a temperature of $T=(9.82\pm0.84)$keV, $M_{g,500c}=(1.17\pm0.04)\times10^{14}h_{70}^{-1}\Msol$ and a corresponding $Y_{X,500c}=(11.49\pm1.33)\times10^{14}h_{70}^{-1}\Msol$keV. From a $Y_{X}$ MOR calibrated with relaxed objects under the assumption of HSE \citep{2010AuA...517A..92A}, they calculate a mass of $M_{500c}=(9.25\pm0.60)\times10^{14}h_{70}^{-1}\Msun$.

The \citet{2011AuA...536A...9P} report the cluster to have a disturbed X-ray morphology.

From the catalogued SZ likelihood of \citet{2013arXiv1303.5089P} we calculate mass estimates using the X-ray $\theta_{500}$ as a fixed size and the self-consistency method as $M_{500c}=10.3^{+0.9}_{-1.0}\times10^{14}$ and $10.6^{+1.1}_{-1.0}\times10^{14}h_{70}^{-1}\Msol$, respectively.

\subsubsection{Weak Lensing Analysis}

Our $\kappa$ map of PLCKESZ G292.5+22.0 shows a rather complex structure. The central peak is not as strong as a second structure towards the north. The latter is approximately centred on a diffuse galaxy $\Delta m_R=0.98$ brighter than the BCG of PLCKESZ G292.5+22.0 and at a redshift that is indiscriminable based on the \emph{R}-\emph{I} colour.

The shear profile (Fig.~\ref{fig:picplckg292}, bottom panel) has clear signs of this structure, the dip in tangential shear at 5~arcmin and counter-excess at 7~arcmin indicating that an overdensity is present at $\approx5$~arcmin projected separation from the BCG.

A single halo fit, fixing the centre at the BCG and marginalizing over concentration, yields $M_{200m}=(6.8^{+3.8}_{-3.0})\times 10^{14}h_{70}^{-1}\Msol$ ($M_{500c}=(3.4^{+1.7}_{-1.4})\times 10^{14}h_{70}^{-1}\Msol$). The observed shear profile and the confidence interval of the NFW fit are shown in Fig.~\ref{fig:picplckg292} (bottom panel).

The best-fitting mass is reduced considerably when modelling multiple nearby structures. We then find $M_{200m}=(5.0^{+3.1}_{-2.5})\times 10^{14}h_{70}^{-1}\Msol$ ($M_{500c}=(2.7^{+1.5}_{-1.3})\times 10^{14}h_{70}^{-1}\Msol$). Two structures (D and E) are assigned masses of close to $M_{200m}=10^{15}h_{70}^{-1}\Msol$ in this combined analysis. Their influence is seen from the tangential shear profile (Fig.~\ref{fig:picplckg292}, bottom panel) as a low outlier around 5~arcmin (and a high outlier at 7~arcmin) radius, where the BCG-centred overdensity is decreased (increased) by matter on the edge of (inside) the respective annulus.

We hypothesize that this vicinity of other massive structures might be the reason for the $\approx4\sigma$ discrepancy between our mass estimate and the X-ray and SZ results.

\subsubsection{Strong Lensing}

\begin{figure}
\centering
\includegraphics[width=0.28\textwidth]{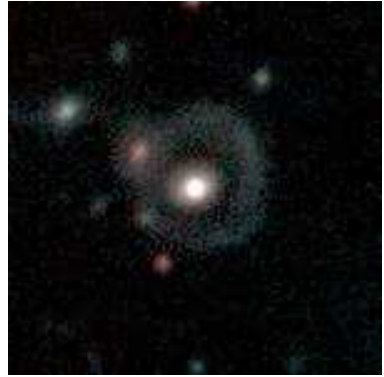}
\caption{\emph{R}-\emph{I} colour image of the candidate galaxy lens in the field of PLCKESZ G292.5+22.0. The cutout size is $0.5\times0.5$~arcmin$^2$, centred on $(\alpha,\delta)=(12^{\rm h}01^{\rm m}21.8^{\rm s},-39^{\circ}51'22'')$ (position B in Fig.~\ref{fig:picplckg292}).}
\label{fig:plckg292sl}
\end{figure}

There is one more interesting feature in the field of PLCKESZ G292.5+22.0 beside the strongly magnified image of a background galaxy lensed between the BCG and another bright foreground galaxy (see Fig.~\ref{fig:picplckg292}). This other feature is a symmetric ring around a red galaxy (\emph{R}-\emph{I} colour of 0.59, redshift not well constrained by this particular colour but consistent with the cluster redshift) with bluer colour (\emph{R}-\emph{I}~$\approx$~0.3) at a radius of $\theta\approx4$~arcsec (see Fig.~\ref{fig:plckg292sl}). The WL analysis shows no significantly massive halo centred at this position, making it doubtable whether the feature is a star-forming physically associated ring or indeed lensed. However, the lensing strength of the red galaxy could also be boosted by the projected dark matter density of the surrounding cluster haloes sufficiently to allow for such a large Einstein radius.

\subsection{MACS J0416.1--2403}

\subsubsection{Visual Appearance}

\begin{figure}
\centering
\includegraphics[width=0.43\textwidth]{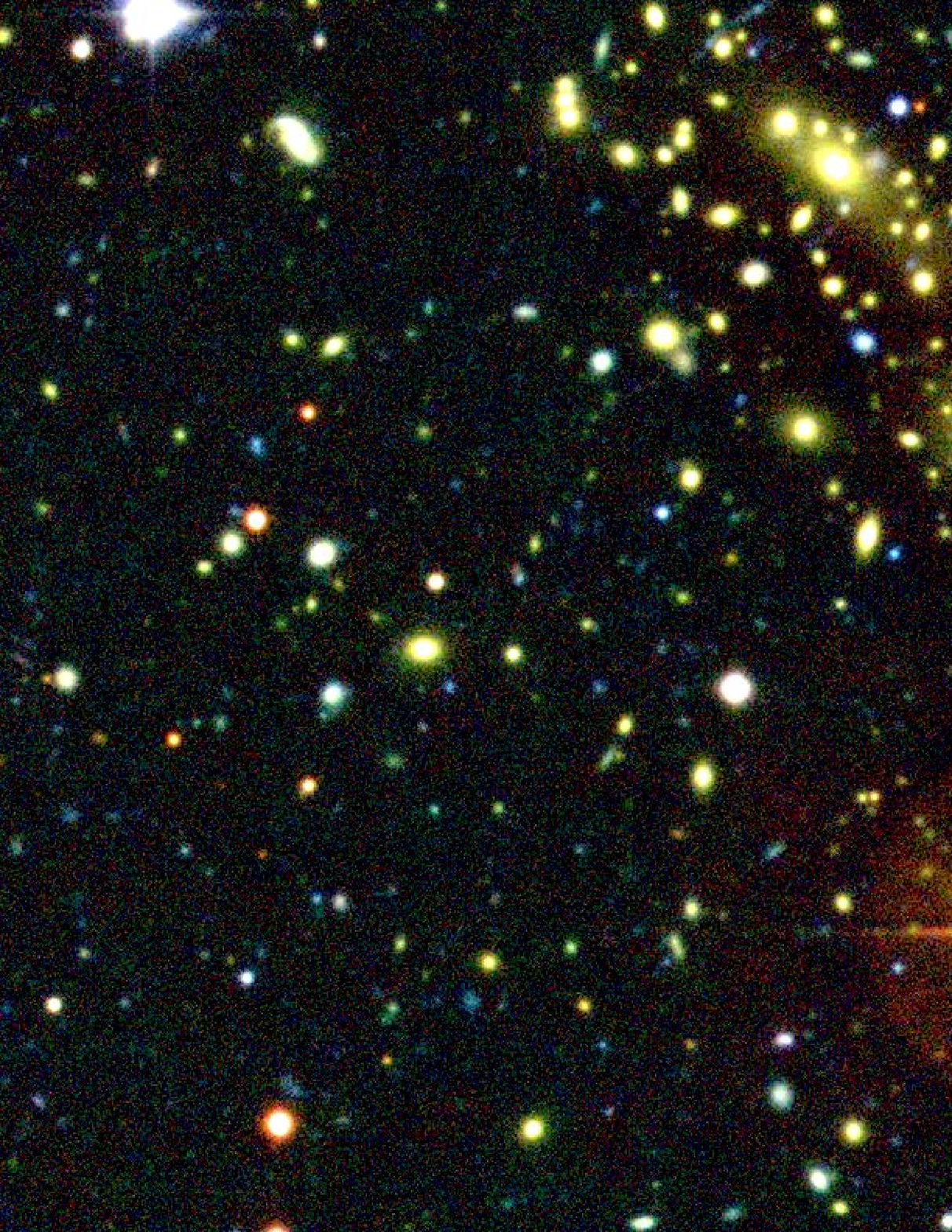}
\includegraphics[width=0.47\textwidth]{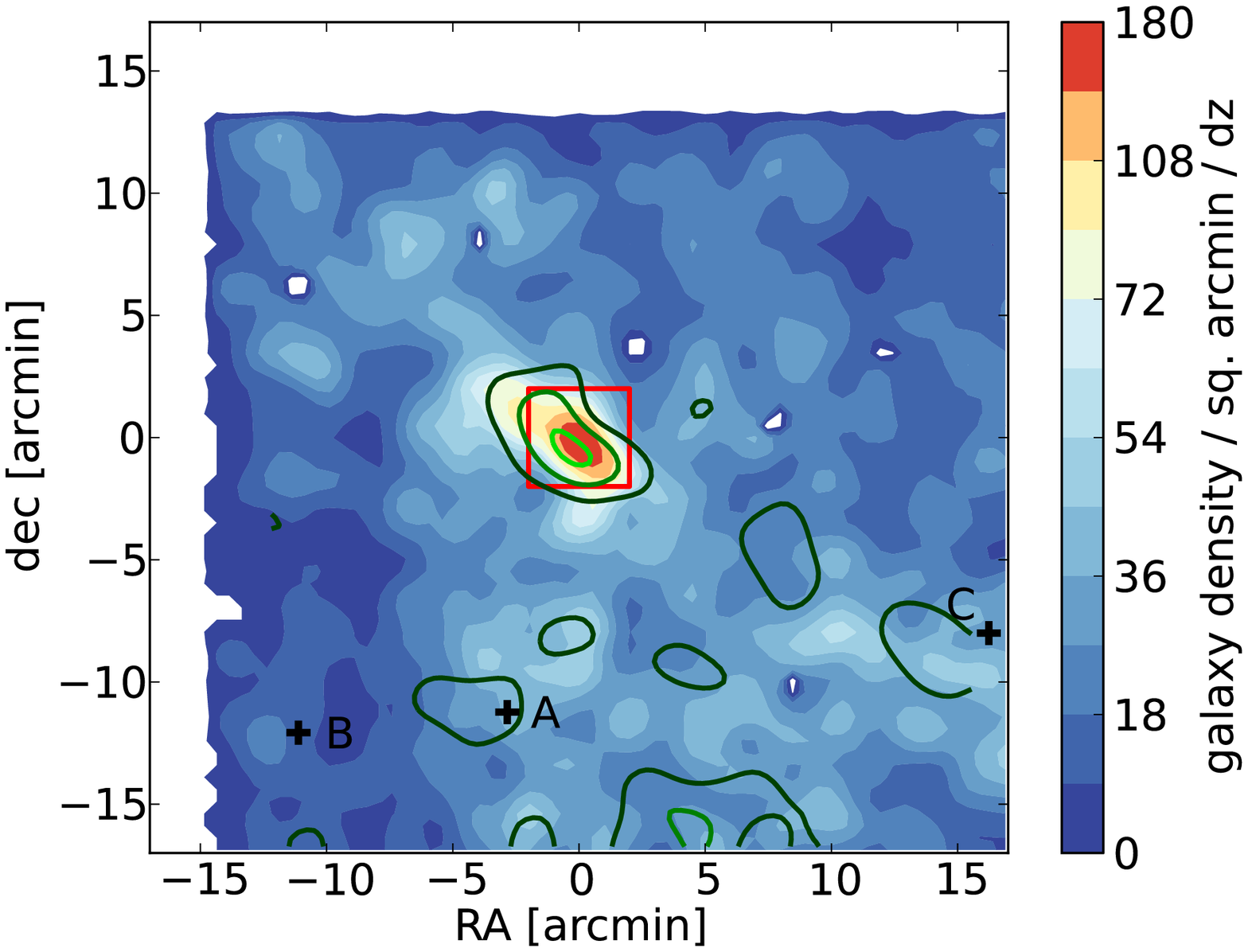}
\includegraphics[width=0.44\textwidth]{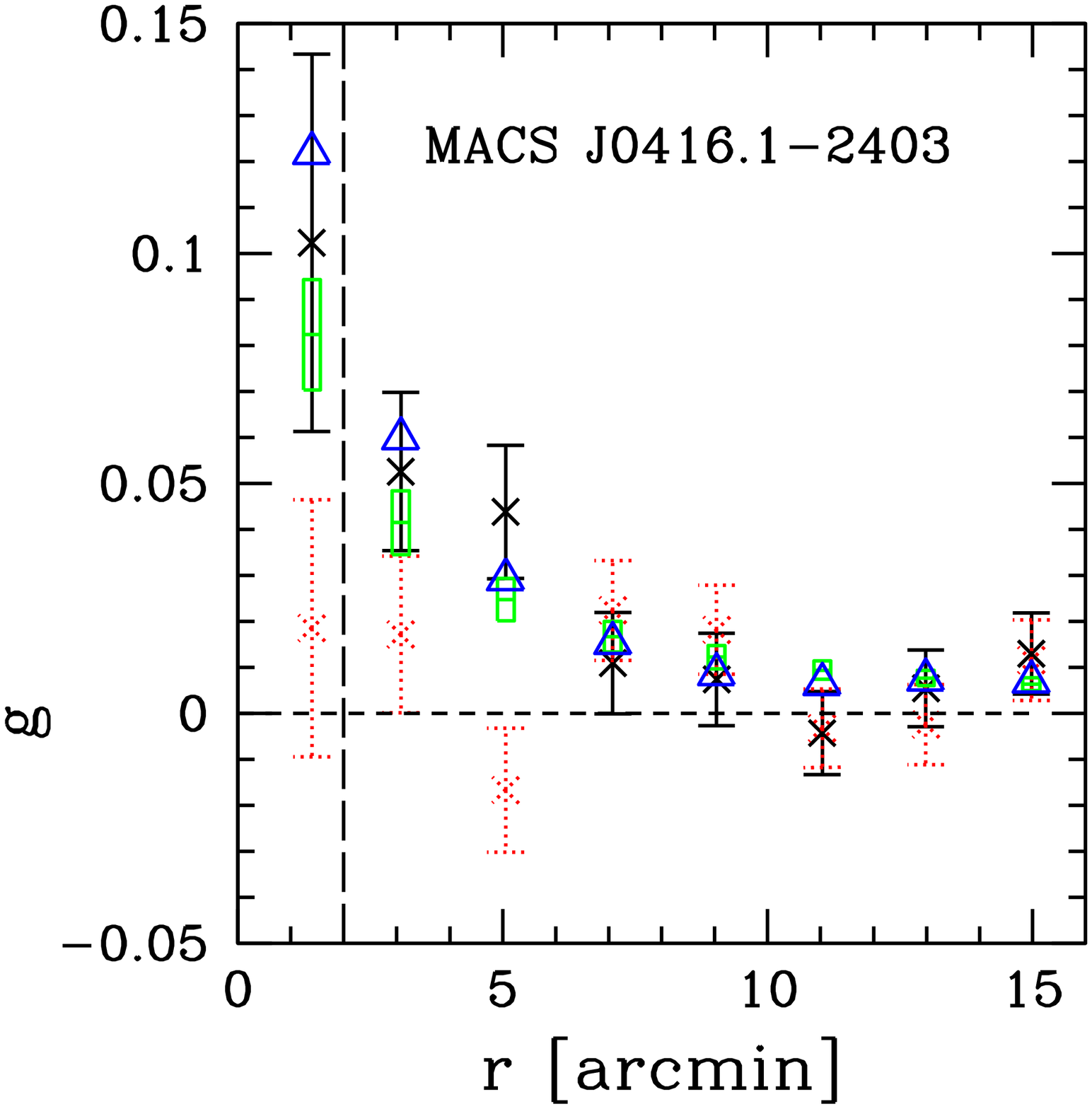}
\caption{Colour image from \emph{BRI} frames (top panel), three-dimensional galaxy density and $\kappa$ contours (central panel) and shear profile (bottom panel) of MACS J0416.1--2403.  The latter includes the prediction of tangential shear (blue triangles) from the SL+WL density profile of \citet[][their Fig.~2]{2013ApJ...762L..30Z}. See Fig.~\ref{fig:picspt4} and Section~\ref{sec:spt4visual} for details.}
\label{fig:picmacs0416}
\end{figure}

A view of the central part of MACS J0416.1--2403 (cf. Fig~\ref{fig:picmacs0416}) reveals its highly elongated shape, with a chain of bright cluster member galaxies along the north-east to south-western direction. Several strong lensing features are visible.

\subsubsection{Previous Work}

This cluster was discovered in the \emph{ROSAT} All-Sky survey as the bright source 1RXS J041609.9-240358 \citep{1999A&A...349..389V}. \citet{2012ApJS..199...25P} estimate its redshift from the \emph{Chandra} X-ray spectrum as $z=0.42\pm0.02$. \citet{2014ApJS..211...21E} provide spectroscopy for 65 galaxies in the field of MACS J0416.1--2403 and determine the cluster redshift as $z=0.397$, the value we use in this work.

\citet{2013ApJ...768..177S} analyse \emph{Chandra} X-ray data for the cluster and find a mass of $M_{500c}=(9.1\pm2.0)\times10^{14}h_{70}^{-1}\Msol$, assuming a redshift of $z=0.42$. They determine this value from the gas mass of $M_{g,500c}=(1.05\pm0.23)\times10^{14}h_{70}^{-1}\Msol$ with $f_{\mathrm{gas},500c}=0.115$ in analogy to \citet{2010MNRAS.406.1773M}.

The cluster is covered by the CLASH project \citep{2012ApJS..199...25P} as part of the high magnification sample. \citet{2013ApJ...762L..30Z} find a highly elongated ($\approx 5:1$) critical area with extraordinarily high density of multiple images. The system has also been selected as an \emph{HST} Frontier Field.

MACS J0416.1--2403 is not listed in the \citet{2013arXiv1303.5089P} SZ catalogue and therefore below the $4.5\sigma$ detection significance limit in the Planck SZ map.

\subsubsection{Weak Lensing Analysis}

Our $\kappa$ map of MACS J0416.1--2403 (see Fig.~\ref{fig:picmacs0416}, central panel) shows a strong elongation of the matter density along the north-east to south-western direction, in line with the galaxy density and the strong lensing model of \citet{2013ApJ...762L..30Z}. The distribution of galaxies near the cluster redshift in the overall field shows higher density in the south-western quadrant, consistent with the numerous small mass peaks detected in this region. Both observations are in line with the hypothesis of MACS J0416.1--2403 being located at the crossing point of two filaments running in a north-south and a north-east to south-west direction.

The centre of MACS J0416.1--2403 is ambiguous, with the two brightest galaxies differing in magnitude by only 0.02. Setting the centre of the system at the midpoint between the two brightest cluster galaxies as proposed by \citet{2013ApJ...762L..30Z} yields a best-fitting mass for the single halo analysis at $M_{200m}=(9.4^{+2.7}_{-2.5})\times 10^{14}h_{70}^{-1}\Msol$ ($M_{500c}=(5.5^{+1.5}_{-1.4})\times 10^{14}h_{70}^{-1}\Msol$), marginalizing over concentration. The observed shear profile and the confidence interval of the NFW fit are shown in Fig.~\ref{fig:picmacs0416} (bottom panel). 

For comparison, we perform a single halo fit fixing the centre at the marginally brighter north eastern galaxy. This yields a consistent mass, yet the shear in the most central bin, which is excluded from the fit, at this position is significantly lower, indicating that the BCG is indeed slightly off-centred from the mass peak.

Subtraction of the maximum-likelihood profiles of neighbouring structures determined in a combined fit yields a slightly larger best-fitting mass of $M_{200m}=(9.8^{+2.8}_{-2.5})\times 10^{14}h_{70}^{-1}\Msol$ ($M_{500c}=(5.9^{+1.5}_{-1.5})\times 10^{14}h_{70}^{-1}\Msol$) \emph{(updated)}.

Both of these estimates are below the X-ray measurement at $\approx1.5\sigma$. This is less of a tension since it is to be expected that both X-ray estimate and spherical WL model fit are significantly influenced by the disturbed morphology of the system in the direction of the observed discrepancy and the X-ray analysis uses a slightly higher value for the cluster redshift.

\subsubsection{Strong Lensing}

MACS J0416.1--2403 is a known prominent strong lensing system, with several of the known multiple images also visible from our data. Thanks to the availability of deeper \emph{HST} and \emph{Subaru} data from the CLASH survey, we can compare our shear measurement with the predicted shear from the \citet{2013ApJ...762L..30Z} projected density profile based on a combined weak and strong lensing analysis (blue triangles in Fig.~\ref{fig:picmacs0416}, bottom panel). We find our shear measurements to be in good agreement with their model. Both our measurement and the \citet{2013ApJ...762L..30Z} model exceed the shear in the central region as predicted by the NFW fit (green). Since the latter is drawn with fixed concentration according to the \citet{2008MNRAS.390L..64D} mass-concentration relation, this shows that the concentration of MACS J0416.1--2403 is indeed higher than average.
Note that our mass measurement, however, is not based on
the fixed concentration profile drawn here for illustration
purposes, but rather marginalizes over concentration with a
prior motivated from simulations \citep{2001MNRAS.321..559B}.

\subsubsection{Non-Detection in Planck}

Since MACS J0416.1--2403 is a large and highly magnifying gravitational lens, selected as such in the CLASH survey \citep{2013ApJ...762L..30Z}, the question is whether its non-detection in the Planck SZ catalogue is in line with expectations. We make a rough estimate of detection probability given the result of the WL analysis here.

MACS J0416.1--2403 is situated 44$^{\circ}$ away from the Galactic plane, in a region of expected average noise levels of the Planck SZ map. At the $\theta_{500c}$ corresponding to its best-fitting mass of $\theta_{500c}\approx3$~arcmin, the Monte Carlo analysis of \citet[][cf. their fig. 9]{2013arXiv1303.5089P} suggests 50\% completeness at a Compton parameter of $Y_{500}\approx6\times10^{-4}$~arcmin$^2$. This corresponds, in turn, to a mass of $M_{500c}\approx7\times10^{14}h_{70}^{-1}\Msol$, above the observed mass of the cluster. We therefore conclude there is no tension with either the Planck calibrations or our WL model, since both the SZ-detection and non-detection of MACS J0416.1--2403 are possible at the mass reconstructed from the shear. The non-detection of this relatively high-redshift system, however, is also made even more likely if there is a yet unaccounted for redshift dependence of the Planck SZ MOR (cf. Section~\ref{sec:planckzdep}).

\subsection{SPT-CL J2248--4431}

We refer the reader to the detailed WL analysis of the system by \citet{2013MNRAS.tmp.1221G}. Here, the system is only used as part of the combined analysis and our results are compared to the Planck results released after this study.

\subsubsection{Comparison with SZ data}

The de-biased significance for SPT at full depth is calculated as $\zeta=32.8$, assuming the mass estimate of \citet{2011ApJ...738..139W} and equation (\ref{eqn:sptmor}). Note that this is the de-biased significance at full SPT depth, while in fact \citet{2011ApJ...738..139W} present the cluster based on less deep SPT observations (and consequently with lower measured significance).

Using the X-ray mass estimate of $M_{500c}=(12.6\pm0.2)\times10^{14}h_{70}^{-1}\Msol$ from \citet{2011AuA...536A..11P} as a fixed size prior and, alternatively, the self-consistency method, we determine the SZ mass estimate for the system at $M_{500c}=(14.5^{+0.7}_{-0.7})\times10^{14}h_{70}^{-1}\Msol$ and $M_{500c}=(14.8^{+0.8}_{-0.8})\times10^{14}h_{70}^{-1}\Msol$, respectively. We note that this result is $\approx20$\% higher than the previous estimate based on the early SZ catalogue in \citet{2013MNRAS.tmp.1221G}, due to the changed MOR which includes a hydrostatic bias of $(1-b)=0.8$. It is still consistent with the previous WL result. The substructure-subtracted WL result is lower than the SZ mass estimate by $1.5\sigma$ ($1.8\sigma$ for the self-consistent method), consistent with the hypothesis that the substructure at $\theta=3.5$~arcmin separation is blended with the main cluster in the Planck SZ signal or that the hydrostatic mass bias assumed in the Planck MOR is not correct in the case of this system.

\subsubsection{Strong Lensing}

SPT-CL J2248--4431 is a prominent strong lensing system with several known arcs also visible in our data. Our lensing mass reconstruction has recently been compared with the strong lensing model of \citet{annamonna}, two independent CLASH re-analyses of the WFI data, which include magnification and strong lensing in addition to shear (\citealt{2014arXiv1404.1376M} and \citealt{2014arXiv1404.1375U}) and a WL analysis of Dark Energy Survey science verification data \citep{2014arXiv1405.4285M}, in all cases with consistent mass estimates.

\subsection{PSZ1 G168.02--59.95}

\subsubsection{Visual Appearance}

\begin{figure}
\centering
\includegraphics[width=0.44\textwidth]{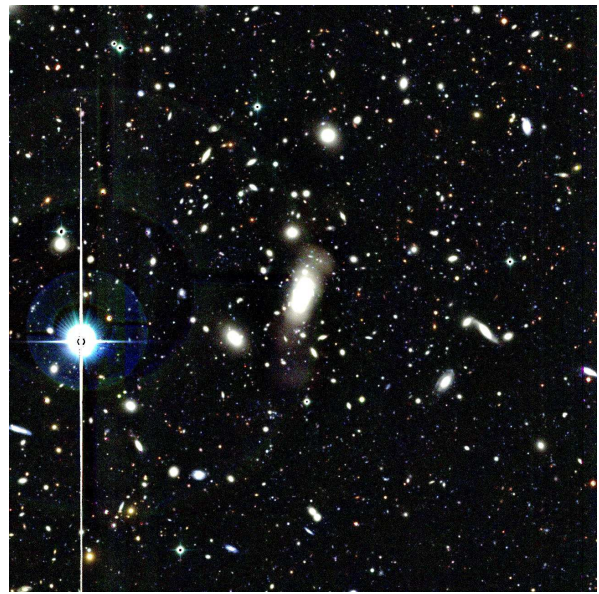}
\includegraphics[width=0.47\textwidth]{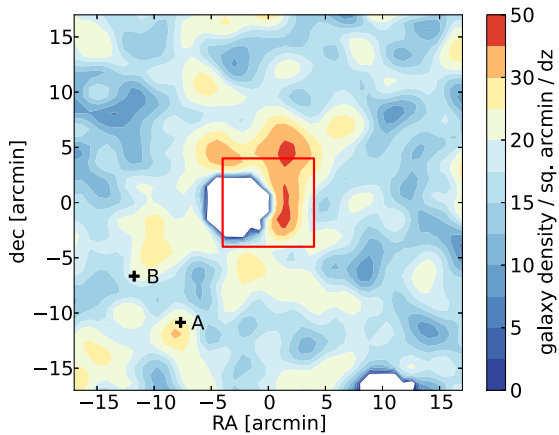}
\includegraphics[width=0.44\textwidth]{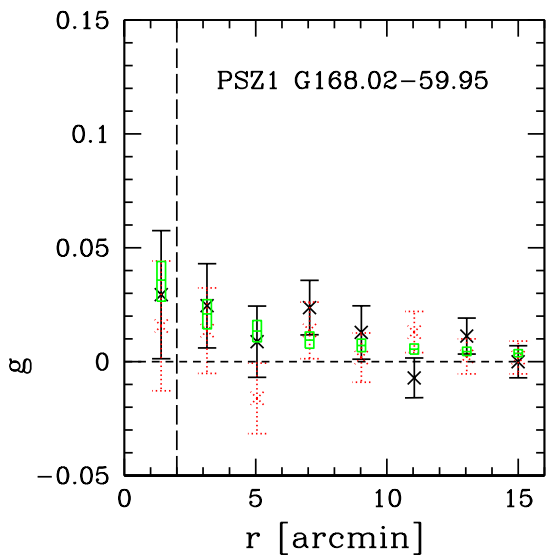}
\caption{Colour image of $8\times8$~arcmin$^2$ central region from \emph{gri} frames (top panel), three-dimensional galaxy density smoothed with 1.5~arcmin radius Epanechnikov kernel (central panel) and shear profile (bottom panel) of PSZ1 G168.02--59.95. See Fig.~\ref{fig:picspt4} and Section~\ref{sec:spt4visual} for details.}
\label{fig:picpsz168}
\end{figure}

The central region of PSZ1 G168.02--59.95 is shown in Fig.~\ref{fig:picpsz168}, on a cutout size of $8\times8$~arcmin due to the smaller distance to the system. One can see the complex and disturbed structure of the bright central galaxy. The three-dimensional density is difficult to interpret due to the large masked area around a number of bright stars.

\subsubsection{Previous Work}

PSZ1 G168.02--59.95 was first discovered by \citet{1958ApJS....3..211A} and classified at the second-lowest of six richness classes. 

Based on a magnitude-redshift relation, \citet{2000ApJS..126....1Q} give a redshift estimate of $z=0.141$. The value adopted here is the spectroscopic $z=0.1456$ from Mirkazemi et al. (in preparation), based on four cluster member galaxies.

The cluster is serendipitously discovered by the RASS as RXC J0214.6--0433 \citep{2002ApJS..140..239C}. Based on its X-ray luminosity and temperature in the \emph{XMM} Cluster Survey, \citet{2012MNRAS.422.1007V} estimate the probability of detecting the system by its SZ signal in Planck to be 76 per cent. The system is also part of the CFHT X-ray cluster sample of Mirkazemi et al. (in preparation).

The system is associated with the \emph{ROSAT} bright X-ray source 1RXS J021439.0-433319 \citep{1999A&A...349..389V}. We apply \emph{ROSAT} band conversion factor, Galactic H\textsc{i} column density correction \citep{2005A&A...440..775K} and $k$-correction from tables 2-4 of \citet{2004A&A...425..367B}. In this procedure, we assume an X-ray temperature $T_{\rm X}=5$~keV, as justified by the relative independence of the conversion factors on $T_{\rm X}$ and the large intrinsic scatter of the $T_{\rm X}-L_{\rm X}$ relation \citep{2009A&A...498..361P} that could be used for self-consistent calibration. The derived X-ray luminosity corresponds to a mass of $M_{500c}=(2.5\pm0.2)\times10^{14}h_{70}^{-1}\Msol$ according to the \citet{2002ApJ...567..716R} $L_{\rm X}-M$ relation, which was calibrated using measurements under the assumption of HSE.

From the catalogued SZ likelihood of \citet{2013arXiv1303.5089P} we calculate mass estimates using the X-ray $\theta_{500}$ as a fixed size and the self-consistency method as $M_{500c}=4.6^{+0.6}_{-0.8}\times10^{14}h_{70}^{-1}\Msol$ and $5.0^{+0.9}_{-0.9}\times10^{14}h_{70}^{-1}\Msol$, respectively.

\citet{1967AuJPh..20..715T} fail to detect radio emission from the system at 1410~MHz.

\subsubsection{Weak Lensing Analysis}

A single halo fit, fixing the centre at the BCG and marginalizing over concentration, yields $M_{200m}=(2.3^{+1.4}_{-1.1})\times 10^{14}h_{70}^{-1}\Msol$ ($M_{500c}=(1.2^{+0.7}_{-0.6})\times 10^{14}h_{70}^{-1}\Msol$). The observed shear profile and the confidence interval of the NFW fit are shown in Fig.~\ref{fig:picpsz168} (bottom panel).

Subtraction of the maximum likelihood signal of neighbouring structures before re-fitting the central halo yields a consistent mass $M_{200m}=(2.5^{+1.5}_{-1.2})\times 10^{14}h_{70}^{-1}\Msol$ ($M_{500c}=(1.3^{+0.7}_{-0.6})\times 10^{14}h_{70}^{-1}\Msol$).

This result is marginally consistent with the X-ray mass estimate ($1.5\sigma$ lower), yet significantly lower than the Planck SZ estimate at $\approx3\sigma$.

\subsubsection{Strong Lensing}

Consistent with the relatively low mass and redshift of the system, we find no evidence for strong lensing in PSZ1 G168.02--59.95.

\subsection{PSZ1 G230.73+27.70}

\subsubsection{Visual Appearance}

\begin{figure}
\centering
\includegraphics[width=0.44\textwidth]{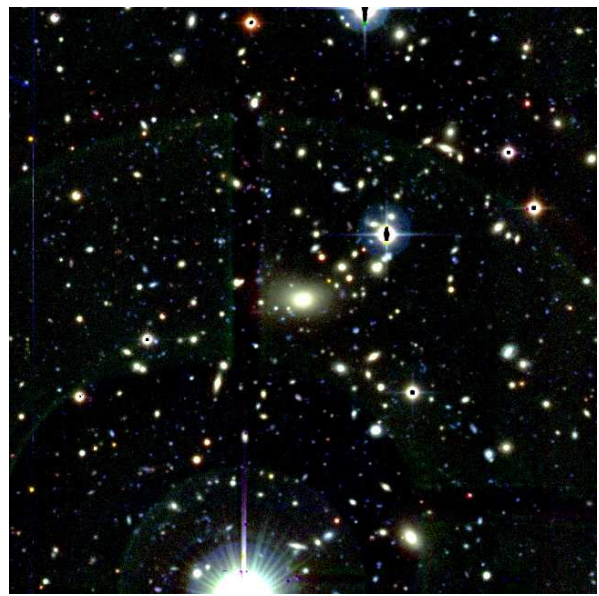}
\includegraphics[width=0.47\textwidth]{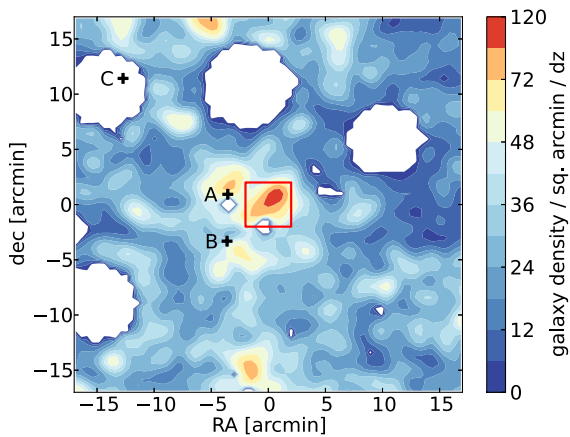}
\includegraphics[width=0.44\textwidth]{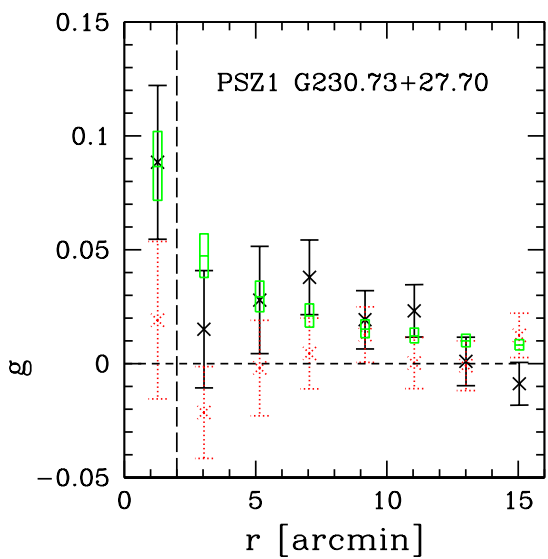}
\caption{Colour image from \emph{gri} frames (top panel), three-dimensional galaxy density (central panel) and shear profile (bottom panel) of PSZ1 G230.73+27.70. See Fig.~\ref{fig:picspt4} and Section~\ref{sec:spt4visual} for details.}
\label{fig:picpsz230}
\end{figure}

Fig.~\ref{fig:picpsz230} shows the central region of PSZ1 G230.73+27.70. A large number of cluster members is visible, although imaging is somewhat obstructed by a bright star in the south of the system. We note that in addition the central BCG, there are several additional diffuse elliptical galaxies of similar colour outside this cutout. Two of them (A and B) are included as halo centres in our multi-halo analysis of the system.

\subsubsection{Previous Work}

In the Planck catalogue, PSZ1 G230.73+27.70 is associated with a system first described in the maxBCG catalog of clusters at a photometric redshift of $z=0.2944$, although the Planck detection is centred on a galaxy with similar colour 4~arcmin away from the maxBCG centre. 

\citet{2007ApJ...660..239K} give the $N_{r_{200}}$ richness of the maxBCG system as 60. \citet{2012ApJ...746..178R} calculate a richness $\lambda=70.2\pm4.7$, to which corresponds a mass of $M_{200m}=(6.6\pm0.5)\times10^{14}h_{70}^{-1}\Msol$ with an intrinsic scatter of $\approx30$\%.

The system is associated with the \emph{ROSAT} faint X-ray source 1RXS J090134.0-013900 \citep{2000IAUC.7432R...1V}. We apply \emph{ROSAT} band conversion factor, Galactic H\textsc{i} column density correction \citep{2005A&A...440..775K} and $k$-correction from tables 2-4 of \citet{2004A&A...425..367B}. In this procedure, we assume an X-ray temperature $T_{\rm X}=5$~keV, as justified by the relative independence of the conversion factors on $T_{\rm X}$ and the large intrinsic scatter of the $T_{\rm X}-L_{\rm X}$ relation \citep{2009A&A...498..361P} that could be used for self-consistent calibration. The derived X-ray luminosity corresponds to a mass of $M_{500c}=(3.6\pm0.8)\times10^{14}h_{70}^{-1}\Msol$ according to the \citet{2002ApJ...567..716R} $L_{\rm X}-M$ relation, which was calibrated using measurements under the assumption of HSE.

From the catalogued SZ likelihood of \citet{2013arXiv1303.5089P} we calculate mass estimates using the X-ray $\theta_{500}$ as a fixed size and the self-consistency method as $M_{500c}=6.0^{+0.9}_{-1.0}\times10^{14}h_{70}^{-1}\Msol$ and $M_{500c}=6.2^{+1.2}_{-1.3}\times10^{14}h_{70}^{-1}\Msol$, respectively.

\subsubsection{Weak Lensing Analysis}

A single halo fit, fixing the centre at the BCG and marginalizing over concentration, yields $M_{200m}=(10.0^{+3.9}_{-3.2})\times 10^{14}h_{70}^{-1}\Msol$ ($M_{500c}=(4.9^{+1.6}_{-1.5})\times 10^{14}h_{70}^{-1}\Msol$). The observed shear profile and the confidence interval of the NFW fit are shown in Fig.~\ref{fig:picpsz230} (bottom panel).

Subtraction of the maximum-likelihood profiles of neighbouring structures yields a lower mass of $M_{200m}=(6.8^{+3.2}_{-2.6})\times 10^{14}h_{70}^{-1}\Msol$ ($M_{500c}=(3.5^{+1.5}_{-1.3})\times 10^{14}h_{70}^{-1}\Msol$).

\subsubsection{Strong Lensing}
We find no evidence for strong lensing in the field of PSZ1 G230.73+27.70.

\subsection{PSZ1 G099.84+58.45}

\subsubsection{Visual Appearance}

\begin{figure}
\centering
\includegraphics[width=0.44\textwidth]{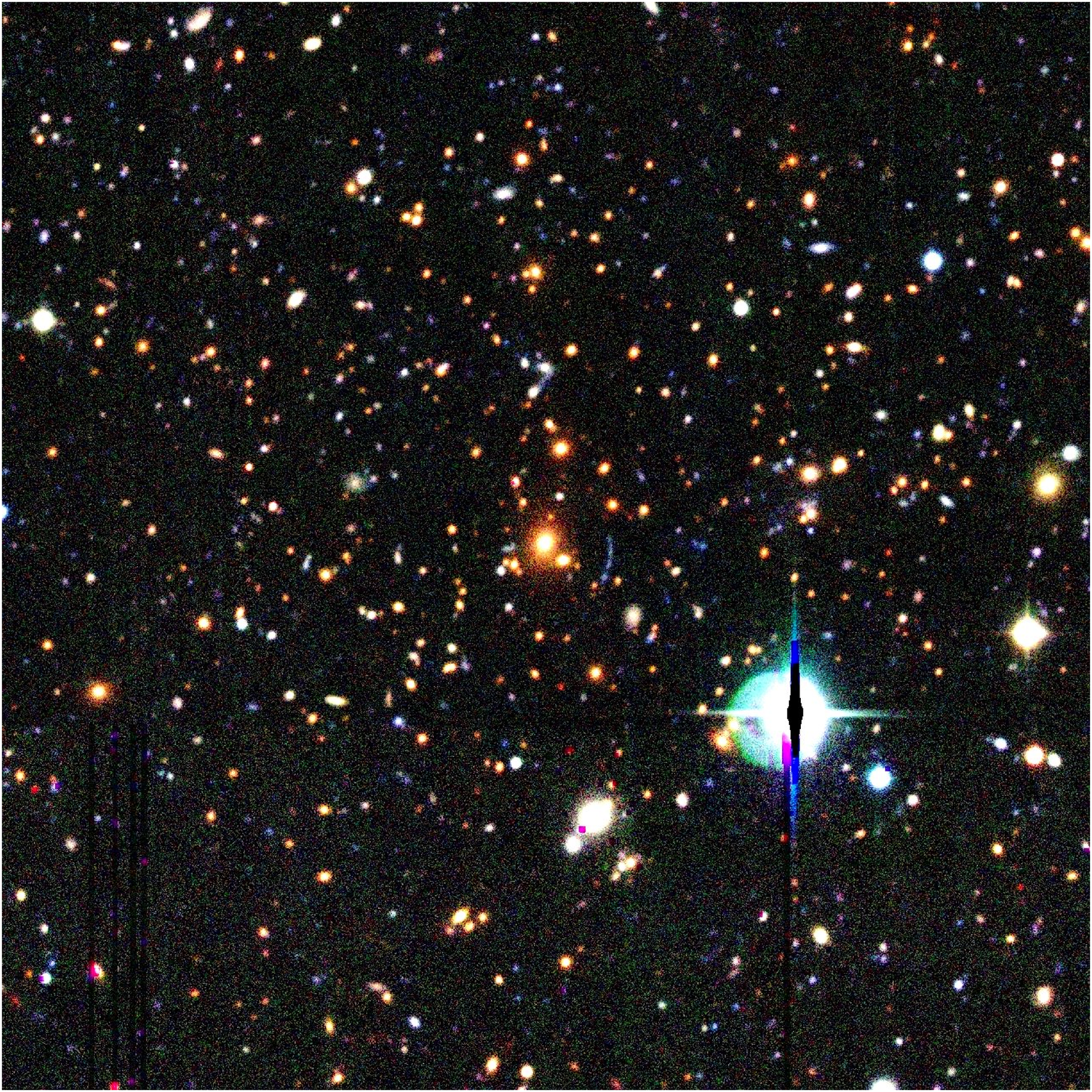}
\includegraphics[width=0.47\textwidth]{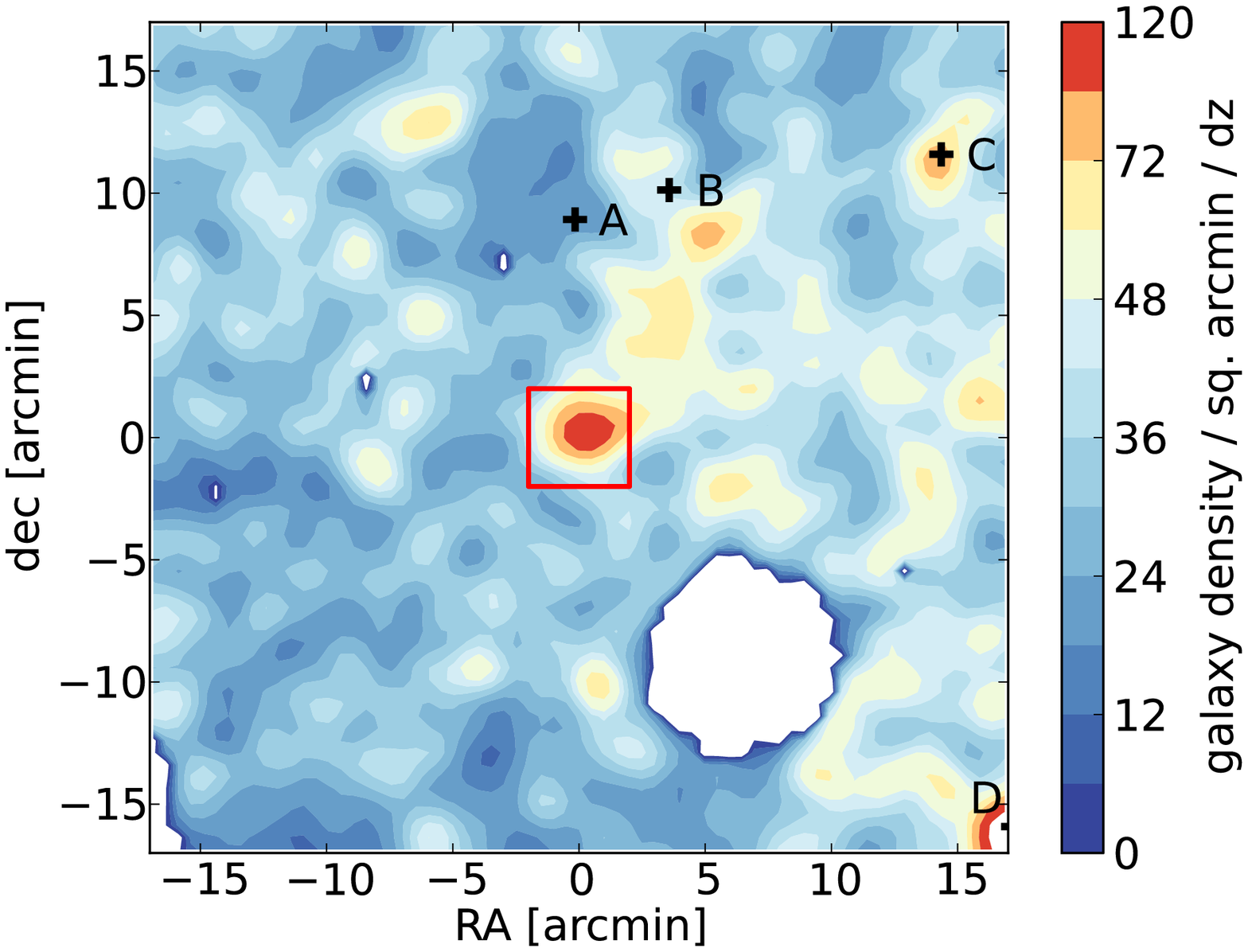}
\includegraphics[width=0.44\textwidth]{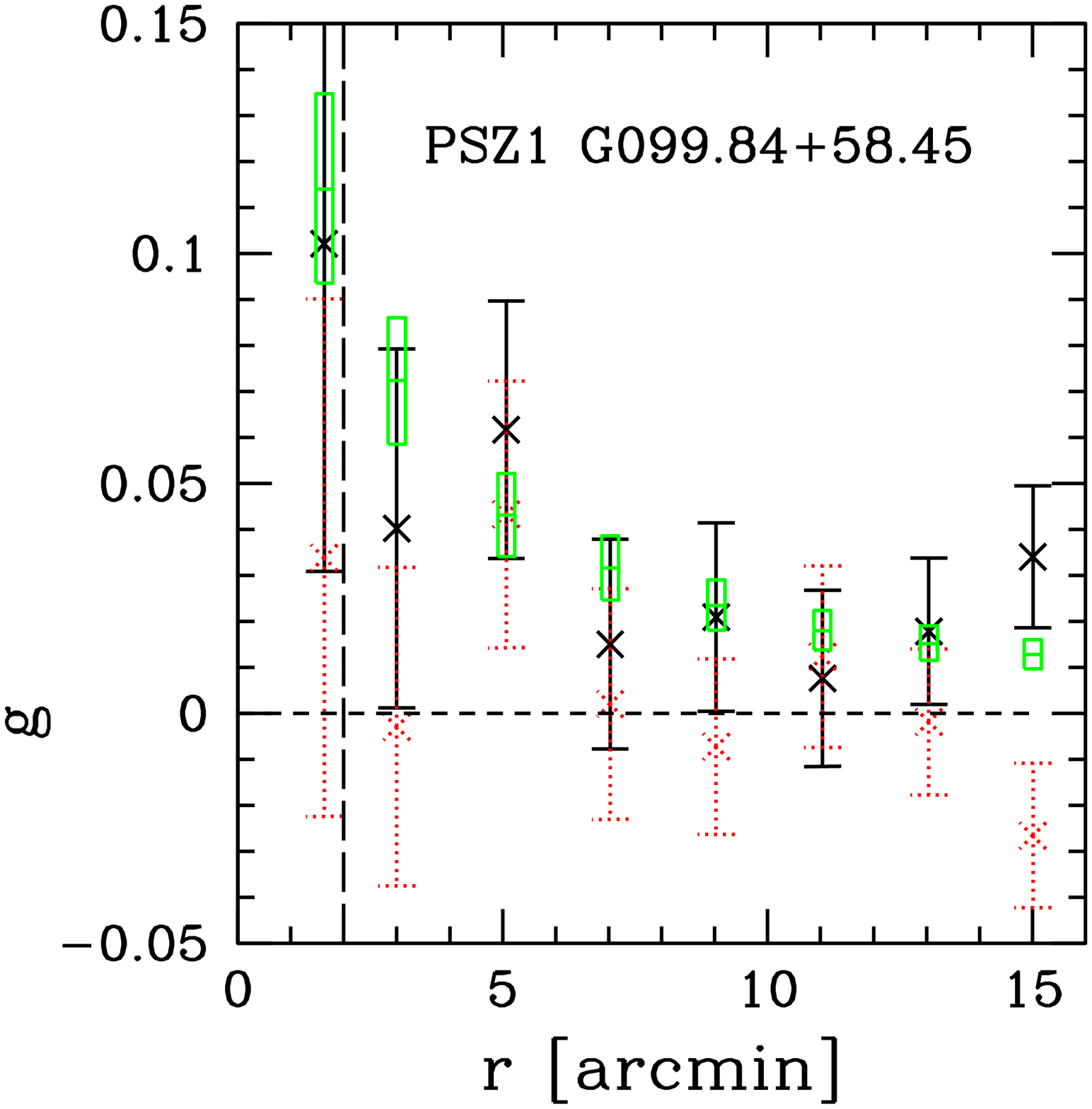}
\caption{Colour image from \emph{gri} frames (top panel), three-dimensional galaxy density (central panel) and shear profile (bottom panel) of PSZ1 G099.84+58.45. See Fig.~\ref{fig:picspt4} and Section~\ref{sec:spt4visual} for details.}
\label{fig:picpsz099}
\end{figure}

Fig.~\ref{fig:picpsz099} shows the central region of PSZ1 G099.84+58.45. The giant arc west of the BCG can be seen along with a very large number of cluster member galaxies. The elongated pattern towards the north-eastern direction in the redshift density map (central panel) is identified visually, although only barely, as a diffuse filamentary structure of high-redshift galaxies.

\subsubsection{Previous Work}

The system was first discovered as a strong lens by the CFHTLS Strong Lensing Legacy Survey (SL2S, \citealt{2007AuA...461..813C}) as SL2S J141447+544703.

\citet{2007AuA...461..813C} give the photometric redshift of the BCG as $z=0.75$, while \citet{2012ApJ...749...38M} find, on the same data, $z=0.63\pm0.02$. The photometric redshift for the central galaxy is $z=0.71$ in our photometric redshift catalogue, from which we also calculate the outlier-clipped median photometric redshift of 32 visually selected cluster members at $z=0.69$, which is the value we adopt in our analysis. 

The system is associated with the \emph{ROSAT} faint X-ray source 1RXS J141443.2+544652 \citep{2000IAUC.7432R...1V}, yet not resolved.  {We apply \emph{ROSAT} band conversion factor, Galactic H\textsc{i} column density correction \citep{2005A&A...440..775K} and $k$-correction in extrapolation of tables 2-4 of \citet{2004A&A...425..367B}. In this procedure, we assume an X-ray temperature $T_{\rm X}=10$~keV, as justified by the relative independence of the conversion factors on $T_{\rm X}$ and the large intrinsic scatter of the $T_{\rm X}-L_{\rm X}$ relation \citep{2009A&A...498..361P} that could be used for self-consistent calibration. The derived X-ray luminosity corresponds to a mass of $M_{500c}=(12.5\pm2.0)\times10^{14}h_{70}^{-1}\Msol$ according to the \citet{2002ApJ...567..716R} $L_{\rm X}-M$ relation, which was calibrated using measurements under the assumption of HSE.

From the catalogued SZ likelihood of \citet{2013arXiv1303.5089P} we calculate mass estimates using the X-ray $\theta_{500}$ as a fixed size and the self-consistency method as 
$M_{500c}=8.6^{+1.1}_{-1.2}\times10^{14}h_{70}^{-1}\Msol$ and $8.3^{+1.2}_{-1.3}\times10^{14}h_{70}^{-1}\Msol$, respectively.

\citet{2013arXiv1308.4674F} follow up the SL2S detections with WL analyses. Their singular isothermal sphere (SIS) fit to the object yields $\sigma_{\rm SIS}=969^{+100}_{-130}$km s$^{-1}$, which we convert to (cf. \citealt{2013MNRAS.tmp.1221G}, their equations 17 and 18) $M_{200m}=(7.1^{+2.5}_{-2.5})\times10^{14}h_{70}^{-1}\Msol$ ($M_{500c}=3.7^{+1.3}_{-1.3}\times10^{14}h_{70}^{-1}\Msol$).

A cluster identified by \citet{2008ApJS..176..414Y} at $(\alpha,\delta)=(14^{\rm h}13^{\rm m}26.3^{\rm s}, 54^{\circ}45'22.0'')$ with spectroscopic redshift $z=0.08278$ lies at $\approx11.5$~arcmin separation from PSZ1 G099.84+58.45. The centre of this system appears to be a galaxy brighter by 0.9 magnitudes than the one given by \citet{2008ApJS..176..414Y} and at a compatible spectroscopic redshift of $z=0.0828$ at a separation of $1.6$~arcmin with coordinates $(\alpha,\delta)=(14^{\rm h}13^{\rm m}27.3^{\rm s}, 54^{\circ}43'46'')$. Due to the low redshift of the system, however, we do not use the position in our combined lensing analysis.

\subsubsection{Weak Lensing Analysis}

A single halo fit, fixing the centre at the BCG and marginalizing over concentration, yields $M_{200m}=(34.4^{+11.6}_{-10.2})\times 10^{14}h_{70}^{-1}\Msol$ ($M_{500c}=(18.1^{+5.8}_{-5.3})\times 10^{14}h_{70}^{-1}\Msol$). The observed shear profile and the confidence interval of the NFW fit are shown in Fig.~\ref{fig:picpsz099} (bottom panel).

Subtraction of the maximum-likelihood profiles of neighbouring structures yields a mass of $M_{200m}=(38.3^{+13.5}_{-11.7})\times 10^{14}h_{70}^{-1}\Msol$ ($M_{500c}=(19.6^{+6.4}_{-5.8})\times 10^{14}h_{70}^{-1}\Msol$), marginally higher than the single-halo value. The high sensitivity of the reconstruction of the high redshift system on matter along the line of sight at lower redshift is due to the fact that the geometrical scaling factor of the lensing signal, $\frac{D_{d}D_{ds}}{D_{s}}$, causes a relative amplification of the foreground signal.

Despite being based on the same observational data and consistency with the strong lensing feature of the system (see following subsection), our results are in conflict with the SIS fit of \citet{2013arXiv1308.4674F} at $\approx3\sigma$ significance (their $\sigma_{\rm SIS}=969^{+100}_{-130}$km s$^{-1}$ is significantly lower than our $\sigma_{\rm SIS}=1540^{+162}_{-190}$km s$^{-1}$ as fitted out to 10~arcmin). 

Several effects can in principle contribute to an underestimation of the shear around rich, high redshift clusters. The method of determining $\beta$ in \citet{2013arXiv1308.4674F} is based on the mean value calculated from the photometric redshift of all sources fainter than $m_{i'}=21$ (except a colour-removed red sequence at the cluster redshift), and all these sources are used for calculating the shear profile. This is suboptimal for rich, high-redshift systems like PSZ1 G099.84+58.45, where most galaxies fainter than this limit are indeed foreground objects or (blue) cluster members. We have verified with the help of Fo{\"e}x (private communication) that the difference between our photo-z based $\beta$ estimated on their background sample and their method of determining $\beta$ is indeed the primary cause of the discrepancy.

We note that the lensing analysis for this cluster includes data from a CFHT pointing (W3m0m0) with strongly ($>10$\%) elliptical PSF along the $\epsilon_1$ direction. Detailed analysis shows that for this particular field we measure significant non-zero mean ellipticity for galaxies along the same direction, particularly at low S/N, indicating an additive bias based on selection effects or insufficient correction for the PSF ellipticity. We verify that also in the CFHTLenS catalogues, S/N dependent non-zero $\epsilon_1$ shear is present in this field. At a highly elliptical PSF, such a bias due to source detection and the simplistic modelling of PSF ellipticity in KSB is not unexpected. 

We choose to model the additive bias by the same functional form used for the multiplicative noise bias in \citet[][their equation 9]{2013MNRAS.tmp.1221G} with $A=0.005$, $B=0.08$ and $C=14$. We have verified that removing the pointing from our analysis or using it without this correction do not significantly alter our results.

\subsubsection{Strong Lensing}

The system was originally discovered as a strong lens with a blue giant arc \citep{2007AuA...461..813C}.

We note that for our best-fitting model with fiducial concentration, the arc redshift would have to be $z_s=1.49$ to give rise to an Einstein radius at the observed arc radius $R_A=(14\pm1)$~arcsec \citep{2007AuA...461..813C}. At the photometric redshift of the arc of $z_s\approx2.0$, the same Einstein radius would be caused by a lens of $M_{200m}=21.6\times 10^{14}h_{70}^{-1}\Msol$, $1.2\sigma$ below our best fit. While spectroscopic redshifts for lens and source would greatly enhance the constraining power of the strong lensing feature, this is still additional evidence for the correctness of our analysis.

\subsection{PSZ1 G099.48+55.62}

\subsubsection{Visual Appearance}

\begin{figure}
\centering
\includegraphics[width=0.44\textwidth]{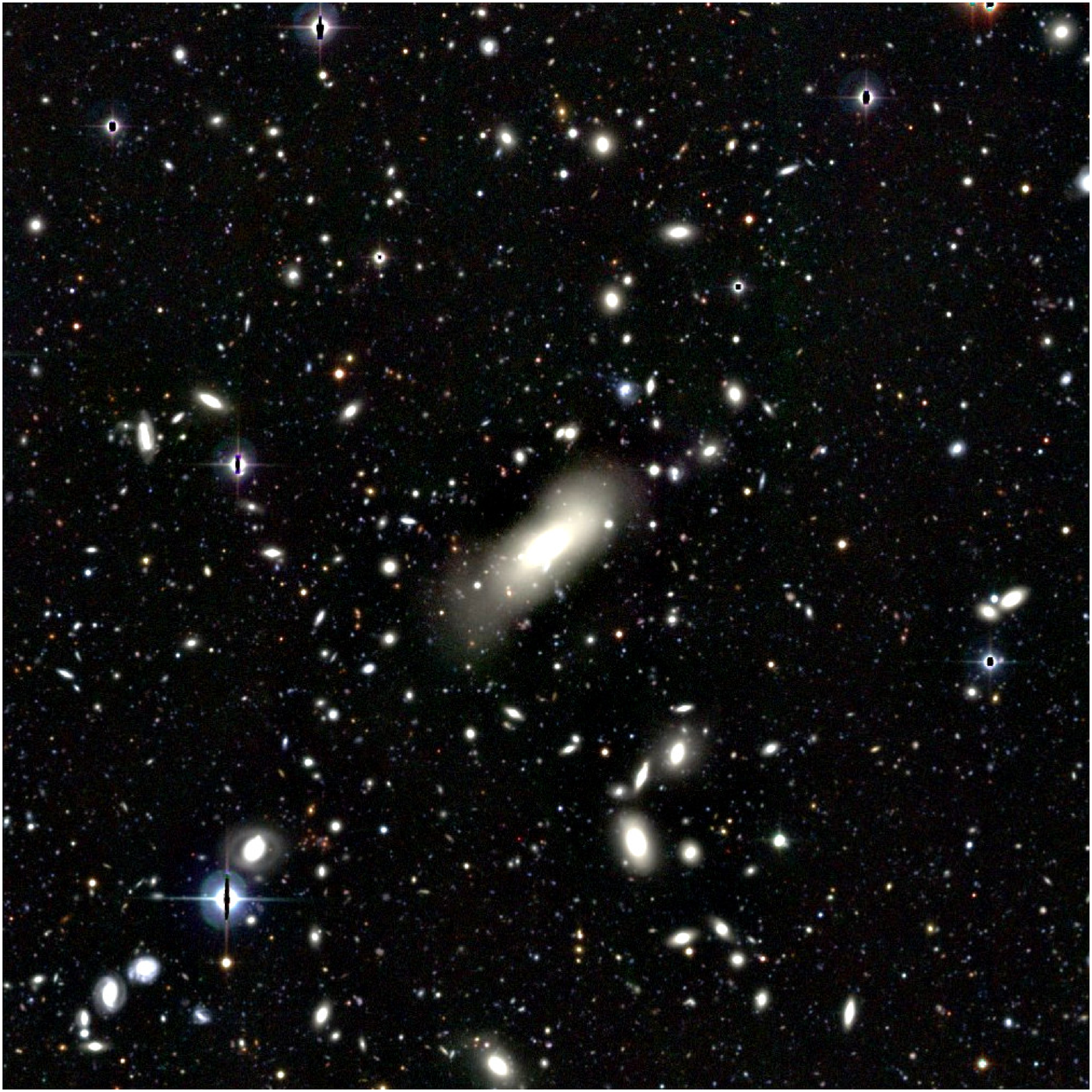}
\includegraphics[width=0.47\textwidth]{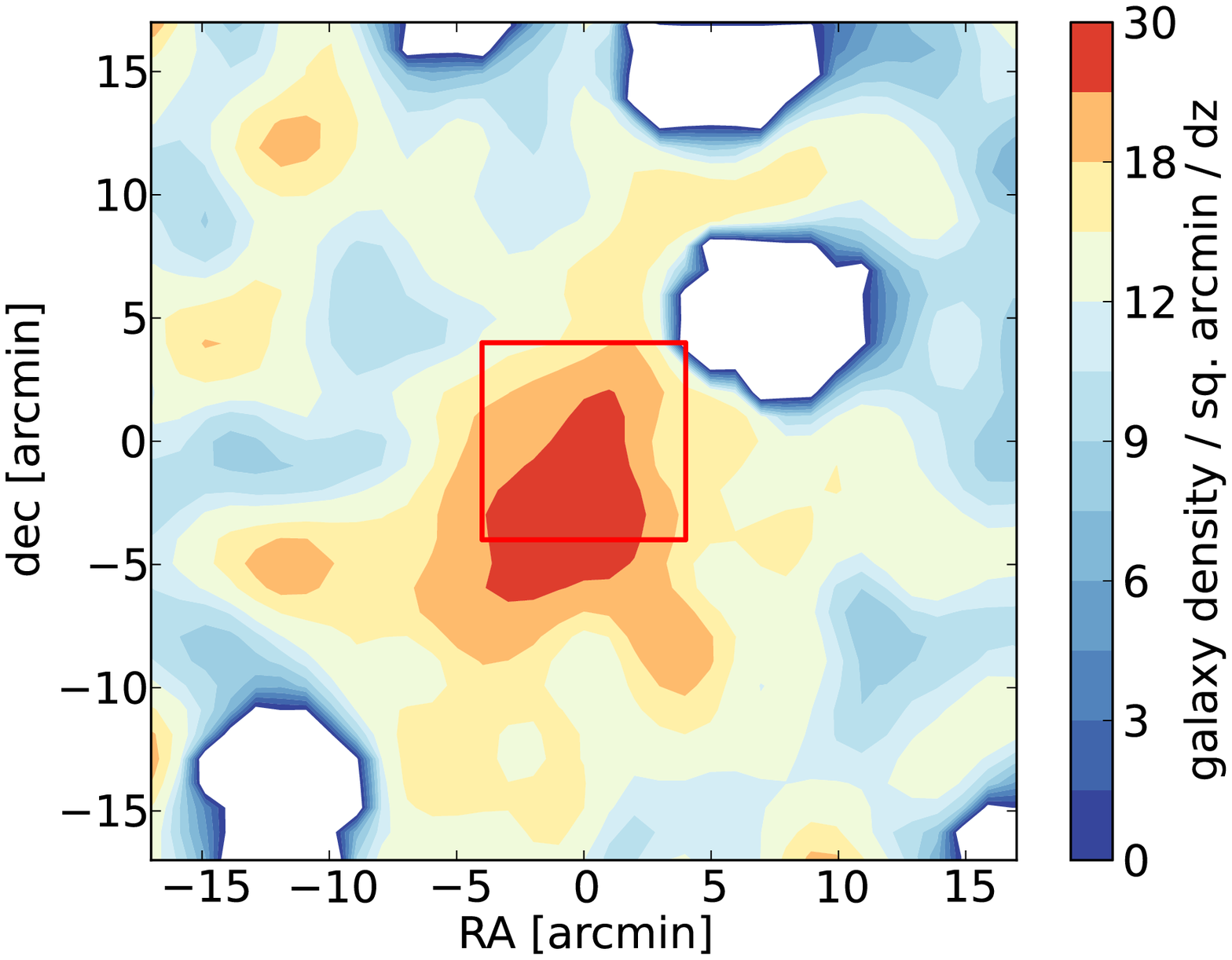}
\includegraphics[width=0.44\textwidth]{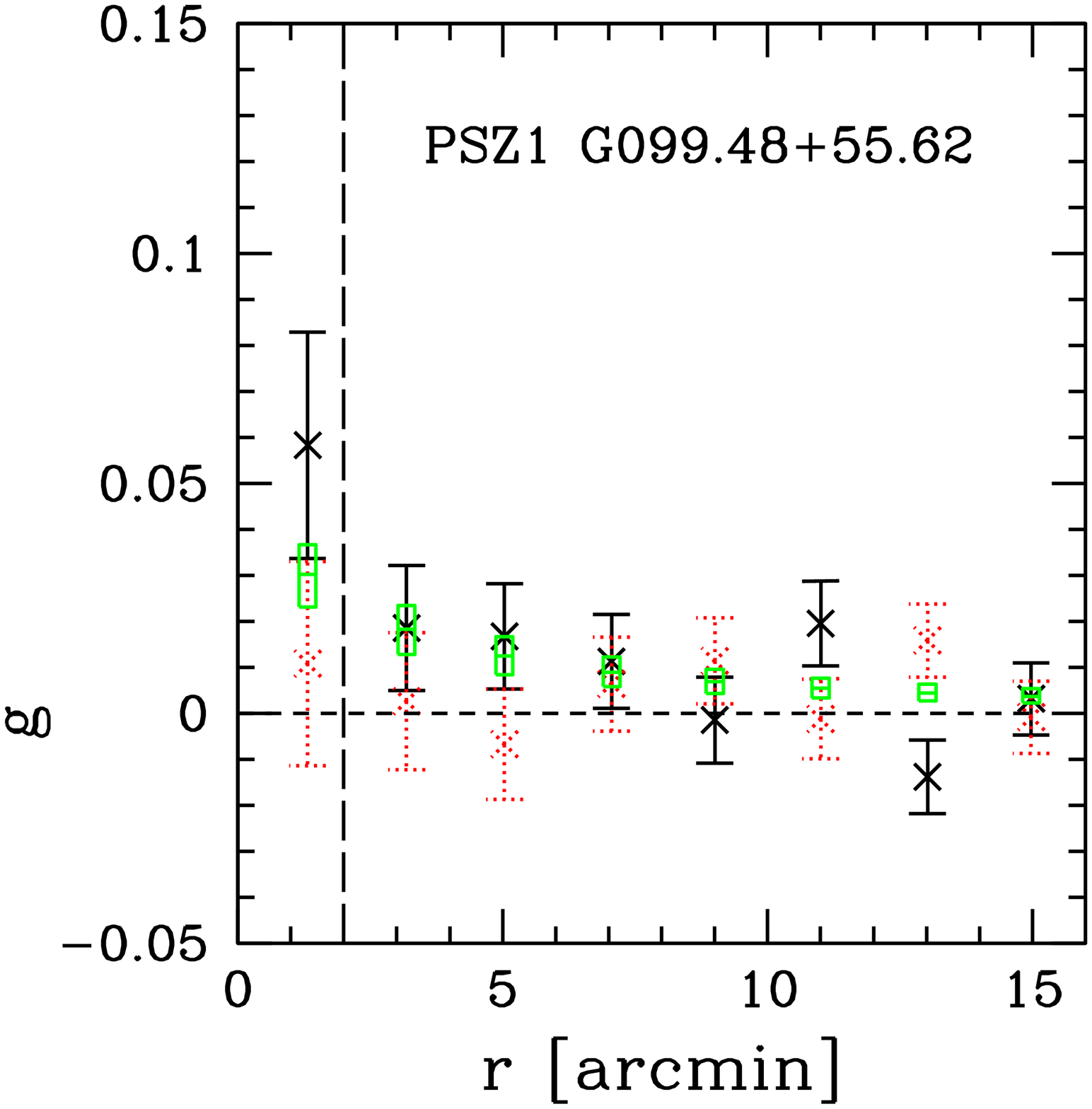}
\caption{Colour image of $8\times8$~arcmin$^2$ central region from \emph{gri} frames (top panel), three-dimensional galaxy density smoothed with 2~arcmin radius Epanechnikov kernel (central panel) and shear profile (bottom panel) of PSZ1 G099.48+55.62. See Fig.~\ref{fig:picspt4} and Section~\ref{sec:spt4visual} for details.}
\label{fig:picpsz099b}
\end{figure}

The cluster image of PSZ1 G099.48+55.62 (cf. Fig.~\ref{fig:picpsz099b}) is dominated by a BCG with bi-modal core and visibly disturbed wings.

\subsubsection{Previous Work}

PSZ1 G099.48+55.62 was first discovered by \citet{1958ApJS....3..211A} and classified at the third-lowest of six richness classes.

\citet{1999ApJS..125...35S} quote a spectroscopic redshift of $z=0.1051$, listing as the source the catalog of \citet{1991AISAO..31...91L}, which is unavailable to the authors of this work.

The system is associated with the \emph{ROSAT} bright X-ray source 1RXS J142830.5+565146 \citep{1999A&A...349..389V}. We apply \emph{ROSAT} band conversion factor, Galactic H\textsc{i} column density correction \citep{2005A&A...440..775K} and $k$-correction from tables 2-4 of \citet{2004A&A...425..367B}. In this procedure, we assume an X-ray temperature $T_{\rm X}=5$~keV, as justified by the relative independence of the conversion factors on $T_{\rm X}$ and the large intrinsic scatter of the $T_{\rm X}-L_{\rm X}$ relation \citep{2009A&A...498..361P} that could be used for self-consistent calibration. The derived X-ray luminosity corresponds to a mass of $M_{500c}=(1.2\pm0.2)\times10^{14}h_{70}^{-1}\Msol$ according to the \citet{2002ApJ...567..716R} $L_{\rm X}-M$ relation, which was calibrated using measurements under the assumption of HSE.

From the catalogued SZ likelihood of \citet{2013arXiv1303.5089P} we calculate mass estimates using the X-ray $\theta_{500}$ as a fixed size and the self-consistency method as $M_{500c}=3.2^{+0.4}_{-0.4}\times10^{14}h_{70}^{-1}\Msol$ and $3.8^{+0.5}_{-0.5}\times10^{14}h_{70}^{-1}\Msol$, respectively, noting that this value is significantly higher than the X-ray mass estimate.

\subsubsection{Weak Lensing Analysis}

A single halo fit, fixing the centre at the BCG and marginalizing over concentration, yields $M_{200m}=(1.9^{+1.2}_{-0.9})\times 10^{14}h_{70}^{-1}\Msol$ ($M_{500c}=(1.0^{+0.5}_{-0.4})\times 10^{14}h_{70}^{-1}\Msol$). The observed shear profile and the confidence interval of the NFW fit are shown in Fig.~\ref{fig:picpsz099b} (bottom panel).

We do not find additional structures in the field of view that would warrant a combined fit. The most prominent feature is a very diffuse concentration of high redshift galaxies towards the south-east of the cluster.

The WL measurement is consistent with our X-ray mass estimate. It is, however, below the Planck SZ value with $3\sigma$ significance.

\subsubsection{Strong Lensing}

Consistent with the relatively low mass and redshift of the system, we find no evidence for strong lensing in PSZ1 G099.48+55.62.

\section{Combined Analysis}
\label{sec:combined}

In the following, will combinedly analyse our WL mass estimates and compare them to the X-ray and SZ measurements. Table~\ref{tbl:masses} gives an overview of the mass estimates used.

\begin{table*}
\begin{center}
\begin{tabular}{|l|r|r|r|r|r|r|r|r|}
\hline
cluster             & $M^{\rm WL,single}_{200m}$ & $M^{\rm WL,single}_{500c}$ & $M^{\rm WL,multi}_{200m}$ & $M^{\rm WL,multi}_{500c}$ & $M^{\rm X}_{500c}$ & $M^{\rm X}_{\mathrm{gas},500c}$ & $M^{\rm SPT}_{200m}$ & $M^{\mathrm{Planck}}_{500c}$ \\ \hline
SPT-CL J0509--5342  & $ 8.4^{+3.4}_{-2.9}$ & $ 4.7^{+1.9}_{-1.7}$ & $ 6.6^{+3.1}_{-2.6}$ & $ 3.8^{+1.7}_{-1.5}$ & $5.6\pm0.6^{\rm C}$  & $0.56\pm0.02$ & $7.3\pm1.5$  & - \\ \hline
SPT-CL J0551--5709  & $11.7^{+5.1}_{-4.2}$ & $ 6.6^{+2.6}_{-2.4}$ & $11.7^{+5.1}_{-4.2}$ & $ 6.6^{+2.6}_{-2.4}$ & $3.4\pm0.4^{\rm C}$  & $0.51\pm0.06$ & $6.9\pm1.5$  & -\\ \hline
SPT-CL J2332--5358  & $14.5^{+4.0}_{-3.5}$ & $ 7.5^{+1.8}_{-1.7}$ & $15.3^{+4.2}_{-3.6}$ & $ 7.9^{+1.8}_{-1.7}$ & $6.7\pm0.5^{\rm X}$  & $0.76\pm0.25$ & $12.1\pm1.4$ & -\\ \hline
SPT-CL J2355--5056  & $ 8.0^{+3.7}_{-3.1}$ & $ 4.1^{+1.7}_{-1.5}$ & $ 3.8^{+2.4}_{-2.1}$ & $ 2.1^{+1.4}_{-1.1}$ & $3.8\pm0.4^{\rm C}$  & $0.39\pm0.15$ & $7.6\pm1.0$  & -\\ \hline
PLCKESZ G287.0+32.9 & $37.7^{+9.5}_{-7.6}$ & $19.5^{+3.3}_{-3.2}$ & $35.4^{+8.9}_{-7.1}$ & $18.7^{+3.2}_{-3.1}$ & $15.7\pm0.3^{\rm X}$ & $2.39\pm0.30$ & - & $17.7^{+0.8}_{-0.9}$ \\ \hline
PLCKESZ G292.5+22.0 & $ 6.8^{+3.8}_{-3.0}$ & $ 3.4^{+1.7}_{-1.4}$ & $ 5.0^{+3.1}_{-2.5}$ & $ 2.7^{+1.5}_{-1.3}$ & $9.3\pm0.6^{\rm X}$  & $1.17\pm0.04$ & - & $10.3^{+0.9}_{-1.0}$ \\ \hline
MACS J0416.1--2403  & $10.0^{+2.9}_{-2.6}$ & $ 5.9^{+1.6}_{-1.5}$ & $10.4^{+3.0}_{-2.6}$ & $ 6.1^{+1.7}_{-1.5}$ & $9.1\pm2.0^{\rm C}$  & $1.05\pm0.23$ & - & - \\ \hline
SPT-CL J2248--4431  & $31.1^{+6.3}_{-5.2}$ & $13.8^{+2.8}_{-2.4}$ & $24.7^{+6.3}_{-6.0}$ & $ 9.9^{+2.6}_{-2.4}$ & $12.6\pm0.2^{\rm X}$ & $1.89\pm0.02$ & $29.0\pm3.7$ & $14.5^{+0.7}_{-0.7}$ \\ \hline
PSZ1 G168.02--59.95 & $ 2.3^{+1.4}_{-1.1}$ & $ 1.2^{+0.7}_{-0.6}$ & $ 2.5^{+1.5}_{-1.2}$ & $ 1.3^{+0.7}_{-0.6}$ & $2.5\pm0.2^{\rm R}$  & - & - & $4.6^{+0.6}_{-0.8}$ \\ \hline
PSZ1 G230.73+27.70  & $10.0^{+3.9}_{-3.2}$ & $ 4.9^{+1.6}_{-1.5}$ & $ 6.8^{+3.2}_{-2.6}$ & $ 3.5^{+1.5}_{-1.3}$ & $3.6\pm0.8^{\rm R}$  & - & - & $6.0^{+0.9}_{-1.0}$ \\ \hline
PSZ1 G099.84+58.45  &$34.4^{+11.6}_{-10.2}$& $18.1^{+5.8}_{-5.3}$ &$38.3^{+13.5}_{-11.7}$& $19.6^{+6.4}_{-5.8}$ & $12.5\pm2.0^{\rm R}$ & - & - & $8.6^{+1.1}_{-1.2}$ \\ \hline
PSZ1 G099.48+55.62  & $ 1.9^{+1.2}_{-0.9}$ & $ 1.0^{+0.5}_{-0.4}$ & $ 1.9^{+1.2}_{-0.9}$ & $ 1.0^{+0.5}_{-0.4}$ & $1.2\pm0.2^{\rm R}$  & - & - & $3.2^{+0.4}_{-0.4}$ \\ \hline
\end{tabular}
\end{center}
\caption{Summary of mass estimates for the WISCy sample. The columns give weak lensing (WL) results for single halo and multi halo reconstructions (cf. Sec.~\ref{sec:central}) in two definitions of spherical overdensity in units of $10^{14}h_{70}^{-1}\Msol$. For Planck SZ, we use the estimates based on the X-ray size $R_X$ (cf. Section~\ref{sec:planckmor}). Details on the X-ray and SZ measurements can be found in the respective part of Section~\ref{sec:individual}. Superscripts C (\textit{Chandra}), X (\textit{XMM-Newton}) and R (\textit{ROSAT}) indicate the instrument used for the X-ray mass and gas mass estimates. Values for SPT-CL J2248--4431 are taken from \citet{2013MNRAS.tmp.1221G} (WL, as in all other cases we use the mass estimate marginalized over the concentration parameter) and \citet{2011AuA...536A..11P} (X-ray).}
\label{tbl:masses}
\end{table*}

\subsection{Comparison of X-ray and weak lensing mass estimates}
\label{sec:xcomp}

\begin{figure*}
\centering
\subfigure[]{
\includegraphics[width=0.48\textwidth]{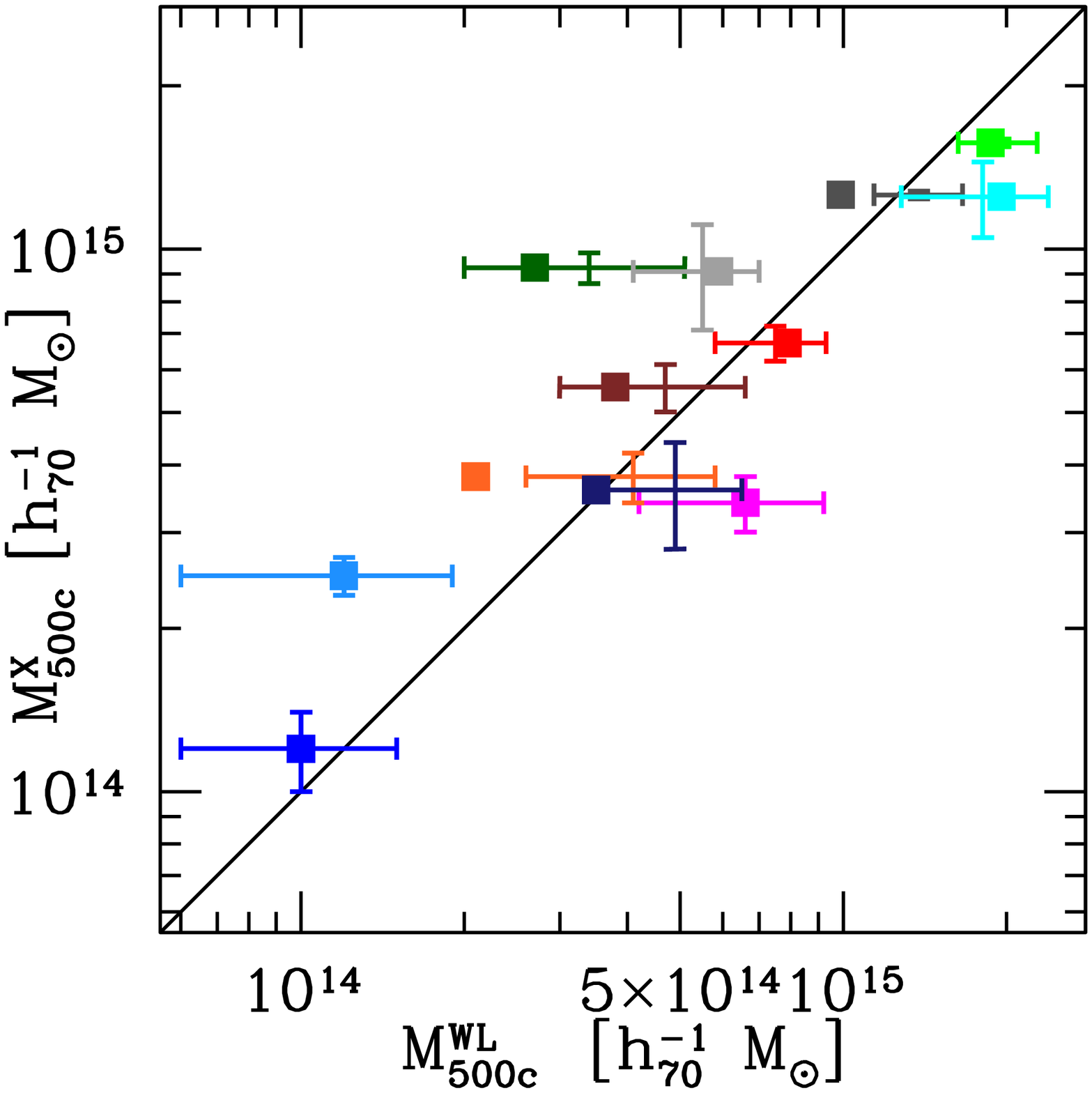}
}
\subfigure[]{
\includegraphics[width=0.48\textwidth]{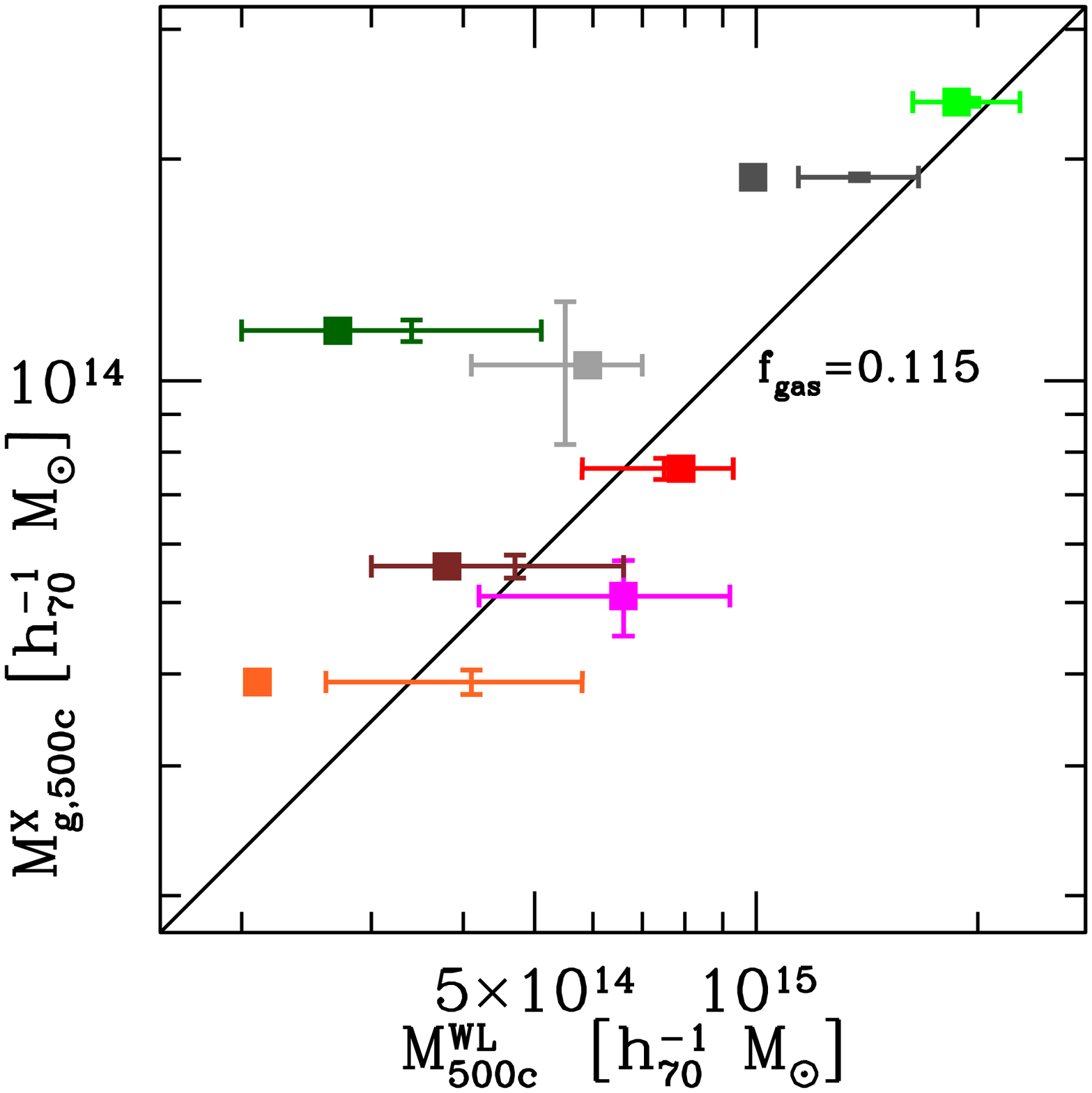}
}
\caption{Comparison of X-ray mass estimates determined under the assumption of hydrostatic equilibrium (a) and X-ray gas masses (b) \textbf{with our} weak lensing result (see individual cluster analysis for details). Error bars indicate best-fitting value and confidence regions of single-halo fit. The solid symbols are best-fitting masses of the central halo in a combined fit to multiple structures in the field of view, to be interpreted with errors of similar size. Colour coding for the individual clusters is as in Fig.~\ref{fig:szcomp}. Our measurements are consistent with no mean bias and no mass dependent bias of the HSE mass and a gas fraction $f_{\mathrm{gas},500c}=0.115$.}
\label{fig:xcomp}
\end{figure*}

We briefly compare the X-ray gas and total mass estimates for the WISCy sample with our WL analysis. Our full sample has the disadvantage that the X-ray mass estimation is far from homogeneous. However, the interesting question of a mean hydrostatic mass bias can still be addressed, since we have taken care to only use X-ray mass estimates calibrated under the assumption of HSE. For the subsample of eight clusters where X-ray gas mass estimates exist, we can make a more homogeneous analysis.

Fig.~\ref{fig:xcomp} shows a comparison of the two mass estimates. There is apparent good agreement between WL masses and X-ray based estimates. The most notable exception is the case of PLCKESZ G292.5+22.0, likely related to its complex structure (cf. Sec.~\ref{sec:p292}).

We are interested in determining the intrinsic scatter, normalization and slope between X-ray mass estimate $M^{\rm X}$ and WL measurement $M^{\rm WL}$. To this end, we assume a relation of the form
\begin{equation}
M_{500c}^{\rm WL}=10^{A} \times \left(\frac{M_{500c}^{\rm X}}{6\times10^{14}h_{70}^{-1}\Msol}\right)^B
\label{eqn:mxm}
\end{equation}
with lognormal intrinsic scatter $\sigma_{\mathrm{int},\log_{10}}$. The pivot mass is chosen such that errors in $A$ and $B$ are uncorrelated. Note that $A=0$ and $B=1$ is the case of agreement with hydrostatic calibration.

In order to determine confidence limits for the parameters, we calculate the likelihood
\begin{eqnarray}
-2\ln \mathcal{L}_i = \frac{(\log M^{\rm WL}_i - \log f(M^{\rm X}_i))^2}{\sigma_{i,\log M^{\rm WL}}^2+(B\sigma_{i, \log M^{\rm X}})^2+\sigma_{\mathrm{int},\log_{10}}^2} \nonumber \\ + \ln(\sigma_{i,\log M^{\rm WL}}^2+(B\sigma_{i, \log M^{\rm X}})^2+\sigma_{\mathrm{int},\log_{10}}^2) \; ,
\label{eqn:sigintlik}
\end{eqnarray}
where $f(M^{\rm X}_i)$ is the relation of equation (\ref{eqn:mxm}). In consideration of \citet{2011MNRAS.416.1392G}, we increase the WL uncertainties by 10\% for all following analyses in order to account for intrinsic variations due to cluster substructure, orientation and projected structures.

For the WL masses based on a single-halo analysis, we get projected single-parameter confidence intervals of $A=0.01^{+0.05}_{-0.05}$, $B=1.06^{+0.18}_{-0.18}$ and $\sigma_{\mathrm{int},\log_{10}}=0.00^{+0.07}_{-0.00}$. Using the estimates after subtraction of secondary haloes, we find $A=-0.03^{+0.06}_{-0.06}$, $B=1.08^{+0.19}_{-0.18}$ and $\sigma_{\mathrm{int},\log_{10}}=0.00^{+0.09}_{-0.00}$. This is consistent with no hydrostatic bias and regular scaling of X-ray observables with mass and marginally inconsistent with the assumption of a 20\% negative bias of hydrostatic masses (as in the calibration of \citealt{p2013cosmology}).

The relation between gas mass and WL estimate of total mass is consistent with a gas fraction $f_{\mathrm{gas},500c}=0.128^{+0.029}_{-0.023}$, in line with the $f_{\mathrm{gas},500c}=0.115$ of \citet{2008MNRAS.383..879A} and \citet{2010MNRAS.406.1773M}. For the multi-halo estimates, the fit is much more uncertain but still consistent.

\subsection{Comparison of SZ signal and mass}
\label{sec:morcal}

\begin{figure*}
\subfigure[]{
\includegraphics[width=0.48\textwidth]{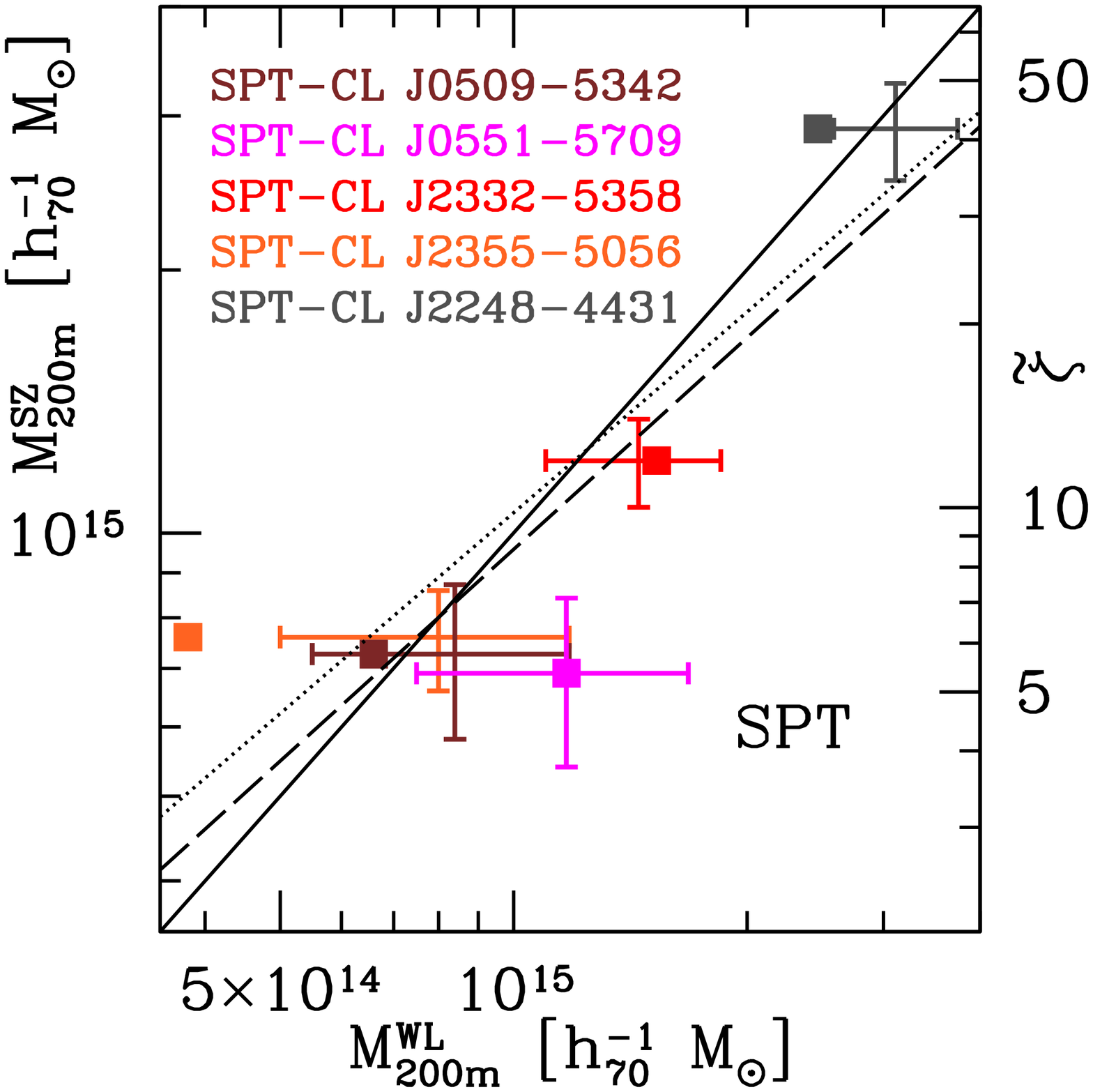}
}
\subfigure[]{
\includegraphics[width=0.48\textwidth]{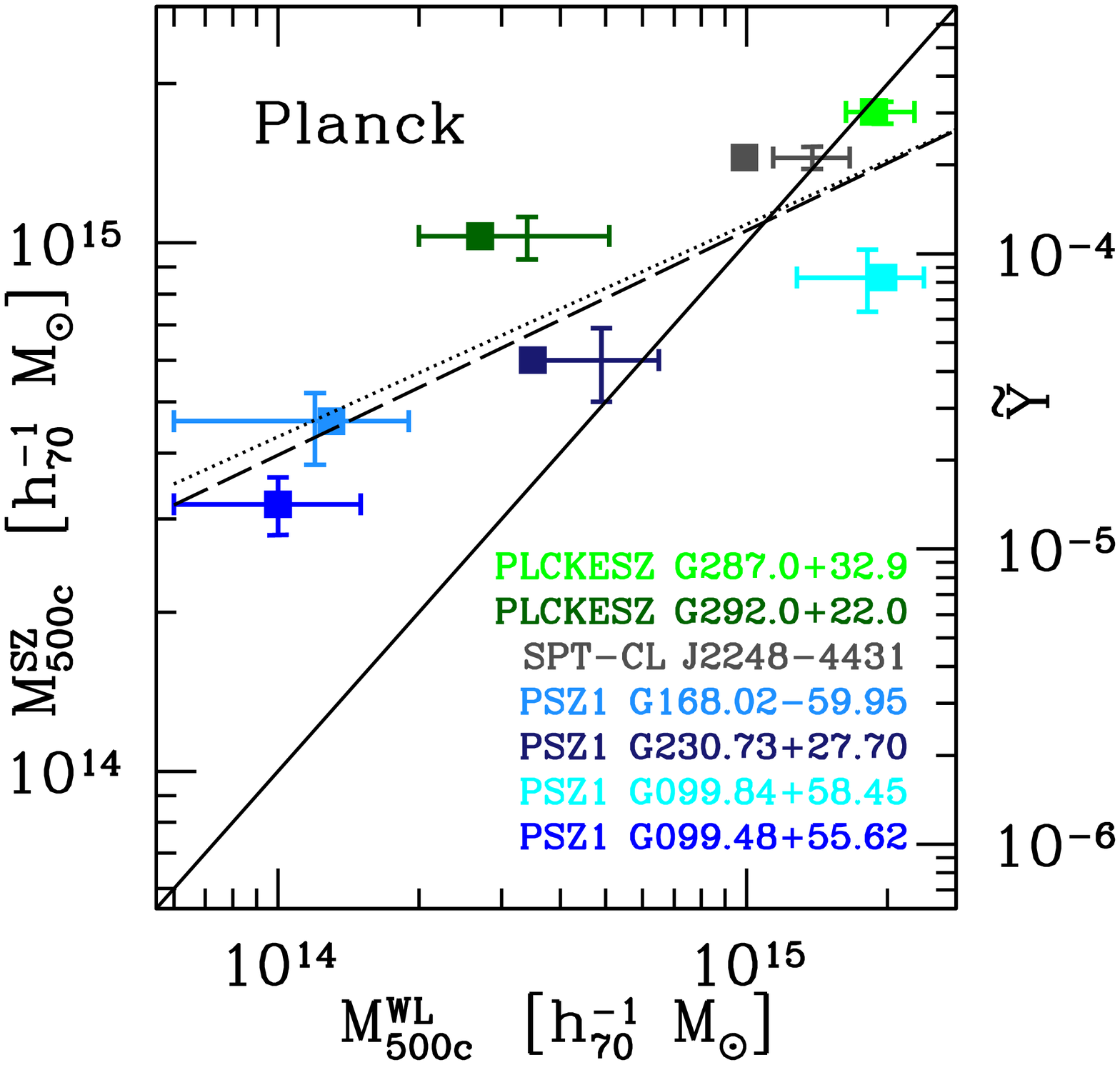}
}
\caption{Comparison of SZ and weak lensing (WL) results for the WISCy clusters. Panel (a) shows SPT, panel (b) \emph{Planck} clusters. The solid line indicates the mass-observable relations of the respective survey given by \citet{2010ApJ...722.1180V} ($A=5.62$, $B=1.43$) and \citet{p2013cosmology} ($A=-4.19$, $B=1.79$, $b=0.2$, cf. equations \ref{eqn:sptmor} and \ref{eqn:pszm}). WL single halo fits and statistical uncertainties are given by the horizontal and vertical error bars. The solid symbols indicate best-fitting WL result after subtracting neighbouring structures, to be interpreted with errors of approximately the same size. Masses are compared to SPT significance $\zeta$ and Planck $Y_{500c}^{R_X}$ inside a sphere of the $r_{500c}$ radius corresponding to the X-ray mass estimate, where we use shorthand notations $\tilde{\zeta}=\zeta[(1+z)/1.6]^{-C}$ and $\tilde{Y}=E(z)^{-2/3}D_A^2 Y_{500c}^{R_X} \rm{Mpc}^{-2}$. Dashed lines indicate the fit based on our single-halo WL analysis, and dotted lines the relation after subtracting the signal of neighbouring structures.}
\label{fig:szcomp}
\end{figure*}

In the following, we will compare our WL measurements of cluster mass to SZ observables, namely the SPT significance and the Planck Compton parameter inside the $\theta_{500c}$ of the X-ray mass measurement. This is meant both as a test of the earlier calibrations of the MORs by \citet{2010ApJ...722.1180V} and \citet{p2013cosmology} and an independent lensing-based determination of the MOR parameters.

In this, we need to account for uncertainties in both the WL mass measurement and the SZ signal, together with an intrinsic scatter. While some studies (e.g. \citealt{2002ApJ...567..716R,2013arXiv1308.4674F,p2013cosmology}) have used line fitting techniques such as the one described in \citet{1996ApJ...470..706A} in the past, this setting favours a Bayesian approach for unbiased estimation of the MOR parameters (cf. e.g. \citealt{Kelly2007}). We therefore determine the parameters from the likelihood, in analogy to e.g. \citet[][their equations 7-9]{2012MNRAS.427.1298H} and our equation (\ref{eqn:sigintlik}), with lognormal uncertainties in all parameters,
\begin{eqnarray}
-2\ln \mathcal{L}_i = \frac{(\log f(M^{\rm WL}_i,z_i) - \log f_i)^2}{(\frac{\mathrm{d} \log f}{\mathrm{d} \log M}\sigma_{i,\log M^{\rm WL}})^2+\sigma_{i, \log f_i}^2+\sigma_{\mathrm{int},\log_{10}}^2} \nonumber \\ + \ln\left(\left(\frac{\mathrm{d} \log f}{\mathrm{d} \log M}\sigma_{i,\log M^{\rm WL}}\right)^2+\sigma_{i, \log f_i}^2+\sigma_{\mathrm{int},\log_{10}}^2\right) \; ,
\label{eqn:morlikelihood}
\end{eqnarray}
where $f_i$ is the SZ observable of cluster $i$ with intrinsic scatter $\sigma_{\mathrm{int},\log_{10}}$, and $f(M,z)$ is the MOR.

Our key results are illustrated in Fig.~\ref{fig:szcomp}, which we discuss subsequently. As for the X-ray comparison (see previous section) we increase the WL uncertainties by 10\% for this procedure.

\subsubsection{South Pole Telescope}

For SPT, we assume the MOR of equation (\ref{eqn:sptmor}) \citep{2010ApJ...722.1180V}. A direct comparison of WL estimate and SZ mass for the best-fitting parameters $A=5.62$ and $B=1.43$ \citep{2010ApJ...722.1180V} shows that while there is consistency within the errors, all our systems are measured with WL to be at a higher mass than predicted with the SZ MOR, particularly at lower mass (cf. Fig.~\ref{fig:szcomp}a). This effect is removed when modelling surrounding structures with WL separately (solid squares), which lowers the mass estimate for two systems considerably. 

For the independent MOR determination (cf. equation \ref{eqn:sptmor}), we fix $C=1.4$ \citep{2010ApJ...722.1180V}, since our SPT sample has little leverage on redshift dependence, and determine the remaining parameters $A$, $B$ and the $\log_{10}$-normal intrinsic scatter $\sigma_{\mathrm{int},\log_{10}}$ from a combined likelihood.

We find parameters $A=5.8^{+1.8}_{-1.8}$ and $B=1.15^{+0.32}_{-0.22}$ from our single halo analysis (Fig.~\ref{fig:szcomp}a, dashed line). These values are in agreement with the calibration of \citet{2010ApJ...722.1180V} (solid line) at the 68\% confidence level and we detect a scaling of SPT significance with mass at $3\sigma$ significance in the combiend three-parametric likelihood. When we instead use the WL mass estimates made after subtraction of neighbouring structures, we find $A=6.8^{+2.6}_{-2.0}$ and $B=1.09^{+0.33}_{-0.47}$ (dotted line), also consistent with \citet{2010ApJ...722.1180V}. The shallower slope is mostly due to the unusually complex structure of SPT-CL J2355-5056, which results in a small mass of the central system in the multi-halo fit.

The intrinsic scatter is consistent with zero at $1\sigma$ confidence with best fits $\sigma_{\mathrm{int},\log_{10}}=0$ and upper limits $\sigma_{\mathrm{int},\log_{10}}<0.13$ (0.23) for the single-halo (the multiple halo) analysis, respectively. These values are consistent with previous results \citep{2009ApJ...701L.114M,2012ApJ...754..119M,2012MNRAS.427.1298H,2013ApJ...763..147B} and in agreement with predictions \citep[e.g.][]{2012MNRAS.422.1999K}.

An analogous analysis using the MOR definition of \citet{2013ApJ...763..127R} and \citet{2013ApJ...763..147B} yields $A=6.0^{+1.9}_{-1.8}$ and $B=1.25^{+0.36}_{-0.28}$ at $\sigma_{\mathrm{int},\log_{10}}<0.15$ for the single-halo and $A=7.6^{+3.0}_{-2.6}$, $B=1.02^{+0.62}_{-0.68}$ at $\sigma_{\mathrm{int},\log_{10}}=0.15^{+0.18}_{-0.15}$ for the multi-halo modelling of neighbouring structures. In both cases, we fix the slope of the redshift dependence $C=0.83$ \citep{2013ApJ...763..127R,2013ApJ...763..147B}. We find consistency with the MOR parameters of \citet{2013ApJ...763..127R} and \citet{2013ApJ...763..147B} within 68\% confidence.

\subsubsection{Planck}
\label{sec:planckfit}
We perform a similar analysis for the Planck $Y_{500c}$, fitting for parameters $A$ and $B$ in equation (\ref{eqn:pszm}) and the intrinsic scatter $\sigma_{\mathrm{int},\log_{10}}$. This returns $A+B\log_{10}(1-b)=-4.09^{+0.09}_{-0.08}$, $B=0.76^{+0.20}_{-0.20}$ and $\sigma_{\mathrm{int},\log_{10}}=0.14^{+0.12}_{-0.14}$.

The amplitude $A$ is consistent with the value from \citet{p2013cosmology} when assuming no hydrostatic mass bias. Their baseline hydrostatic bias of $b=0.2$ (corresponding to $A+B\log(1-b)=-4.36\pm0.02$ in \citealt{p2013cosmology}) is disfavoured from our data at $\approx1.2\sigma$ significance in three-dimensional parameter space. Their combination of $A$, $B$, $b$, and $\sigma_{\mathrm{int},\log_{10}}$ is excluded at $3\sigma$ significance ($p<0.0028$). This is due mostly to the shallower slope $B$ in our measurement, with their value $B=1.79$ excluded at $2.5\sigma$ significance in the three-dimensional parameter space. The discrepancy cannot be relieved by allowing a higher intrinsic scatter - even with the optimal $\sigma_{\mathrm{int},\log_{10}}=0.38$ for the Planck $A$, $b$ and $B$ calibration, the latter is still excluded with high significance ($p<0.004$) relative to our best fit. The maximum likelihood point with self-similar slope $B=5/3$ (cf. e.g. \citealt{2008ApJ...675..106B}) is similarly unlikely in $A$-$B$-$\sigma_{\mathrm{int},\log_{10}}$ space ($p<0.01$ for the best fit under the $B=5/3$ constraint).

Our multi-halo analysis yields a consistent $A+B\log_{10}(1-b)=-4.06^{+0.10}_{-0.10}$, $B=0.72^{+0.20}_{-0.19}$ and $\sigma_{\mathrm{int},\log_{10}}=0.19^{+0.11}_{-0.12}$. It is inconsistent with the \citet{p2013cosmology} calibration with similar significance.

The cautionary remark has to be made that our disagreement with \citet{p2013cosmology} is based on a relatively small sample. The majority of the objects are at a comparatively low Planck SZ S/N $\leq 7$, where the simple treatment of Malmquist bias likely does not capture selection biases completely. In addition, even one of the systems with higher significance, PLCKESZ G292.5+22.0 (see Section~\ref{sec:p292}), is likely to be influenced strongly by neighbouring structures. It would therefore be of great benefit to increase the sample size, either with pointed lensing observations or a large area survey.

\subsection{Hypothesis tests}

For testing the dependence of deviations from the MOR on several cluster properties, we use Spearman rank correlation coefficients, which have been applied previously in comparable cluster studies \citep{2004ApJ...610..745L,2012A&A...546A.106F}. After determining the unique ranking of a set of $N$ entities by two properties $A$ and $B$, the Spearman rank correlation coefficient $\rho$ is defined as
\begin{equation}
 \rho=1-\frac{6\sum_{i=1}^N(r^A_{i}-r^B_i)^2}{N(N^2-1)} \; ,
\end{equation}
where $r^A_i=1,\ldots,n$ and $r^B_i=1,\ldots,N$ are the ranks of item $i$ in the respective property. This definition ensures that the mean $\rho$ for unrelated rankings is 0, while $\rho=1$ ($\rho=-1$) corresponds to properties A and B which are a monotonically increasing (decreasing) function of each other. For the small sample size used in our work, the probability of random rankings to exceed some $|\rho|$ can be easily calculated by means of calculating all permutations. In the following, we will always quote the probability of the null hypothesis $p$.

The Spearman rank coefficient between the $M^{\rm SZ}/M^{\rm WL}$ ratio and redshift for the five Planck clusters is $\rho=-0.93$,  formally a 99\% confidence ($p<0.01$) for rejecting the null hypothesis (for both self-consistent and X-ray size prior SZ mass estimates). Indeed, the data suggest that for high redshift systems, our mass measurement with WL exceeds the expectations from the Planck SZ signal (and vice versa). 
We note that this is similar to testing for a connection between $M^{\rm SZ}/M^{\rm WL}$ and angular size (for which we also find $p<0.01$), since $\theta_{500c}$ for our sample is ranked almost inversely as redshift (cf. Table~\ref{tbl:psz}). From a comparison of WL and X-ray mass estimates (cf. Fig~\ref{fig:xcomp}), it appears that the highest redshift system PSZ1 G099.84+58.45 and the second-to-lowest redshift system PSZ1 G168.02--59.95 are outliers in the same direction, although not as strongly. This indicates that our particular sample might overestimate the effect. Yet also a comparison of $M^{\rm SZ}/M^{\rm X}$ rank with redshift yields evidence for an interdependence ($p<0.01$). Interestingly, the likelihood analysis of \citet{p2013cosmology} also shows a redshift dependent tilt (cf. their fig.~7), where at redshifts $z>0.5$ the majority of redshift-binned counts are below the predictions of any of the models, including their own best-fit to the SZ data.

For the five SPT clusters, no significant rank correlation of mass ratio with redshift is less significant ($\rho=0.7$), consistent with no dependence at $p>0.10$. We note that for this sample, the comparatively low dynamic range in redshift and mass make the ordering very noisy.

We combine mass ratios from \emph{Planck} and SPT clusters into one by taking the arithmetic mean in the case of SPT-CL J2248--4431. Determining several Spearman rank coefficients, we find the following.
\begin{itemize}
 \item Mass ratios are related to the ellipticity of the brightest cluster galaxy (BCG) with $\rho=0.65$ (higher $M^{\rm SZ}/M^{\rm WL}$ with higher BCG ellipticity with confidence $p<0.05$). This is in line with the observation of \citet{2012ApJ...754..119M}, although we do not make a distinction here between relaxed and unrelaxed systems.
 \item Mass ratios are not found to be significantly related to the magnitude gap between the brightest and second-to-brightest cluster galaxy ($p>0.68$). While the magnitude gap is a tracer of assembly history and has significant predictive power for richness at fixed mass \citep{2013MNRAS.430.1238H}, we do not find evidence for any impact on SZ measurements.
\end{itemize}

\subsection{Planck redshift dependence}

\label{sec:planckzdep}

As described in the previous sections, our data shows evidence for a redshift-dependent systematic offset of Planck Compton parameters from their values expected according to the WL measurement and the MOR (equation \ref{eqn:pszm}). In this section, we attempt to derive a modified MOR that includes a redshift dependent term.

We assume a MOR of the form
\begin{eqnarray}
 E(z)^{-2/3}\times\left(\frac{D_A^2\times Y_{500c}}{\mathrm{Mpc}^2}\right)= \nonumber \\10^A\times\left(\frac{M_{500c}}{6\times10^{14}h_{70}^{-1}\Msol}\right)^B\times\left(\frac{1+z}{1.31}\right)^C \; .
\label{eqn:pszmc}
\end{eqnarray}
The normalization of the redshift term has been chosen such as to null the effect for the medium redshift clusters, for which our analysis above showed good agreement. 

The projected 68\% confidence interval in each of the parameters according to the likelihood of equation \ref{eqn:morlikelihood} is $A=-4.08^{+0.06}_{-0.08}$, $B=0.92^{+0.20}_{-0.18}$ with a $1\sigma$ indication of a redshift dependence of $C=-2.1^{+1.7}_{-2.0}$. The same analysis with the multi-halo based estimates of mass yields a result that is consistent within the errors and in terms of its interpretation. Here we find 68\% confidence intervals $A=-4.05^{+0.08}_{-0.08}$, $B=0.93^{+0.21}_{-0.31}$, $C=-2.4^{+2.4}_{-2.0}$. In both cases, the combined (four-parametric) 68\% confidence region allows $C=0$.

If we assume a fixed slope $B=1.79$ and intrinsic scatter $\sigma_{\mathrm{int},\log_{10}}=0.07$ from \citet{p2013cosmology} and determine only the normalization $A$ and the redshift slope $C$, we get $A=-4.23^{+0.13}_{-0.12}$, in agreement with the \citet{p2013cosmology} parameters and a $b=0.2$ hydrostatic bias, and a larger $C=-6.9^{+2.7}_{-2.7}$. The two parameter combined 68\% confidence region still includes a wide range of values $C=-6.9^{+4.1}_{-4.1}$. For the multi-halo mass estimates, we find $A=4.15^{+0.12}_{-0.13}$ and $C=-7.4^{+2.6}_{-2.8}$ (individual parameters) and $C=-7.4^{+4.0}_{-4.2}$ (combined).

Attempts to constrain $C$ are made difficult by the fact that mass and redshift, and therefore $B$ and $C$, are highly degenerate in the WISCy sample. The combined 68\% confidence region of the four parameters allows for a redshift independent MOR with $C=0$ at the expense of an even shallower mass slope $B$. In Fig.~\ref{fig:bc} we show the likelihood, based on single-halo mass estimates, in $B$-$C$ space when fixing $A$ and $\sigma_{\mathrm{int},\log_{10}}$ to their maximum likelihood values (cf. Section~\ref{sec:planckfit}).

Finally, in the four-parametric likelihood of $A$, $B$, $C$ and $\sigma_{\mathrm{int},\log_{10}}$, a self-similar slope $B=5/3$ is only marginally excluded compared to the best fit, ($p<0.13$) at the expense of a non-zero $C$. 

We note that in the overlap of the sample of \citet{2012MNRAS.427.1298H} with the Planck SZ catalogue, all except one system are at relatively low redshifts around $z\approx0.2$. The only higher redshift system is MACS J0717.5+3745, a strong lens \citep{2013arXiv1304.1223M} at $z=0.548$. For this cluster, the WL mass measurement by \citet{2012MNRAS.427.1298H} is $2.5\ldots3\sigma$ above either of their fitted MORs, adding evidence to an unaccounted for redshift dependence of the MOR.

We conclude by noting that the degeneracy between mass and redshift slope prohibits a conclusion on the origin of the MOR discrepancy from our WL measurement of masses alone. 

\begin{figure}
\centering
\includegraphics[width=0.48\textwidth]{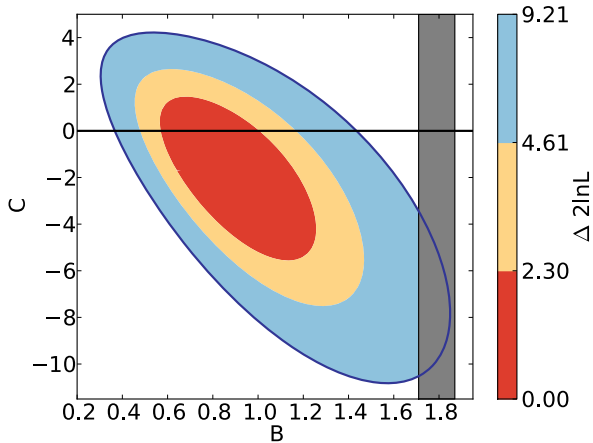}
\caption{Likelihood of mass slope $B$ versus redshift slope $C$ when fixing amplitude and intrinsic scatter of Planck SZ MOR to their best-fitting values (cf. Section~\ref{sec:planckfit}). The \citet{p2013cosmology} value of $B=1.79\pm0.08$ (shaded grey region) is excluded with high significance unless in the case of a large redshift dependence. The physically expected $C=0$ (black line) requires that the mass slope is significantly shallower.}
\label{fig:bc}
\end{figure}

\subsubsection{Comparison with other SZ surveys}
\label{sec:szcomp}
A problem with signal extraction or calibration of the Planck SZ would likely lead to differences with other SZ observations of the same clusters, and if the hypothesis of a redshift dependence should hold these differences should in turn be a function of redshift.

\citet{2013arXiv1301.0816H} compare SZ mass estimates of 11 clusters detected and published by both ACT and SPT with good agreement (cf. their fig.~21). \citet{2013A&A...550A.128P} have compared Planck and Arcminute Microkelvin Imager (AMI) measurements of 11 clusters, finding significant differences between the two even when assuming the same fixed pressure profile (PP) or a profile fitted to the individual objects based on X-ray observations. Neither of them make a redshift distinction in these analyses, however.

Following the recent releases of Planck (The \citealt{2013arXiv1303.5089P}), ACT \citep{2013arXiv1301.0816H} and SPT \citep{2011ApJ...738..139W,2013ApJ...763..127R} catalogues, we match objects detected both in Planck and either of the other catalogues. Rejecting systems with redshift differences $\Delta z>0.03$ between the two respective catalogues and ones not successfully extracted with the Planck MMF3 algorithm, we find 13 matches with \citet{2013arXiv1301.0816H}, 11 with \citet{2013ApJ...763..127R} and 19 with \citet{2011ApJ...738..139W} (3 of which are also in \citealt{2013ApJ...763..127R}).

Fig.~\ref{fig:szmatch} shows the comparison of SZ based mass estimates for these 43 detections. For SPT and ACT, we use the published values of $M_{500c}$, while for Planck we apply the self-consistency technique (see Section~\ref{sec:planckmor}) to estimate the mass. The deviation from mean agreement is significant with $\langle\log_{10} M^{\rm Planck}/M^{\rm ACT/SPT}\rangle=0.10\pm0.02$. The binned geometric mean of the mass ratios is consistent with a 20\% excess in the Planck estimates (corresponding to the hydrostatic bias factor applied there) with no significant indication for redshift dependence. For ACT, we have used the $M_{500c}^{\rm UPP}$ values based on the \citet{2010AuA...517A..92A} PP. Use of the alternative simulation and X-ray based \citet{2012arXiv1204.1762B} MOR increases the ACT masses significantly, yet with no strong trend in redshift.

\begin{figure}
\centering
\includegraphics[width=0.48\textwidth]{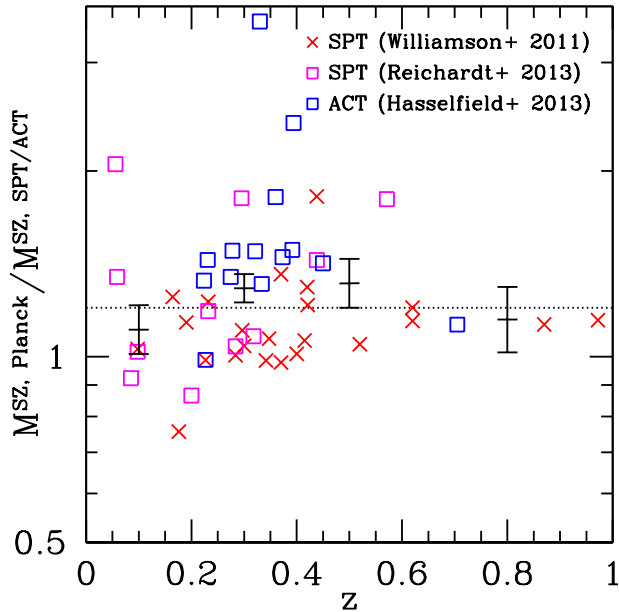}
\caption{Comparison of Planck self-consistent mass estimates with ACT and SPT measurements \citep{2011ApJ...738..139W,2013arXiv1301.0816H,2013ApJ...763..127R}. The dotted line indicates a 20\% excess, the black error bars indicate log-mean values in redshift bins.}
\label{fig:szmatch}
\end{figure}

While this comparison therefore yields no evidence for a redshift dependence in the Planck catalogues that would support our WL based findings, the number of sometimes strong outliers indicates that the statistical uncertainties according to our interpretation of the Planck likelihoods might not be appropriate. A significantly larger uncertainty due to assumptions made during signal extraction, for instance, could reconcile the two strong outliers in our sample (PSZ1 G168.02--59.95 and PSZ1 G099.84+58.45) with the WL estimate and alleviate the tension with the assumption of redshift independence of the MOR.

\subsection{Centring and shear}
\label{sec:centralshear}

We briefly test the appropriateness of the NFW profile and our background selection scheme by comparing the tangential reduced shear measured in the innermost 2~arcmin around the cluster centre to the model prediction. The latter is derived from the confidence region in mass according to our shear profile fit (that only uses galaxies outside 2~arcmin radius) and assuming the \citet{2008MNRAS.390L..64D} concentration-mass relation.

Deviations between model and data could be explained, for instance, by incorrect treatment of the abundance of cluster members near the centre, offsets of the assumed from the true centre (for instance due to multimodality of the density profile), a deviation of the central mass profile slope from the NFW prediction or second-order shear bias (cf. Young et al., in preparation).

Fig.~\ref{fig:central} shows the comparison of the model and measurement, which are consistent within the errors. The ratio of mean shears between model and data is $1.02\pm0.10$.

\begin{figure}
\centering
\includegraphics[width=0.48\textwidth]{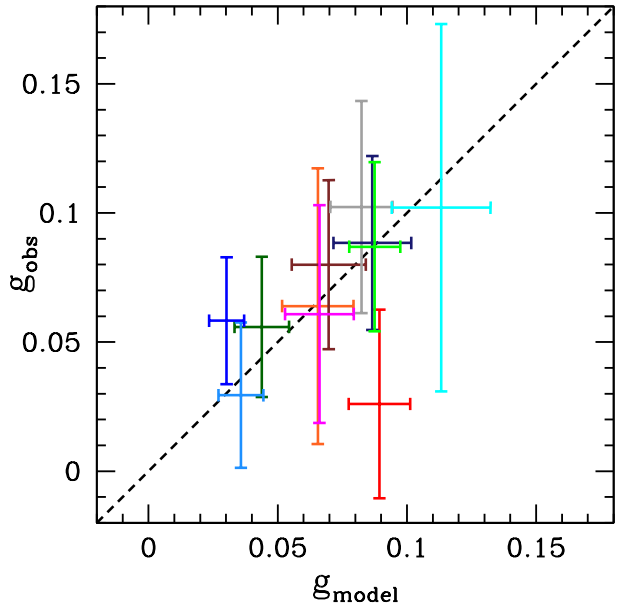}
\caption{Central ($r\leq2$~arcmin) reduced tangential shears as measured plotted against values according to NFW model fitted outside that region. The agreement between the two is consistent with the claim that the central slope of density profiles is as predicted by the model and our background selection is correct.}
\label{fig:central}
\end{figure}

\section{Conclusions}
\label{sec:conclusions}

We have performed a weak lensing (WL) analysis of twelve clusters of galaxies selected by their Sunyaev-Zel'dovich (SZ) effect. 

Among the methods for cluster WL measurements developed and used in the process are a magnitude-based background galaxy selection procedure, along with an account for contamination with cluster members and an optimised background cut (Section~\ref{sec:beta}).

Several of the systems in our sample are very interesting in their own right, among them the \emph{HST} Frontier Field clusters RXC J2248.7--4431 and MACS J0416.1--2403, but also SPT-CL J2355--5056 with its complex structures of neighbouring haloes and PLCKESZ G287.0+32.9 with a very large strong lensing cross-section as based on a large number of candidate strong lensing arcs.

The WL analysis of our cluster sample is consistent with zero mean bias in hydrostatic X-ray masses and no dependence of bias on mass. The gas fraction of our clusters is $f_{\mathrm{gas},500c}=0.128^{+0.029}_{-0.023}$, in line with previous results and expectations.

Our main scientific goal, however, was the comparison of WL measurements with published SZ significances and Compton parameters according to mass-observable relations (MORs) for the SPT and \emph{Planck} surveys.  

We have found agreement with the SPT calibration, calibrated both from simulations \citep{2010ApJ...722.1180V,2013ApJ...763..127R} and a combined cosmology fit to X-ray observations and CMB \citep{2013ApJ...763..147B}, both with and without simultaneous modelling of surrounding projected structures.

In our combined sample, we have confirmed a trend of excess SZ mass estimate with respect to the WL measurement increasing with the projected ellipticity of the BCG, a proxy for projected halo elongation, as noticed before by \citet{2012ApJ...754..119M}.

Finally, our analysis of five Planck SZ-detected clusters shows significant disagreement with the \citet{p2013cosmology} SZ mass-observable relation as calibrated from X-ray observations. Multiple explanations appear possible, among them are as follows.
\begin{itemize}
\item A shallower mass slope of the MOR -- we find a projected 68\% confidence interval of $B=0.76^{+0.20}_{-0.20}$, significantly shallower than the \citet{p2013cosmology} value of $B=1.79$, which our data disfavours w.r.t. the best fit at $2.5\sigma$ significance in the three-dimensional parameter space. It appears difficult to find a physical reason for such a strong deviation from the self-similar $B=5/3$. This is the case even more so as WL analyses of other SZ samples (including the WISCy SPT sample, \citealt{2012ApJ...754..119M} and \citealt{2012MNRAS.427.1298H}) or X-ray samples (including WISCy) are compatible with the self-similar slope.
\item An unaccounted for redshift dependence of the Compton parameter measurement -- this could either be unphysical and related to the low resolution of the Planck beam compared to the angular size of moderate to high redshift clusters, or physical and related to a deviation of the pressure profiles of high redshift clusters from the locally determined \citet{2010AuA...517A..92A} profile. Speaking against the first hypothesis is the fact that \emph{Planck} measurements and SZ observations of higher resolution agree without a strong trend in redshift (cf. Sec.~\ref{sec:szcomp}).
\item Noise bias -- the simplified treatment of Malmquist bias (cf. equation \ref{eqn:malm}) does not cover the full effect of biased object selection near the noise limit of the Planck SZ map. What speaks against this explanation is the fact that, in our sample, objects with relatively low SZ S/N ratio (all PSZ1 clusters) are found to be outliers in different directions.
\item Sample variance -- while these findings are of high formal significance, even given the small sample, one must be careful not to over-interpret the data. For instance, effects of blending with substructure (in the case of PLCKESZ G292.5+22.0) and halo orientation (in the cases of PSZ1 G099.48+55.62, PSZ1 G168.02--59.95 and PSZ1 G099.84+58.45, with large (small) BCG ellipticity at the lowest (highest) redshifts in our sample, respectively) could contribute to the deviations we observe. Blending might be a particularly severe problem in Planck due to the comparatively large beam size and is investigated in more detail in Kosyra et al. (in preparation). It would cause an effect different from intrinsic scatter, since blending can only \emph{add} to the observed signal at fixed mass of the central halo.
\end{itemize}
The first two points especially are intrinsically hard to disentangle, since there is a strong correlation of mass and redshift in our sample and the Planck SZ catalogue in general. The most likely interpretation of our data, in light of Section~\ref{sec:planckzdep}, may be a moderate size/redshift dependent bias that would marginally reconcile our measurements with the self-similar $B$. Additional lensing measurements of SZ clusters, particularly at high redshift, would be of great value for testing the underlying assumptions on the pressure profile and the mass-observable relations themselves.

\section*{Acknowledgments}

The authors thank Bradford Benson, Lindsey Bleem, Gael Fo{\"e}x, Eduardo Rozo, Jochen Weller and Julia Young for helpful discussions.

This work was supported by SFB-Transregio 33 'The Dark Universe' by the Deutsche Forschungsgemeinschaft (DFG) and the DFG cluster of excellence 'Origin and Structure of the Universe'.

For the ESO Deep Public Survey data used in this work, observations have been carried out using the MPG/ESO 2.2m Telescope and the ESO New Technology Telescope (NTT) at the La Silla observatory under Program-ID No. 164.O-0561.

This work is based in part on observations obtained with MegaPrime/MegaCam, a joint project of CFHT and CEA/IRFU, at the Canada-France-Hawaii Telescope (CFHT) which is operated by the National Research Council (NRC) of Canada, the Institut National des Sciences de l'Univers of the Centre National de la Recherche Scientifique (CNRS) of France, and the University of Hawaii. This research used the facilities of the Canadian Astronomy Data Centre operated by the National Research Council of Canada with the support of the Canadian Space Agency. CFHTLenS data processing was made possible thanks to significant computing support from the NSERC Research Tools and Instruments grant program.

Part of this work has used spectroscopic data from SDSS-III. Funding for SDSS-III has been provided by the Alfred P. Sloan Foundation, the Participating Institutions, the National Science Foundation, and the U.S. Department of Energy Office of Science. The SDSS-III website is http://www.sdss3.org/.

SDSS-III is managed by the Astrophysical Research Consortium for the Participating Institutions of the SDSS-III Collaboration including the University of Arizona, the Brazilian Participation Group, Brookhaven National Laboratory, Carnegie Mellon University, University of Florida, the French Participation Group, the German Participation Group, Harvard University, the Instituto de Astrofisica de Canarias, the Michigan State/Notre Dame/JINA Participation Group, Johns Hopkins University, Lawrence Berkeley National Laboratory, Max Planck Institute for Astrophysics, Max Planck Institute for Extraterrestrial Physics, New Mexico State University, New York University, Ohio State University, Pennsylvania State University, University of Portsmouth, Princeton University, the Spanish Participation Group, University of Tokyo, University of Utah, Vanderbilt University, University of Virginia, University of Washington, and Yale University. 

This research has made use of the SIMBAD data base, operated at CDS, Strasbourg, France.

\section*{Appendix A: Secondary components}

The following table lists secondary clusters and groups in the cluster fields that were visually identified from colour images and redshift-sliced density maps or are listed in the SIMBAD database.

\begin{table*}
\begin{center}
\begin{tabular}{|l|r|r|r|r|r|r|}
\hline
ID & R.A.     & dec & r [arcmin] & z & $M_{200m} [10^{14} h_{70}^{-1} \Msol]$ & external reference \\ \hline \hline
SPT-CL J0509--5342 \\ \hline
A & 05:09:11 & -53:39:29 & 3.2 & 0.45$^{b}$ & $2.4^{+2.6}_{-2.0}$ & -$^{1}$ \\ \hline
B & 05:07:50 & -53:48:30 & 15.4 & 0.41$^{b}$ & $0^{+2.3}_{-0.0}$ & - \\ \hline \hline
SPT-CL J0551--5709 \\ \hline
A & 05:51:33 & -57:14:21 & 5.7 & 0.09$^{c}$ & $0^{+0.8}_{-0.0}$ & ACO S 552 \citep{1989ApJS...70....1A} \\ \hline
B & 05:51:52 & -57:18:15 & 9.8 & 0.59$^{b}$ & $0^{+3.2}_{-0.0}$ & - \\ \hline \hline
SPT-CL J2332--5358 \\ \hline
\multirow{2}{*}{A} & \multirow{2}{*}{23:32:36} & \multirow{2}{*}{-54:01:59.5} & \multirow{2}{*}{3.7} & \multirow{2}{*}{0.33$^{a}$} & \multirow{2}{*}{$0^{+1.0}_{-0}$} & SCSO J233231.4--540135.8 \\
& & & & & & \citep{2009ApJ...698.1221M} \\ \hline
B & 23:33:32 & -54:00:53 & 9.7 & 0.2$^{b}$ & $0^{+0.8}_{-0.0}$ & - \\ \hline
C & 23:32:41 & -54:08:20 & 10.0 & 0.71$^{b}$ & $1.2^{+4.8}_{-1.2}$ & - \\ \hline
D & 23:30:31 & -53:51:50 & 18.4 & 0.24$^{b}$ & $5.0^{+2.9}_{-1.8}$ & - \\ \hline \hline
SPT-CL J2355--5056 \\ \hline
A & 23:55:26 & -50:53:26 & 4.0 & 0.3$^{b}$ & $6.7^{+2.9}_{-2.1}$ & - \\ \hline
B & 23:55:10 & -50:56:26 & 6.0 & 0.3$^{b}$ & $1.2^{+1.7}_{-1.2}$ & - \\ \hline
C & 23:56:33 & -50:57:12 & 7.4 & 0.24$^{b}$ & $4.5^{+1.7}_{-1.7}$ & - \\ \hline
D & 23:55:38 & -51:08:34 & 13.0 & 0.75$^{b}$ & $24.3^{+6.9}_{-12.8}$ & - \\ \hline
E & 23:55:57 & -51:10:15 & 14.8 & 0.11$^{b}$ & $0.4^{+1.2}_{-0.4}$ & APMCC 936 \citep{1997MNRAS.289..263D} \\ \hline 
F & 23:54:20 & -51:09:10 & 19.3 & 0.42$^{b}$ & $2.2^{+4.5}_{-2.2}$ & - \\ \hline \hline
PLCKESZ G287.0+32.9 \\ \hline
A & 11:50:57 & -28:07:38 & 3.1 & 0.39$^{b}$ & $0.8^{+2.4}_{-0.8}$ & - \\ \hline \hline
PLCKESZ G292.5+22.0 \\ \hline
A & 12:01:11 & -39:54:44 & 2.5 & 0.3$^{b}$ & $0^{+2.9}_{-0.0}$ & - \\ \hline
B & 12:01:22 & -39:51:22 & 3.3 & 0.3$^{b}$ & $0^{+1.0}_{-0.0}$ & - \\ \hline
C & 12:01:17 & -39:49:02 & 4.1 & 0.3$^{b}$ & $0.9^{+2.7}_{-0.9}$ & - \\ \hline
D & 12:01:19 & -39:46:20 & 6.6 & 0.3$^{b}$ & $10.0^{+4.2}_{-4.4}$ & - \\ \hline
E & 12:01:35 & -40:04:39 & 13.5 & 0.43$^{b}$ & $7.0^{+4.0}_{-3.1}$ & - \\ \hline \hline
MACS J0416.1--2403  \\ \hline
A & 04:16:22 & -24:15:16 & 11.6 & 0.42$^{b}$ & $3.9^{+2.0}_{-1.9}$ & - \\ \hline
B & 04:16:55 & -24:16:04 & 16.4 & 0.23$^{b}$ & $0^{+0.8}_{-0.0}$ & - \\ \hline
C & 04:14:58 & -24:12:01 & 18.1 & 0.41$^{b}$ & $0.6^{+1.4}_{-0.6}$ & - \\ \hline \hline
PSZ1 G168.02--59.95 \\ \hline
A & 02:15:12 & -04:44:51 & 13.3 & 0.2950$^{d}$ & $3.0^{+1.2}_{-1.4}$ & - \\ \hline
B & 02:15:28 & -04:40:41 & 13.5 & 0.3503$^{d}$ & $1.8^{+1.4}_{-1.3}$ & - \\ \hline
\end{tabular}
\end{center}
\caption{Sources for redshifts: (a) \citet{2009ApJ...698.1221M}, (b) red galaxy colour method of Section~\ref{sec:redz}, (c) \citet{2010ApJ...723.1736H}, (d) spectroscopic redshift from SDSS, (e) photometric redshift (this work), (f) \citet{2007ApJ...660..239K}. Additional notes: (1) this component is close to an extended X-ray source reported by \citet{2011ApJ...738...48A}.}
\label{tbl:companions}
\end{table*}

\begin{table*}
\begin{center}
\begin{tabular}{|l|r|r|r|r|r|r|}
\hline
ID & R.A.     & dec & r [arcmin] & z & $M_{200m} [10^{14} h_{70}^{-1} \Msol]$ & external reference \\ \hline \hline
PSZ1 G230.73+27.70 \\ \hline
\multirow{2}{*}{A} & \multirow{2}{*}{09:01:45} & \multirow{2}{*}{-01:38:22} & \multirow{2}{*}{3.7} & \multirow{2}{*}{0.2944$^{f}$} & \multirow{2}{*}{$0.0^{+1.3}_{-0.0}$} & MaxBCG J135.43706--01.63946 \\ & & & & & & \citep{2007ApJ...660..239K} \\ \hline
B & 09:01:45 & -01:42:36 & 4.9 & 0.25$^{e}$ & $2.9^{+2.3}_{-2.9}$ & ZwCl 0859.2-0130 \citep{1961cgcg.book.....Z} \\ \hline
\multirow{2}{*}{C} & \multirow{2}{*}{09:02:22} & \multirow{2}{*}{-01:27:32} & \multirow{2}{*}{17.1} & \multirow{2}{*}{0.2998$^{f}$} & \multirow{2}{*}{$0.0^{+1.6}_{-0.0}$} & MaxBCG J135.59007--01.46456 \\ & & & & & & \citep{2007ApJ...660..239K} \\ \hline \hline
PSZ1 G099.84+58.45 \\ \hline
A & 14:14:48 & 54:55:59 & 8.9 & 0.2285$^{d}$ & $0.9^{+1.3}_{-0.9}$ & - \\ \hline
B & 14:14:22 & 54:57:10 & 10.7 & 0.1581$^{d}$ & $2.6^{+1.4}_{-1.4}$ & - \\ \hline
C & 14:13:07 & 54:58:36 & 18.4 & 0.77$^{e}$ & $6.4^{+7.5}_{-5.7}$ & - \\ \hline
D & 14:12:49 & 54:31:04 & 23.4 & 0.69$^{e}$ & $1.8^{+6.2}_{-1.8}$ & - \\ \hline 
\end{tabular}
\end{center}
\caption{Continued from Table~\ref{tbl:companions}.}
\end{table*}

\addcontentsline{toc}{chapter}{Bibliography}
\bibliographystyle{mn2e}
\bibliography{literature}

\label{lastpage}

\end{document}